\documentclass[12pt]{iopart}
\usepackage{iopams}
\usepackage{dcolumn}
\usepackage{float}
\usepackage{graphicx}
\usepackage{color}

\definecolor{b}{rgb}{0.3,0.3,0.9}
\definecolor{g}{rgb}{0.3,0.9,0.3}
{\color{b} }

\def \LSCO {La$_{2-x}$Sr$_x$CuO$_4$ }

\def \LBCO {La$_{2-x}$Ba$_x$CuO$_4$ }
\def \LBCOa {La$_{2-x}$Ba$_x$CuO$_4$}

\def \YBCO {YBa$_2$Cu$_3$O$_{6+x}$ }
\def \YBCOa {YBa$_2$Cu$_3$O$_{6+x}$}

\def \BISCCO {Bi$_2$Sr$_2$CaCu$_2$O$_{8+\delta}$ }
\def \BISCCOa {Bi$_2$Sr$_2$CaCu$_2$O$_{8+\delta}$}

\def \TL2201 {Tl$_2$Ba$_2$CuO$_{6+\delta}$ }

\def \etal {{\it et al.\ }}
\def \cm {cm$^{-1}$}
\begin{document}

\title{Bosons in high temperature superconductors: an experimental survey}

\author{Jules P. Carbotte$^a$, Thomas Timusk$^a$ and Jungseek Hwang$^b$}

\address{$^a$ Department of Physics and Astronomy, McMaster University, Hamilton ON L8S 4M1, Canada\\
$^b$ Department of Physics, Pusan National University, Busan 609-735, Republic of Korea}

\ead{timusk@mcmaster.ca, jhwang@pusan.ac.kr}

\begin{abstract}
We review a number of experimental techniques that are beginning to reveal fine details of the bosonic spectrum $\alpha^2F(\Omega)$ that dominates the interaction between the quasiparticles in high temperature superconductors. Angle-resolved photo emission (ARPES) shows kinks in electronic dispersion curves at characteristic energies that agree with similar structures in the optical conductivity and tunnelling spectra. Each technique has its advantages.  ARPES is momentum resolved and offers independent measurements of the real and imaginary part of the contribution of the bosons to the self energy of the quasiparticles.  The optical conductivity can be used on a larger variety of materials and with the use of maximum entropy techniques reveals rich details of the spectra  including their evolution with temperature and doping.  Scanning tunnelling spectroscopy offers spacial resolution on the unit cell level.  We find that together the various spectroscopies, including recent Raman results, are pointing to a unified picture of a broad spectrum of bosonic excitations at high temperature which evolves, as the temperature is lowered into a peak in the 30 to 60  meV region and a { featureless }high frequency background in most of the materials studied. This behaviour is consistent with the spectrum of spin fluctuations as measured by magnetic neutron scattering.  However, there is evidence for a phonon contribution to the bosonic spectrum as well.
\end{abstract}

\maketitle

\section{Introduction}
At the 25th anniversary of the discovery of high temperature superconductivity (Bednorz and M\"uller 1986) we still lack a complete understanding of this astonishing phenomenon. However, we can take a look back at the history of low temperature superconductivity to get some perspective on this apparent lack of progress. The original formulation of the Bardeen-Cooper-Schrieffer (BCS) theory (Bardeen \etal 1957a,1957b) treated the interaction between a pair of electrons with up and down spin of equal and opposite momentum at the Fermi energy  by a constant average matrix element. While it was clear from the isotope effect on the transition temperature $T_c$ that phonons were involved, it took high resolution tunnelling experiments by McMillan and Rowell (1965) to accurately resolve the underlying bosonic structure of the electron-phonon spectral density denoted by $\alpha^2F(\Omega)$ and relate it to the phonon spectrum as determined by inelastic neutron scattering (Brockhouse \etal 1962). A further important step was to calculate this function from band theory and phonon dynamics (Leung \etal 1976, Tomlinson \etal (1976) and Carbotte (1990) and more recently in MgB$_2$ by Choi \etal (2002a, 2002b), Golubov \etal (2002) and Brinkman \etal (2002)). Nevertheless, even with all this detailed understanding of low temperature superconductivity, we are {still} treated to surprising new superconductors, not predicted by theory, such as the recent discovery of MgB$_2$ with a $T_c$ of the order of 40 K (Nagamatsu \etal 2001).

The situation in the high $T_c$ cuprates is not at all clear. In a recent Science article,  P. W. Anderson (2007) raised the issue of the very existence of pairing "bosonic glue" in this case, be it phonons (Johnston \etal 2010a), perhaps modified by strong correlation physics (Kulic 2000) or some other boson exchange mechanism such as spin fluctuation (Chubukov \etal 2008) within a nearly antiferromagnetic fermi liquid model (Millis \etal 1990, Monthoux \etal 1994 and Branch \etal 1999).

Returning to the question of whether or not the Cooper pair binding may involve only high energy dynamics on the scale of the Hubbard $U$ ($>$2 eV) or the antiferromagnetic exchange coupling $J$ ($\sim 0.12$ eV) or rather on the smaller scale of the spin fluctuations, the work of Maier \etal (2008) is instructive. The critical question is one of dynamics { and the energy scale} associated with the pairing interaction. These authors find that for both Hubbard and t-J models the dominant contribution to the pairing reflect this smallest energy scale and so one can speak of a spin-fluctuation glue. It is useful to quantify their approach here.

We introduce a gap function $\phi(k, \omega)$ which depends on momentum $k$ as well as energy $\omega$ and has both real $\phi_1(k,\omega)$ and imaginary part $\phi_2(k,\omega)$. These are related by the Cauchy relation
\begin{equation}
\phi_1(k, \omega) = \frac{1}{\pi}\int_{-\infty}^{+\infty}\frac{\phi_2(k,\omega ')}{\omega '-\omega}d\omega '.
\label{Cauchy}
\end{equation}
A useful measure of the frequency dependence of the pairing is to take $\omega$ = 0 in equation \ref{Cauchy} and introduce
\begin{equation}
I(k,\Omega) =\frac{\frac{2}{\pi}\int_0^{\Omega}\frac{\phi_2(k,\omega ')}{\omega '}}{\phi_1(k,0)}.
\label{glueF}
\end{equation}
This quantity measures the contribution to the pairing coming from excitation energies in the range 0 to $\Omega$ and by arrangement, $I(k,\Omega)$  will saturate to value one at $\Omega \rightarrow \infty$.

\begin{figure}
        \begin{center}
         \vspace*{-1.5 cm}%
               \leavevmode
                \includegraphics[origin=c, angle=0, width=15cm, clip]{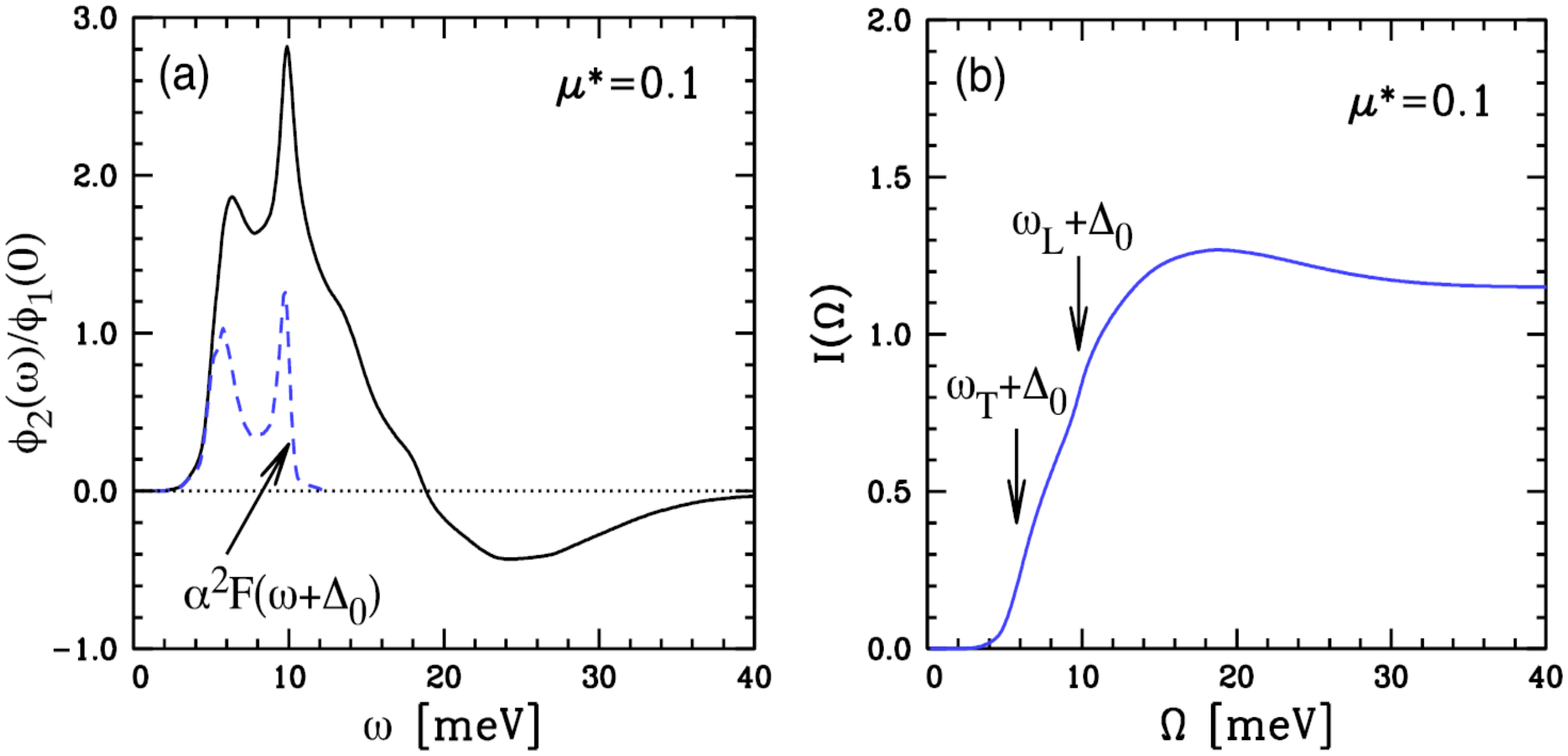}
         \vspace*{-2.5cm}%
        \end{center}
\caption{The role of  transverse $\omega_T$ and longitudinal $\omega_L$ phonons in mediating superconductivity in lead from Maier \etal (2008).  (a) Here  $\phi_2(\omega)$ is the imaginary part of the Pb gap function versus $\omega$ (solid curve). The dashed curve shows the bosonic spectral function $\alpha^2F(\omega)$ determined from tunnelling  but with the peaks in $\phi_2(\omega)$ shifted up by the gap $\Delta_0$.  (b) The cumulative  pairing interaction  $I(\Omega)$ versus $\Omega$ for Pb. $I(\Omega)$ reflects the transverse and longitudinal phonon contributions to the pairing. At larger values of $\Omega$, $I(\Omega)$ decreases  because $\phi_1(0)$ is reduced from the value that it would have just due to the phonons by the presence of the non-retarded screened Coulomb pseudopotential $\mu^*$.}
\label{fgIntro1}
\end{figure}

\begin{figure}
        \begin{center}
         \vspace*{-0.5 cm}%
               \leavevmode
                \includegraphics[origin=c, angle=0, width=15cm, clip]{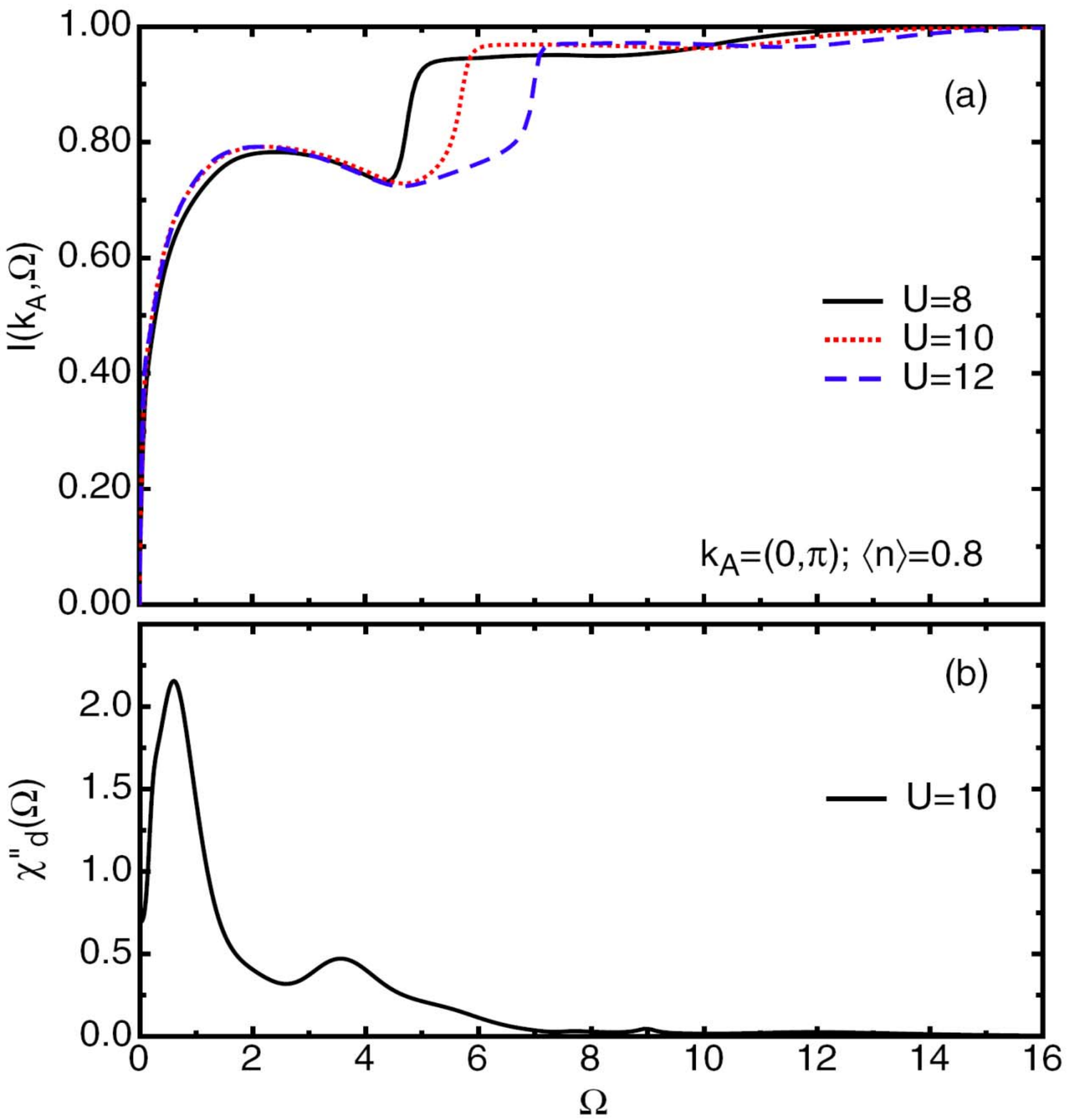}
         \vspace*{-1.5 cm}%
        \end{center}
\caption{(a) The cumulative pairing interaction spectral weight $I(k_A,\Omega)$ versus $\Omega/t$ for the Hubbard model from Maier \etal (2008). (b) The d-wave projected spin susceptibility $\chi''_d(\Omega)$ versus $\Omega/t$ for the same model.  We note that the major contribution to the gap function comes from low frequency excitations with an additional attractive interaction at the 5 \% level from high frequency excitations. }
\label{fgIntro2}
\end{figure}

figure \ref{fgIntro1} shows the results of Maier \etal (2008) for the case of the BCS superconductor Pb which can be described by an isotropic electron-phonon spectral density $\alpha^2F(\Omega)$ shown as the dashed spectrum in the left frame and a coulomb repulsion $\mu^*$ of 0.1 which accounts for the electron-electron interactions. This involves the introduction of a pseudopotential rather than a bare Coulomb repulsion as described by Morel and Anderson (1962). Solutions of the Eliashberg equations for the complex gap gives the solid curve for $\phi_2(\omega)/\phi_1(0)$. Results for $I(\Omega)$ are presented in the right side frame (b). We see structure at the sum of the gap $\Delta_0$ plus the main transverse phonon frequency $\omega_T$ in $\alpha^2F(\Omega)$ as well as at $\Delta_0+\omega_L$ with $\omega_L$ the main longitudinal frequency. Most of the glue occurs in this energy range. In fact $I(\Omega)$ becomes larger than the one before it gradually decreases to one because  of the nonretarded repulsive coulomb term $\mu^*$.

Turning to the cuprates in figure  \ref{fgIntro2}, we reproduce from Maier \etal (2008) their results for the Hubbard model. In this case the complex gap depends on momentum as well as energy $\omega$ and in the top frame (a) $k=k_A=(0,\pi)$ with a doping of $<\!\!n\!\!> =$ 0.8 where $<\!\!n\!\!> =$ 1 corresponds  to half filling. Three values of $U$ are used as labeled. It is clear that all cases show pairing glue with $I(k_A, \Omega)$ reaching $\sim 95$ \% of its saturated value for $\Omega/t \sim$ 7 where $t$ is the first nearest neighbour hopping parameter of the model. In the lower frame the d-wave projection of the spin susceptibility is plotted for $U =$ 10 and we note that its energy scale corresponds closely to the energy scale for the contribution of the pairing glue to the pairing strength  which suggests that spin fluctuations form the glue. There have been other similar calculations, among these the work of Kyung \etal (2009). They use a cellular dynamical mean-field theory for the two-dimensional Hubbard model to study retardation effects in d-wave superconducting pairing. They confirm that the appropriate energy scale involved coincides with the scale of the short-range spin fluctuation and correlates with the spin susceptibility. Earlier cluster dynamical mean-filed computations by Haule \etal (2007) had already produced a phase diagram for the d-wave order parameter which follows the dome behaviour as a function of doping seen for the value of the critical temperature $T_c$. Additional work by Kancharla \etal (2008) gives more details on the competition between superconductivity and antiferromagnetism in doped Mott insulators. These results provide in part, a motivation for the experimental investigation of possible boson effects in the high $T_c$ oxides as will be reviewed here.

The role of the electron-phonon interaction in the cuprates and its effect on boson structure has been addressed in many theoretical papers. Here again it is not our aim to provide a comprehensive review. Kulic (2000) presents the case for the electron-phonon interaction in strongly correlated systems such as the cuprates. Due to the reduction in screening, a forward scattering peak appears in the electron-phonon interaction while at the same time backward scattering is suppressed. Kulic concludes that under such conditions the mass enhancement factor $\lambda$ associated with the d-channel which is the important quality for estimates of the value of the critical temperature, is of the same order as for the s-wave or renormalization channel. Also, the transport electron-phonon coupling is much smaller than its quasiparticle counterpart which provides a direct explanation of the observation that the resistivity is linear in $T$ in the normal state above $T_c$ (Zeyher 1996). Recent local density approximation (LDA) calculations in YBCO (Heid \etal 2008, 2009) and in LSCO (Giustino \etal 2008) have found however, very small values of the coupling in the d-channel and conclude that the electron-phonon interaction is much too small to provide the glue for Cooper pair formation. Also quasiparticle mass enhancement factor renormalizations are small as is its transport counterpart. Renormalization due to electron correlations beyond LDA are not expected to greatly enhance these values (Zeyher \etal 1996) although some have argued that this need not necessarily be the case (Reznik \etal 2008). A review emphasizing the role of phonons in ARPES has been given by Johnston \etal (2010b). A discussion of a possible combined role of spin fluctuations plus a phonon part is found in Johnston \etal (2010a). Another approach to the origin of kinks in ARPES spectra is that of Byczuk \etal (2007) who discuss the role of Hubbard subbands in strongly correlated electronic systems. Bauer and Sangiovanni (2010) consider dispersion kinks in a Hubbard model supplemented with electron-phonon coupling in a Holstein model using dynamical mean-field theory.  They also consider the effect of including antiferromagnetic correlations on the electron phonon interaction and find it enhances the phonon effect in the electronic dispersions. As described by Datta \etal (2007), kinks also exist in the "effective"  dispersion curves defined by the peaks in the hole spectral density at fixed $\omega$ in the momentum distribution curve of ARPES within Tomonaga-Luttinger liquid models which exhibit spin-charge separation. In view of all these possible effects, it is important to turn to experiment to understand better what signatures of boson structure are present and how they might be interpreted. This is the main aim of this review.

To get fine details of the electron-boson interaction tunnelling has been the spectroscopy of choice in the past. However, the same spectral density can be obtained by other experimental techniques. For example, for the elemental BCS superconductor lead, optics has given (Joyce \etal 1970 and Farnworth \etal 1974, 1976) results for $\alpha^2F(\Omega)$ in excellent agreement with calculations of $\alpha^2F(\Omega)$ (Tomlinson \etal 1976) and with tunnelling (McMillan \etal 1965). In addition to optics a spectroscopic technique that has been particularly important in the study of the cuprate high temperature superconductors is angle resolved photoemission spectroscopy (ARPES). More general reviews of this powerful technique already exist (Damascelli \etal 2003, Campuzano \etal (2004), Eschrig 2006, Chubukov \etal 2008, Garcia and Lanzara 2010 and Kordyuk \etal 2010). While optics and ARPES will be our main focus, scanning tunnelling spectroscopy (STS) (for review see Fischer \etal 2007), break junction tunnelling will also be considered as well as the very recent Raman scattering data (Muschler \etal 2010).

\section{Angle Resolved Photoemission Spectroscopy}

\begin{figure}
        \begin{center}
         \vspace*{-2.5 cm}%
               \leavevmode
                \includegraphics[origin=c, angle=0, width=15cm, clip]{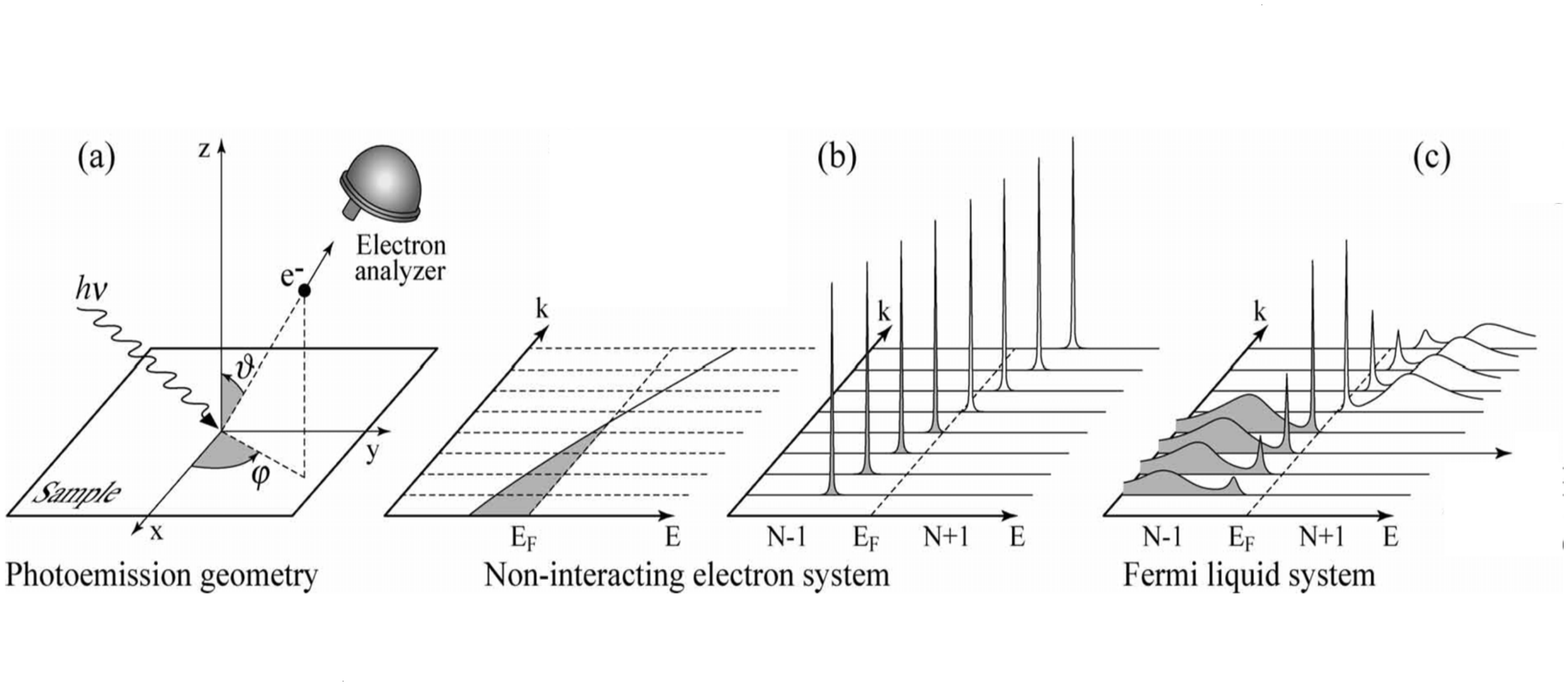}
         \vspace*{-3.5 cm}%
        \end{center}
\caption{ARPES spectroscopy from Damascelli  \etal (2003); a) a photon is absorbed by the sample and emits a photoelectron in the direction specified by the angles
$\theta$ and $\phi$. Its momentum parallel to the surface equals the momentum of the hole in the filled band (grey area in (b)). An analyzer has a slit that accepts electrons with different $k$ vectors and spreads them in energy with an electric field perpendicular to the slit. The electrons are collected in a fluorescent screen that will display a two-dimensional plot of the energy vs. momentum along the slit. b) shows  a non-interacting electron system (with a single energy band dispersing across $E_F$ ); (c) an interacting Fermi liquid system. Note the broad sideband due to bosonic interactions in addition to the sharp peak.}
\label{fgARPES0}
\end{figure}

\begin{figure}
        \begin{center}
         \vspace*{-0.5 cm}%
               \leavevmode
                \includegraphics[origin=c, angle=0, width=15cm, clip]{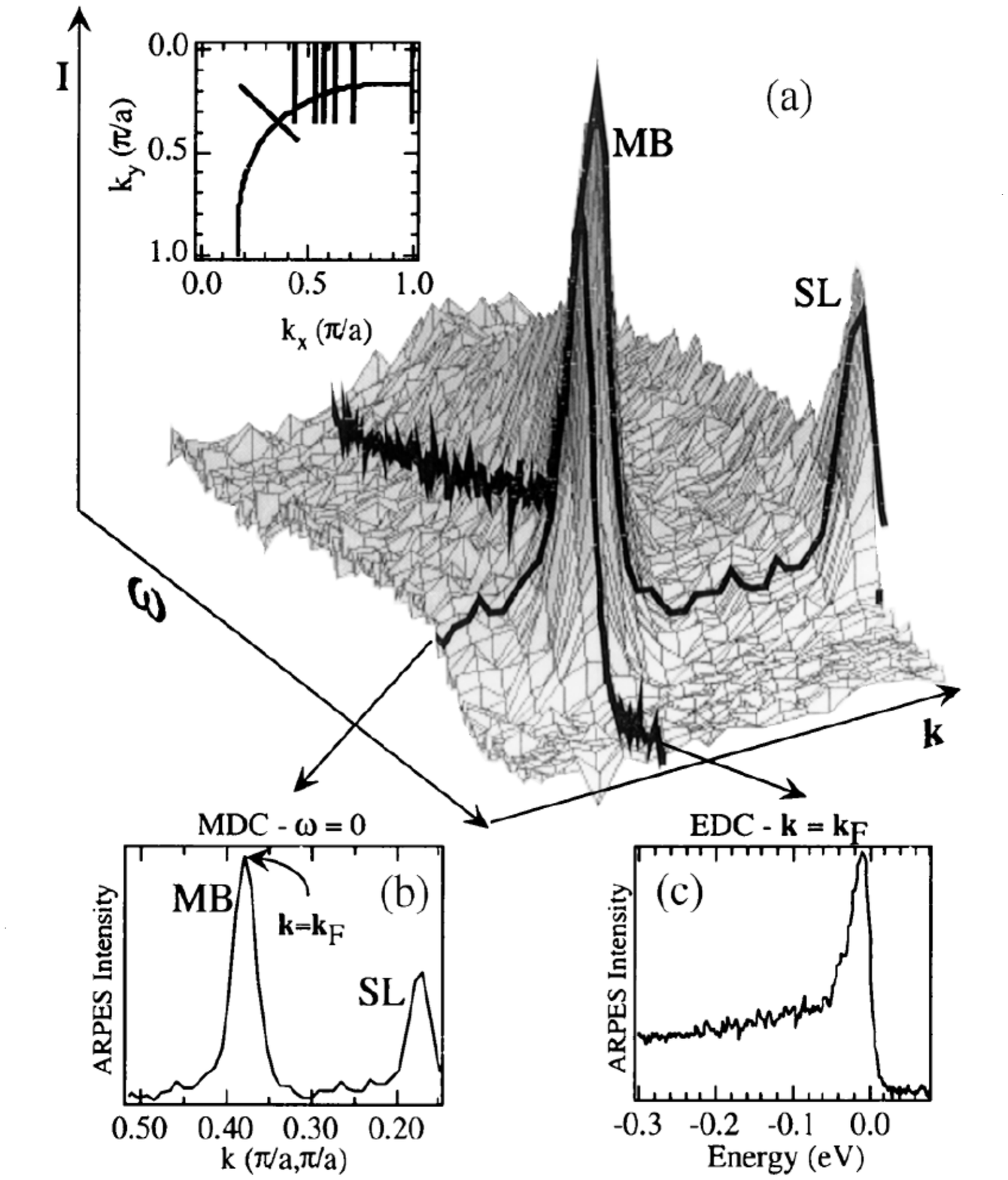}
         \vspace*{-1.0 cm}%
        \end{center}
\caption{(a)  The ARPES intensity $I({\bf k},\omega)$ for optimally doped \BISCCO  from Kaminski \etal (2001) for ${\bf k}$ along the diagonal line in the $k_x,k_y$ plane of the Brillouin zone shown in the inset. The curved line is the Fermi surface. MB is the main band and SL a superlattice image. (b) and (c) show respectively  a constant $\omega$ cut (MDC) and constant ${\bf k}$ cut (EDC) from (a).}
\label{fgARPES1}
\end{figure}

Modern angle resolved photoemission spectroscopy is based on the classical photoelectric effect. A high energy photon impinges on the surface of a single crystal cleaved {\it in situ} in ultrahigh vacuum. The energy of the photo-ejected electron as well as the direction of its velocity are determined. Conservation of momentum demands that the momentum component  parallel to the crystal surface of the emerging electron must be equal and opposite to the momentum of the hole created in the sample. In this way the energy and momentum relationship of the occupied states below the Fermi surface of a two-dimensional system can be mapped out. This is illustrated in figure  \ref{fgARPES0} from Damascelli \etal (2030). At present a resolution in energy of better than 1 meV and 0.1 \% in angle is
possible (Garcia and Lanzara 2010). The photocurrent magnitude depends on the product of three factors. The first is a dipole matrix element $M({\bf k},\omega)$ that depends on the initial and final electronic states, incident photon energy and polarization. While some attention needs to be paid to this factor, the other two factors are of most general interest and combine to make the technique a powerful tool that yields a detailed picture of the energy and momentum dependence of the electronic structure of the occupied states below the Fermi surface. The first of these factors is the Fermi Dirac (FD) distribution function $f(\omega)$ which states that at zero temperature only states below the Fermi energy $E_F$ are accessible and at finite $T$ the thermal tail of the FD distribution provides some limited information on unoccupied states. The second factor, which is the most important one, is the spectral density of the charge carriers $A({\bf k},\omega)$.  It is related to the imaginary part of the electron Green's function $G({\bf k},\omega)$.  The photo current intensity  $I({\bf k},\omega)$ is given by the product of the three factors:
\begin{equation}
I({\bf k},\omega) = |M({\bf k},\omega)|^2 f(\omega) A({\bf k},\omega)
\label{Intensity}
\end{equation}
In terms of the electronic self energy $\Sigma({\bf k},\omega)$ the spectral density is written as
\begin{equation}
A({\bf k},\omega) = -{1 \over \pi} {\Sigma''({\bf k},\omega) \over {[\omega - \epsilon_{\bf k} - \Sigma'({\bf k},\omega)]^2 + \Sigma''({\bf
k},\omega)^2}}
\label{SpectFunc}
\end{equation}
where $\Sigma'$ and $\Sigma''$ are respectively the real and imaginary parts of $\Sigma({\bf k},\omega)$. For free electrons, $A({\bf
k},\omega)$ reduces to a Dirac delta function on the band structure dispersion curves, $\epsilon_{{\bf k}}$ vs. ${\bf k}$ as shown in figure \ref{fgARPES0}b). When interactions are included, the quasiparticle energies are renormalized and they acquire a finite lifetime and an incoherent side band develops figure \ref{fgARPES0}c). The new energies are given by the solution of the equation
\begin{equation}
E_{\bf k} - \epsilon_{{\bf k}} - \Sigma'({\bf k},\omega)=0
\label{EnergyCons}
\end{equation}
and these can be measured in ARPES experiments. In figure  \ref{fgARPES1} we show results (Kaminski \etal 2001) for the ARPES intensity $I({\bf k},\omega)$ as a function of ${\bf k}$ and $\omega$ for incident photon energy $h\nu = 22$ eV and temperature $T= 40$ K. Frame (a) gives a 3-D plot while frames (b) and (c) give respectively a constant energy curve called the Momentum Distribution Curve (MDC) and a constant ${\bf k}$ cut called the Energy Distribution Curve (EDC). MB denotes the main band SL is a superlattice image. For (b) the frequency $\omega=0$ and hence the peak in the curve is at ${\bf k}={\bf k}_F$ (the Fermi momentum in this direction). Data on ${\bf k}_F$ in different directions in the Brillouin zone will determine the Fermi surface. The line in the MDC spectrum is symmetric and close to Lorentzian in shape with only a small background. By contrast (c) is far from Lorentzian and has a large tail extending to a high binding energy (negative values). {These tails, also called the "dip-hump" structure reflect inelastic processes where, in addition to the creation of a hole in the valence band, a bosonic excitation has been created.  An analysis of these tails can yield information on the bosonic spectrum that interacts with the holes. figure \ref{fgARPES0} (c) gives a schematic illustration of these processes. The widths of the MDC's can be analyzed to yield frequency dependent scattering rates $1/\tau(\omega)$ which are proportional to the imaginary part of $\Sigma({\bf k},\omega)$. Modelling of the self energy with a synthetic bosonic spectrum to get a fit to the dispersion curves can be used to yield the real part of $\Sigma({\bf k},\omega)$.}

\begin{figure}
        \begin{center}
         \vspace*{-0.5 cm}%
                \leavevmode
                \includegraphics[origin=c, angle=0, width=20 cm, clip]{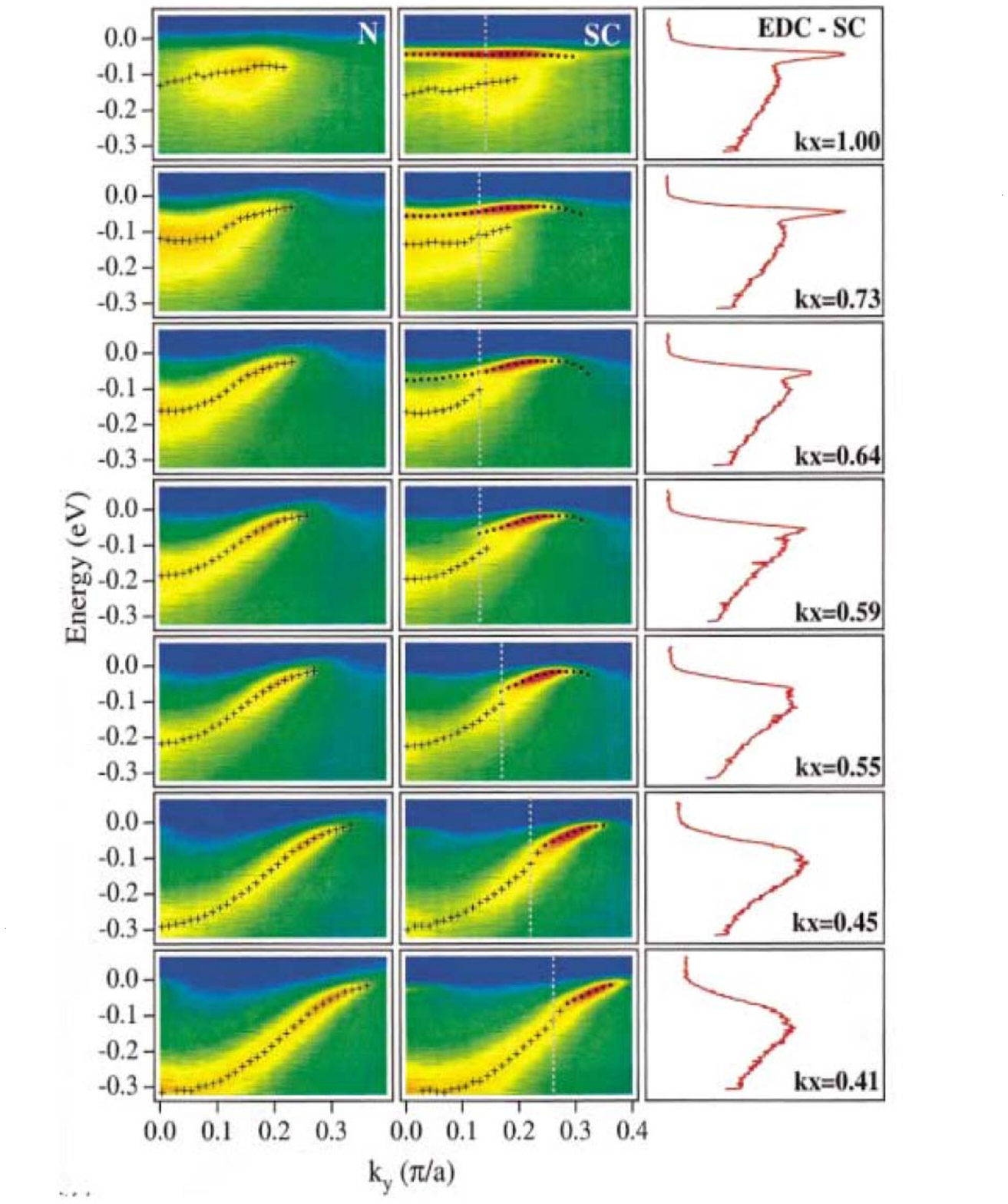}
         \vspace*{-1.3 cm}%
        \end{center}
\caption{The variation of ARPES  intensity along selected cuts in the ${\bf k_y}$ direction for optimally doped \BISCCO (shown in the zone inset of figure \ref{fgARPES1}) Left panel, the normal state ($T=140$ K), middle panel the superconducting state  ($T = 40$ K). Right panels: EDCs at locations marked by the vertical lines in the middle panels. From Kaminski \etal (2001). These data show the increase in coupling to a bosonic mode at $\approx 70$ meV as one moves from the nodal direction (lower panels) towards the antinodal direction, (upper panels) { both in the normal and superconducting states}.}
\label{fgARPES2}
\end{figure}

While the MDCs are more straight forwardly interpretable using equation \ref{SpectFunc} with the inverse quasiparticle  lifetime $1/\tau$ related to $\Sigma''({\bf k},\omega)$, historically the  EDCs  were first used in the interpretation of ARPES dispersion curves.  In figure \ref{fgARPES2} results for the dispersion curves are reproduced from Kaminski \etal (2001). From top to bottom, we go from the
antinodal to the nodal direction along lines shown in the inset of figure  \ref{fgARPES1} (a) where one quarter of the Brillouin zone is shown along with the Fermi contour (solid black curve). { The left panels show the normal state dispersion curves. Note the changes in slope at low energy which are signatures of a kink in dispersion, a signature of interaction with bosonic modes.} In the middle panels for the superconducting state {  these features become stronger and} we see two dispersion branches which overlap in momentum with the lower energy branch showing almost no dispersion (change of energy) with ${\bf k}$ while the higher energy branch varies more, although it too remains quite flat. These dispersions correspond to the energy of the sharp peaks seen in the EDC (right panel) while the second branch corresponds to the "hump" separated from the larger sharp peak at lower energy by the "dip" (Engelsberg and Schrieffer (1963)). We note that the evolution of the dispersion curves in the superconducting state (SC): there is a gradual closing of the gap between the two branches until a single branch emerges but with a distinct "kink" remaining in the nodal direction.  Two observations are made by Kaminski \etal (2001). First, the energy of the gap structure which evolves into the kink is reasonably fixed along the Fermi surface. Interpreting the kink as a signature of an interaction with a bosonic excitation suggests that this excitation has a fixed and well defined energy. Second, the normal state dispersions are comparatively smooth which is taken as evidence that the mechanism involved is coupling to a collective mode which vanishes at temperature $T = T_c$, as does the well known spin resonance seen in inelastic neutron scattering at 41 meV in optimally doped \YBCOa. It is also clear from this figure and this is noted by Kaminski \etal (2001) that the coupling to the collective mode increases as one progresses from the nodal to the antinodal direction. We need to keep this in mind in the discussions of the origin of the boson involved. However, as we have already noted, a more careful examination of the normal state dispersions, and as we will see below in more detail, shows that strong bosonic interactions are present in the normal state.

The peak-dip-hump structure in the antinodal direction (at $(\pi,0)$ in the Brillouin zone) has a long history (Dessau \etal 1991, 1992,
Shen and Dessau 1995, and Ding \etal 1996). It was interpreted by many to be distinctly related to the coupling to the spin-one resonance
seen in inelastic magnetic neutron scattering experiments. An early formulation was given by Norman \etal (1997) and a detailed comparison between the charge carrier self energy due to a sharp boson mode of momentum $(\pi,\pi)$ and the results in figure \ref{fgARPES2} was made (Eschrig and Norman 2000) with excellent qualitative agreement noted. An independent but closely related formulation of the relation between ARPES data and the neutron resonance peak was given by Abanov and Chubukov \etal (1999) with much the same conclusions. This work was questioned by Kee \etal (2002) on the grounds that the spin resonance had too small a spectral weight and coupling to charge carriers to have a strong effect on the electronic self energy.  A response by Abanov \etal (2002) provides results of calculations that show that even though the resonance has a small spectral weight, this weight is confined to a small region in momentum space and energy and consequently this resonance can provide clear signatures in electronic properties. More details can be found in Eschrig and Norman (2003). An important fact underlying these calculations is the disappearance of the resonance at $T_c$ in many systems although for underdoped materials a remanence remains up to a higher temperature $T^*$, the pseudogap energy scale. The evolution of the self energy as one moves from the $(\pi,0)$ to the $(\pi,\pi)$ direction is also an important feature of the data.  Later we will return to this issue and
discuss how eventually it was recognized that bilayer splitting can have a strong effect on ARPES results in the antinodal direction and this needs to be accounted for before the dispersion curves can be analyzed for self energy effects. Modifications of the picture just described are needed but some elements of the explanation given remain valid. Finally, we note the experimental study of Campuzano \etal (1999) extended the work to doping as well as momentum and temperature dependencies and found that the hump scales with the peak and persists above  $T_c$ in the pseudogap state. Also the inferred mode energy has the same doping dependence as does the magnetic resonance peak position seen in inelastic neutron scattering (Fong \etal 1999 and He \etal 2001). They conclude that the peak-dip-hump structure arises from the electronic interaction with a collective mode with a wave vector ($\pi,\pi$). As we will see below,  the sharp resonance that appears at $T_c$ is only a part of the bosonic spectrum.  In addition, there is a broad background of excitations that is present in the normal state and will be the ultimate trigger of superconductivity at $T_c$. However, before we review further the recent ARPES studies of the bosonic spectra, we must address  the problem of bilayer splitting which needs to be considered as it could mask intrinsic self energy effects.

\begin{figure}
        \begin{center}
         \vspace*{-1.0 cm}%
                \leavevmode
                \includegraphics[origin=c, angle=0, width=13cm, clip]{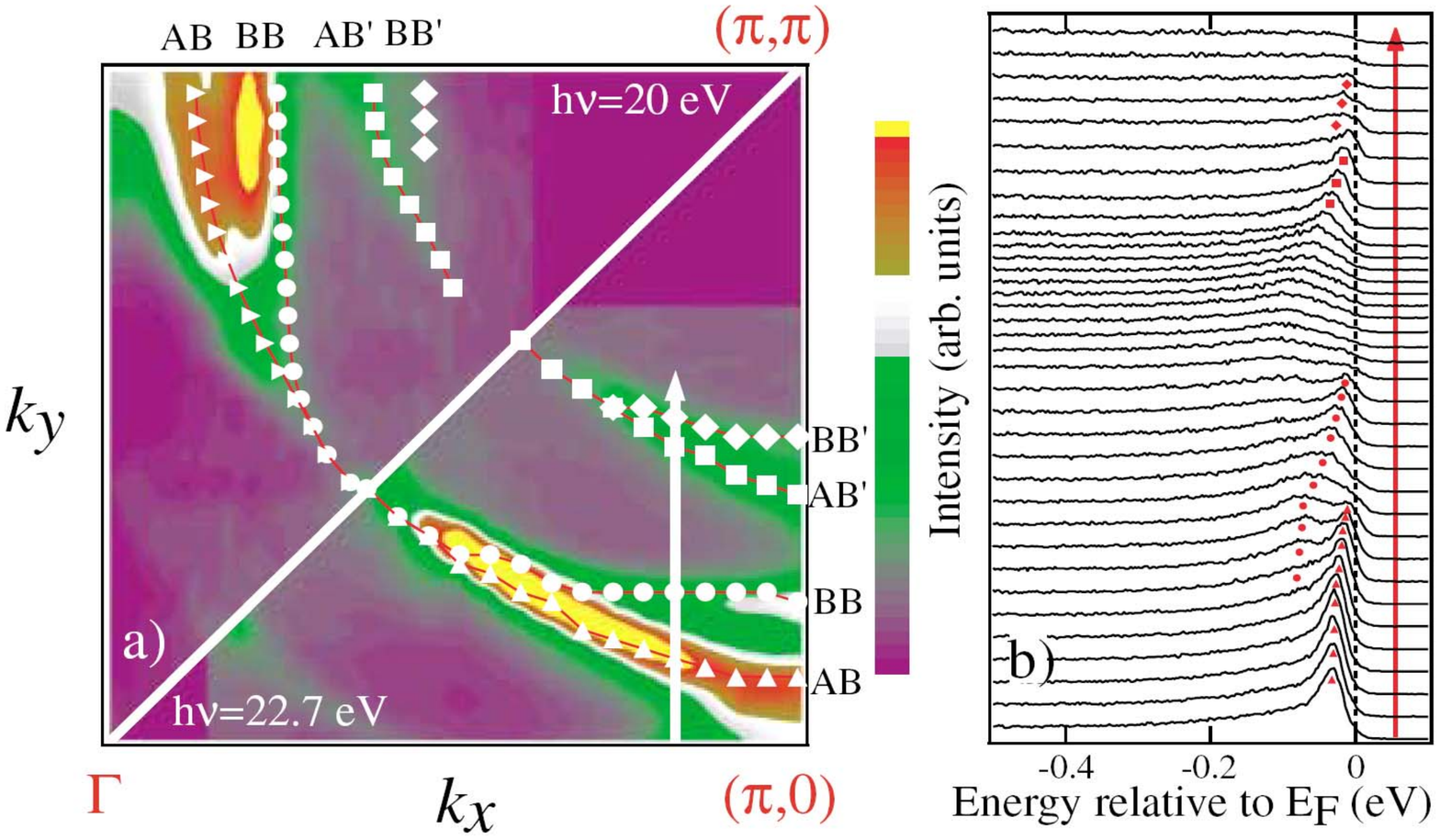}
         \vspace*{-2.0 cm}%
        \end{center}
\caption{The effect of bilayer splitting on ARPES data from Feng \etal (2001) on overdoped \BISCCO . (a) False colour plot of the spectral weight mapping near $E_F$ ([-20 meV, 10 meV]) of OD65 taken at 22.7 eV (lower right half, $T = 75$ K) and 20 eV (upper left half, $T= 80$ K) (note they are from different experiments). The Fermi surface is plotted for antibonding states (AB, triangles), bonding states (BB, circles). A second set of states arise from superstructure with antibonding states (AB', squares), and bonding states (BB', diamonds). (b) ARPES spectra along the cut indicated by the arrow in (a).}
\label{fgARPES17}
\end{figure}

In systems with two copper oxygen layers such as \BISCCOa, the family most studied with ARPES, the degeneracy of the two bands is split into a bonding and an antibonding band. An unfortunate coincidence is that this splitting energy is of the same order as the energy of the bosonic excitations discussed previously. Illustrated in figure  \ref{fgARPES17} are the ARPES results by Feng \etal (2001) on the Fermi surface of overdoped Bi2212 ($T_c=65$ K). We see clearly the bonding (BB) and the antibonding (AB) bands as well as their superstructure replicas. Note that in this sample no splitting is observed in the nodal direction.

\begin{figure}
        \begin{center}
         \vspace*{-1.5 cm}%
                \leavevmode
                \includegraphics[origin=c, angle=0, width=15cm, clip]{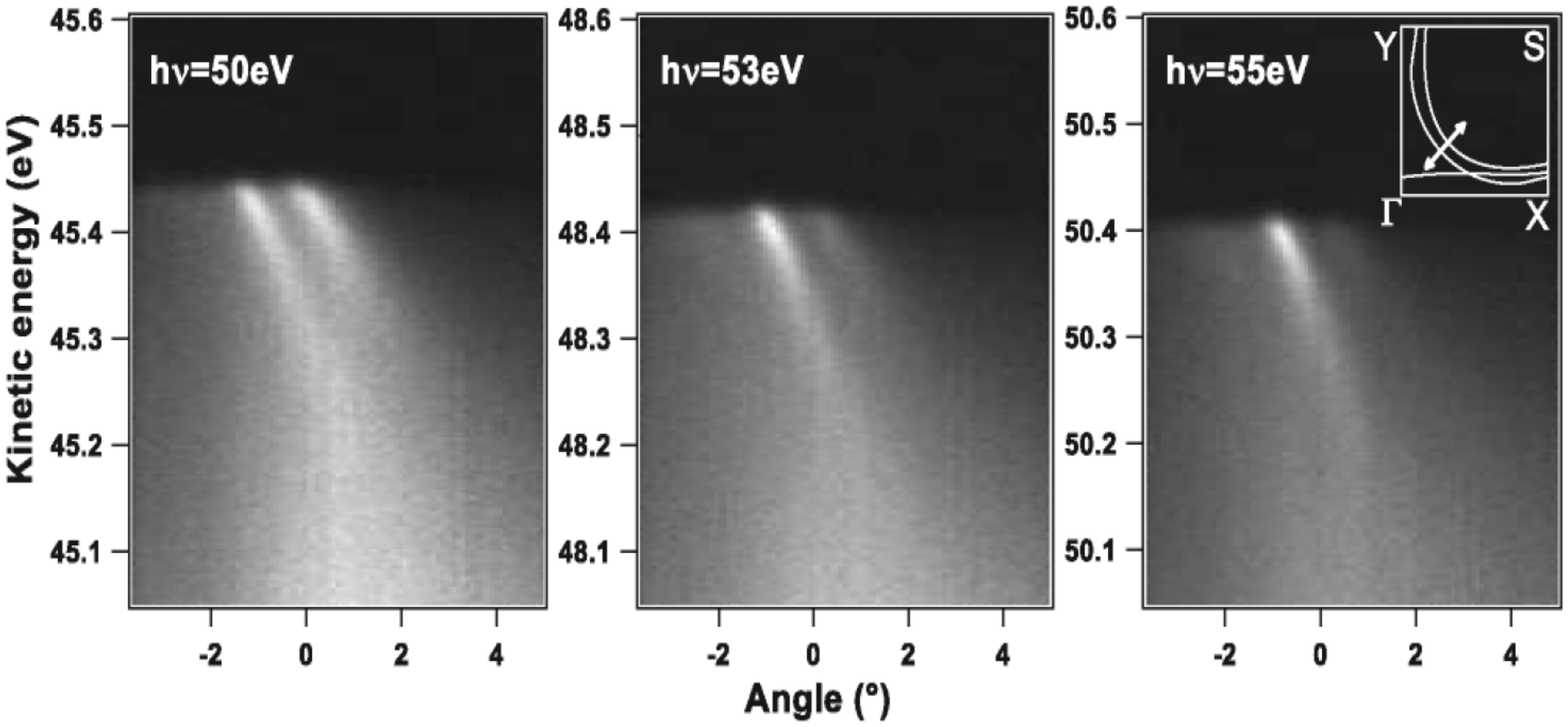}
         \vspace*{-3.0 cm}%
        \end{center}
\caption{The effect of incident photon energy on ARPES photoemission intensity in YBCO. Note the decrease in the antibonding band intensity at 53 and 55 eV incident photon energy.  The inset schematically shows the LDA-predicted Fermi surface and a cut in the $k$ space along which the data have been taken. From Borisenko \etal (2006).}
\label{fgARPES18}
\end{figure}

However, in figure \ref{fgARPES18} we show nodal direction renormalized electronic dispersion curves as a function of angle (instead of momentum) for
\YBCO from the work of Borisenko \etal (2006) for an underdoped sample UD35 with $T_c \approx 35$ K at a temperature of 30 K. Three photon energies $h\nu =50$ eV, 53 eV, and 55 eV are employed. In the left frame (photon
energy 50 eV) we clearly see two different dispersion curves, bonding and antibonding. With increasing photon energy  a single bonding band dominates  in the right hand frame. This shows that photon energy has a significant effect on the photoemission matrix
elements that determine the admixture of the antibonding vs. the bonding bands. It should be clear that different admixtures of bonding and antibonding bands
could be mistaken for self energy effects when in reality the structure is due to bilayer splitting.  For additional discussion of bonding and antibonding bands  see among others Chuang \etal (2001), Yamasaki \etal (2007), Kordyuk \etal (2002), Borisenko \etal (2003), Kim \etal (2003) and for a review Kordyuk \etal (2010).

\begin{figure}
        \begin{center}
         \vspace*{-2.0 cm}%
                \leavevmode
                \includegraphics[origin=c, angle=0, width=15 cm, clip]{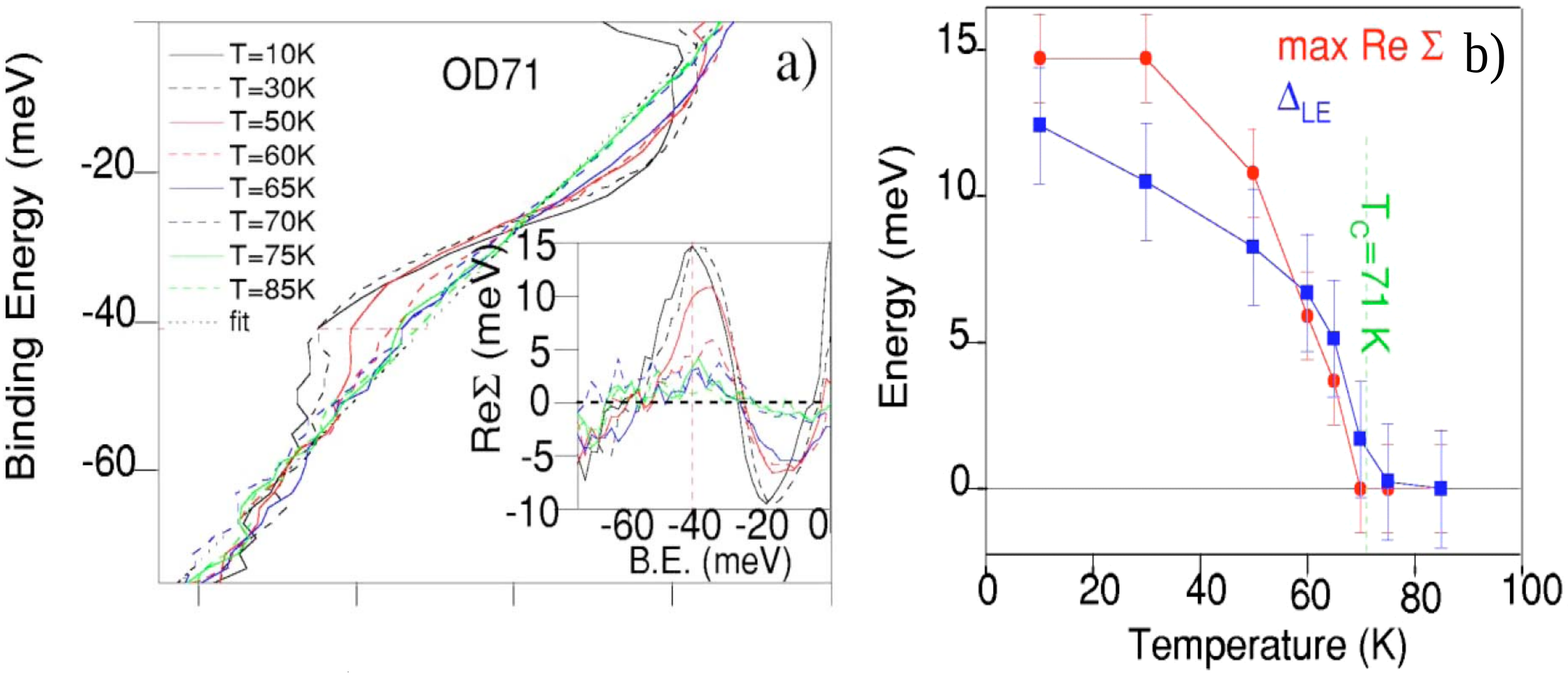}
         \vspace*{-3.1 cm}%
        \end{center}
\caption{The $(\pi,0)$ kink strength from Gromko \etal (2003) and its variation with temperature. (a) MDC dispersion from overdoped \BISCCO along $(\pi,0)-(\pi,\pi)$. The inset shows  Re$\Sigma$ after a linear term has been subtracted from the data showing a kink energy is $\approx40$ meV. (b) Temperature dependence of the amplitude of the maximum in Re$\Sigma$ (red circles) from panel (a) and the superconducting gap $\Delta_{LE}(T)$ (blue squares).}
\label{fgARPES21}
\end{figure}

\begin{figure}
        \begin{center}
         \vspace*{-2.2 cm}%
                \leavevmode
                \includegraphics[origin=c, angle=0, width=15 cm, clip]{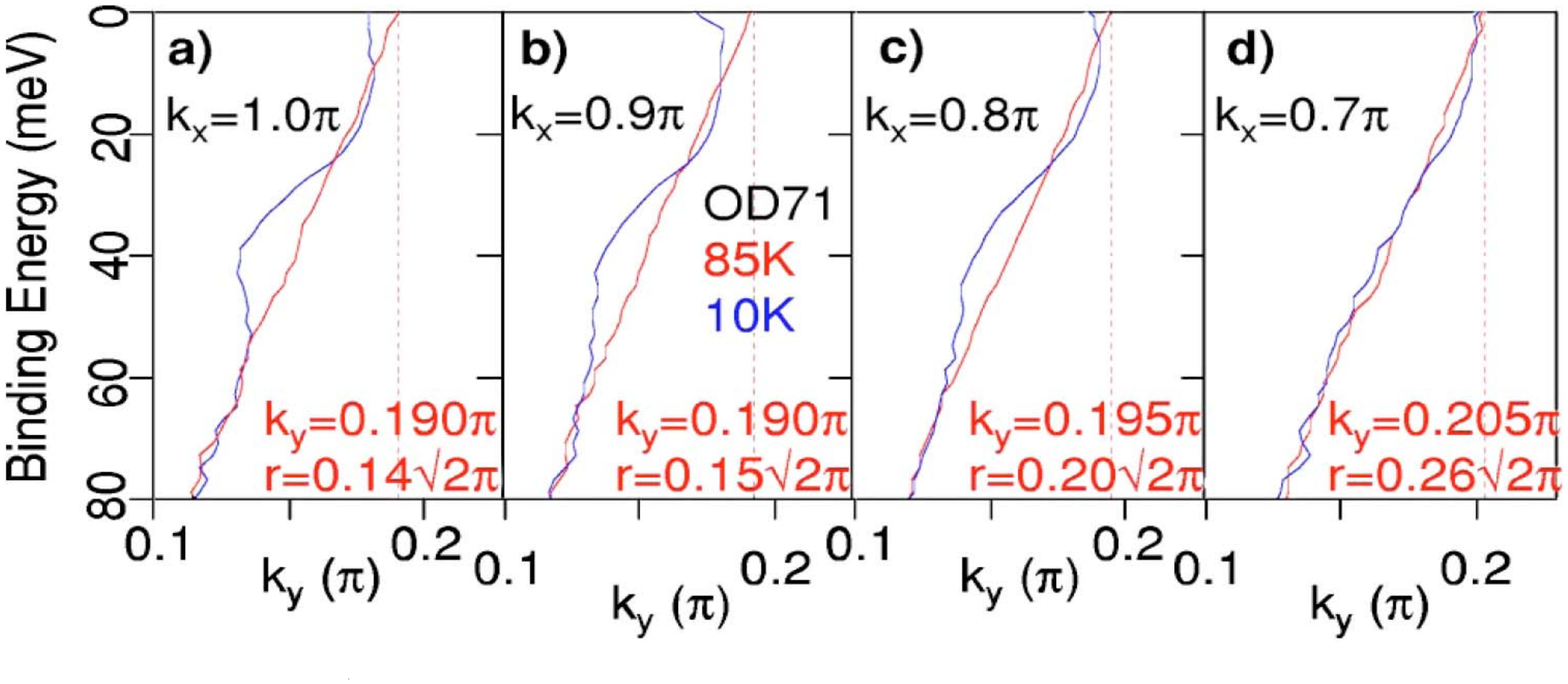}
         \vspace*{-3.5 cm}%
        \end{center}
\caption{The change of the kink strength with ${\bf k}$ from Gromko \etal (2003) of a sample of overdoped Bi-2212.   Here $r$ is the radial distance from $(\pi, 0)$ which increases as we move from panel a) to panel d).  The kink weakens as we move towards the node.}
\label{fgARPES22}
\end{figure}

Gromko \etal (2003) provide a detailed analysis of the bilayer effect in \BISCCO yielding information on self energy effects without
contamination by bilayer issues. They find that the kink in the bonding band in overdoped  \BISCCO is stronger in the $(\pi,0)$ direction and appears at a lower energy $\approx 40$ meV for overdoped samples, and is only seen in the superconducting state as expected if the mode structure is due to the magnetic resonance mode. In figure \ref{fgARPES21} we display their  results for the temperature evolution for the bonding band energy in their overdoped sample with $T_c = 71$ K of the $(\pi,0)$ kink and the inset gives the self energy derived from these data using equation \ref{EnergyCons}  (Re$\Sigma(\omega)$). The amplitude of the maximum in Re$\Sigma(\omega)$ as a
function of temperature is plotted in figure  \ref{fgARPES21} (b) as red circles. Also shown for comparison is the leading edge superconducting gap $\Delta_{LE}$, blue squares. We see that both sets of data track each other and the peak in the self energy difference between the superconducting state and the normal state, modeled by a straight line(see frame a) vanishes at $T_c$ with the vanishing of the gap $\Delta_{LE}$. The evolution of the superconducting state structures with momentum for the same overdoped $T_c=71$ K sample is shown in figure \ref{fgARPES22} for four points in the Brillouin zone with the left panel near $(\pi,0)$ and the right panel near $(\pi,\pi)$ as labeled in the figure. It is clear that the dispersion curve structure remains around 40 meV but becomes much weaker as one moves away from the antinodal direction and strong $(\pi,\pi)$ scattering such as is the case for the coupling to the spin one resonance.

\begin{figure}
        \begin{center}
         \vspace*{-0.7 cm}%
                \leavevmode
                \includegraphics[origin=c, angle=0, width=15 cm, clip]{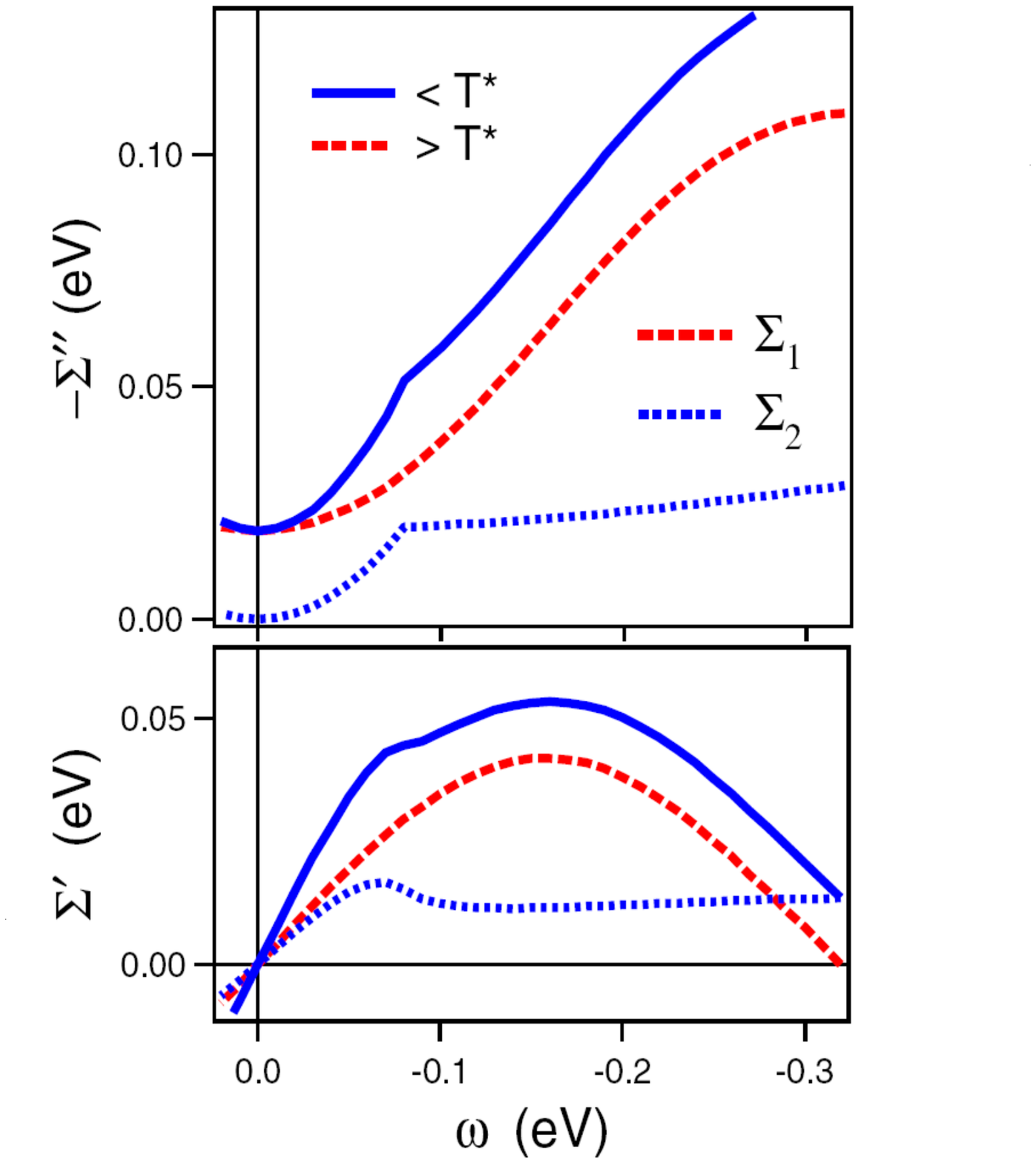}
         \vspace*{-1.2 cm}%
        \end{center}
\caption{Splitting of the nodal self energy into two components by Kordyuk \etal (2006) $\Sigma_1$ and $\Sigma_2$ based on the data
for the optimally doped sample. Top panel, the real part, bottom panel the imaginary part. The authors associate the $\Sigma_1$  component with a temperature independent coulomb repulsion and the $\Sigma_2$ component with a bosonic excitation.}
\label{fgARPES23}
\end{figure}

A different method of analysis aimed at eliminating artifacts arising from bilayer splitting effects has been developed which employs  Kramers-Kronig (KK) consistent (Kordyuk \etal 2002, 2004a, 2004b, 2005) procedures as a prominent element. The physical idea behind this procedure is simple: since the real and imaginary parts of the self energy can be measured independently by EDC and MDC, they should be related by KK transformations. The method allows the simultaneous extraction of self energy and the bare dispersion. This technique has been applied by Kordyuk \etal (2006) to the study of the 70 meV kink in Bi2212 in the nodal direction. The authors separate the self energy obtained from experiment into two components ("primary" and "secondary") as shown in figure  \ref{fgARPES23} by thick red and blue dashed lines for $\Sigma_1$ and $\Sigma_2$ respectively. The coupling in $\Sigma_2$ is to a single boson mode while the structure in $\Sigma_1$ represents coulomb repulsion. An important conclusion made by these authors is that the primary $\Sigma_1$ is structureless and largely independent of temperature and doping, while the secondary $\Sigma_2$ depends strongly on doping and temperature in agreement with coupling to spin fluctuations.

\begin{figure}
        \begin{center}
         \vspace*{-0.5 cm}%
                \leavevmode
                \includegraphics[origin=c, angle=0, width=15cm, clip]{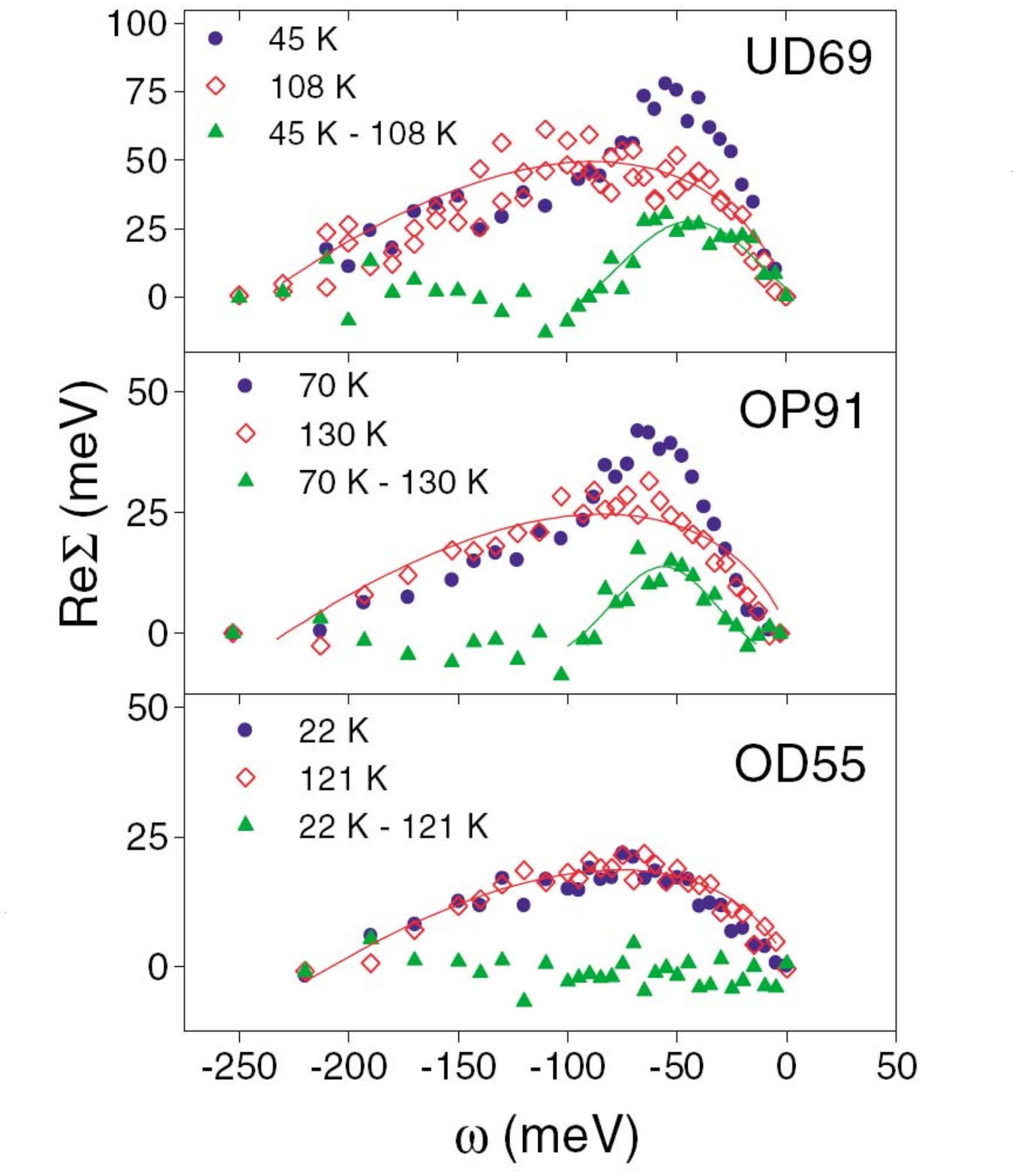}
         \vspace*{-1.2 cm}%
        \end{center}
\caption{The real part of the self energy  of \BISCCO as measured by Johnson \etal (2001) for the superconducting (blue dots) and normal states (open red diamonds) for three doping levels starting with underdoped, top panel, optimally doped middle panel and overdoped, bottom panel. The solid lines through the normal state data represent fits to the data. The difference between the superconducting and the normal state self energies for each level of doping is shown as green triangles. Gaussian fits to extract the peak energy $\omega_{0}^{sc}$ (green line) show that there is well defined peak at 40 to 50 meV that is not present in the normal state and is absent in the overdoped sample at all temperatures. In all samples there is a second component to the self energy in the form of a relatively temperature independent background.}
\label{fgARPES3}
\end{figure}

Returning to equation \ref{EnergyCons} we see that subtracting a bare dispersion curve $\epsilon_{\bf k}$ from the renormalized energy $E_{\bf k}$ of ARPES gives the real part of the quasiparticle self energy $\Sigma'({\bf k},\omega)$ vs. $\omega$ for a fixed value of ${\bf k}$. Johnson \etal (2001) have done this for the nodal direction and several doping levels in \BISCCOa. Their results are shown in figure  \ref{fgARPES3}, from top to bottom three doping levels, underdoped with $T_c=69$ K optimally doped with $T_c = 91$ K and overdoped with $T_c = 55$ K designated UD69, OP91, and OD55 respectively. In each case they show two temperatures, one in the normal state (open red diamonds), and the other in the superconducting state (solid blue circles) as well as the difference in $\Sigma'({\bf k},\omega)$ shown as solid green triangles. It is clear that  {the self energy is dominated by a broad background at all temperatures but there } is a peak in the superconducting state which does not exist in the normal state for the UD and OP samples. For the overdoped sample the peak is absent. The authors also plot the position in energy of the peaks in the self energy $\Sigma'({\bf k},\omega)$ at $\omega=\omega_0$ and in the difference  $\omega=\omega_{sc}$ as a function of $T_c^{max} - T_c$ and find that these go approximately as $\omega \approx 6 k_B T_c$. This law is close to the spin one resonance mode seen in neutron scattering which gives $E_n \approx 5.4 k_B T_c$ (Fong \etal 1999 and He \etal 2001). This correspondence lends support to the idea that the peak seen in figure \ref{fgARPES3} is associated with coupling to the spin resonance mode. This interpretation is further supported by the fact that the real part of the self energy from ARPES, Re$\Sigma(\omega=\omega_0^{sc})$  for the UD69 sample tracks closely the intensity of the neutron mode in a similar sample of \YBCO with $T_c = 74$ K (Dai \etal 1999) as a function of temperature. The most rapid change in intensity occurs for temperatures below the superconducting critical temperature $T_c$ and some intensity remains up to a higher temperature $T^*$ (the pseudogap temperature). Finally, we note that the doping dependence of the mass renormalization $\lambda$ is found to decrease with increasing doping. The coupling $\lambda$ follows from the self energy by $\lambda=\partial \Sigma'(\omega)/ \partial \omega$ evaluated at the Fermi energy $E_F$. We have neglected any small momentum dependence in $\Sigma({\bf k},\omega)$.

\begin{figure}
        \begin{center}
         \vspace*{-0.5 cm}%
                \leavevmode
                \includegraphics[origin=c, angle=0, width=15cm, clip]{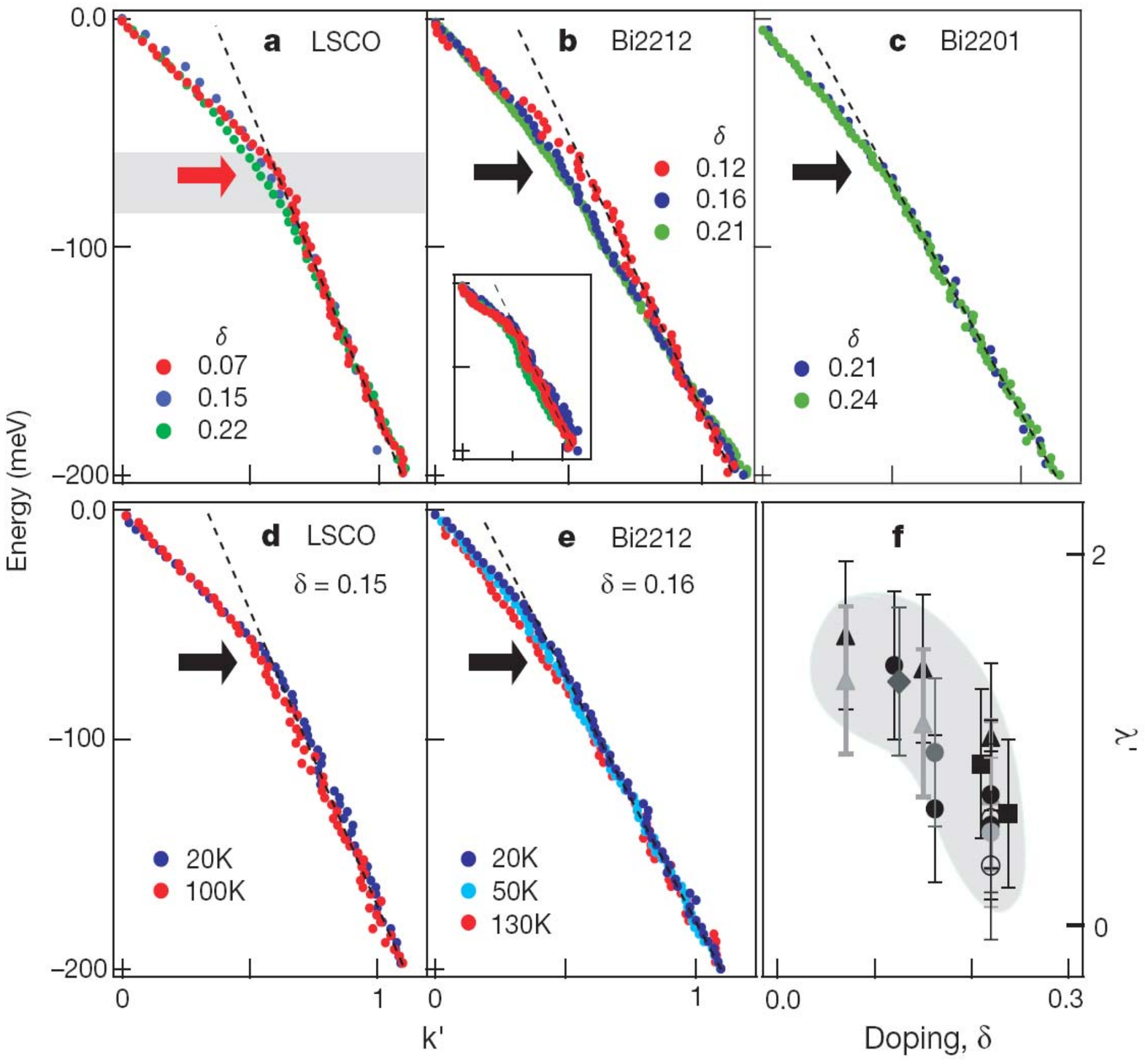}
         \vspace*{-1.2 cm}%
        \end{center}
\caption{The kink in ARPES dispersion in the nodal direction (except panel b inset, which is off this line) for a range of materials (panels a to c) from Lanzara \etal (2001).  The doping level is denoted by $\delta$. The red arrow in panel a) denotes the frequency of an optic phonon.  Panels d) and e) show the temperature dependence of the dispersions for optimally doped LSCO  and Bi2221. Panel f) shows doping dependence of $\lambda'$ the coupling constant to the bosonic excitations in different materials as a function of doping for LSCO (filled triangles) and NdLSCO (1/8 doping; filled diamonds), Bi2201 (filled squares) and Bi2212 (filled circles in the first Brillouin zone, and unfilled circles in the second zone). Note presence of the kink in all the materials at roughly the same energy, denoted by the black arrow. }
\label{fgARPES6}
\end{figure}

\begin{figure}
        \begin{center}
         \vspace*{-1.5 cm}%
                \leavevmode
                \includegraphics[origin=c, angle=0, width=13cm, clip]{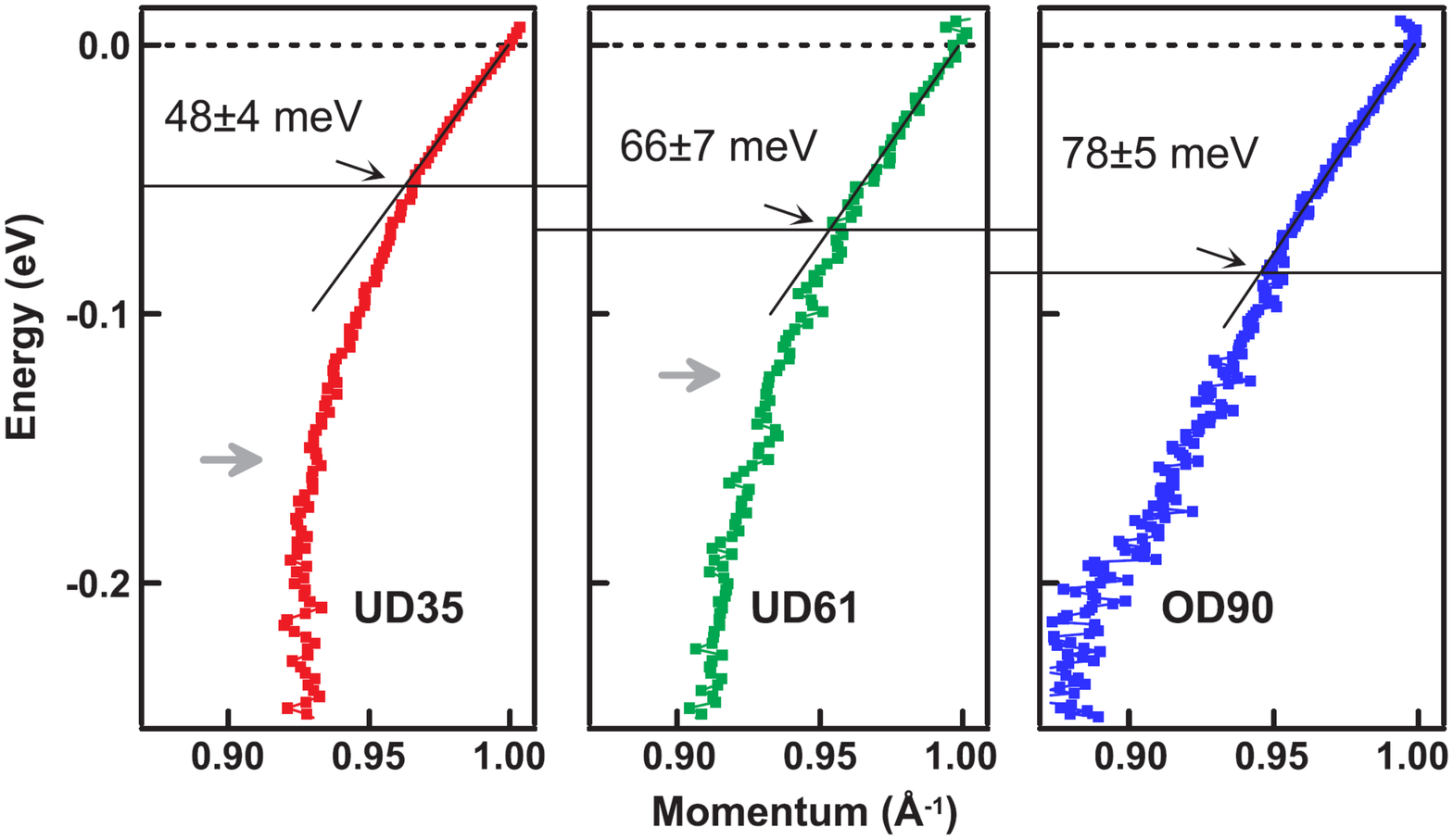}
         \vspace*{-1.7 cm}%
        \end{center}
\caption{ARPES dispersion in \YBCO as a function of doping from Borisenko \etal (2006). Horizontal lines mark the energies of the kink, where dispersion starts to deviate from the straight line. Gray arrows show the position of the second high energy kink. Note the shifting of the kink to higher frequency as the doping is increased}
\label{fgARPES7}
\end{figure}

A different interpretation of the ARPES nodal kink as it has become known since its initial discovery (Bogdanov \etal 2000) has been put forward by Lanzara \etal (2001). These results are summarized in figure  \ref{fgARPES6} for a variety of materials (see also the review by Garcia and Lanzara 2010). Lanzara \etal (2001) note that the energy of the kink, indicated by the heavy horizontal arrow is close to 70 meV in all the different materials. Also, as shown in frame (f) the coupling constant $\lambda'$ is found to decrease with increasing doping. Here $\lambda'$ is estimated from the ratio of the group velocities above and below the kink energy. As further emphasized by Garcia and Lanzara (2010), the continued existence of the kink above the superconducting $T_c$ seen in the data of panels (d - e) leads them to interpret the kink in the self energy as due to the coupling to a zone boundary in-plane oxygen stretching longitudinal optical (LO) phonon. They point to a drop in the quasiparticle scattering rate seen below the kink energy as further evidence for this interpretation (Lanzara \etal 2006).  This interpretation has been disputed by Kodyuk \etal (2006) who argue, on the basis of recent high resolution \YBCO| ARPES data, that there is a strong dependence of the frequency of the kink with doping and temperature which rules out any phonon scenario as an explantation of the kink.  An example of this is shown in figure  \ref{fgARPES7} where the position of the kink, indicated by an arrow, is shown for three doping levels of \YBCO.  A related point is that Zhou \etal (2003) found that the slope of the nodal quasiparticle renormalized dispersion at the Fermi energy remained remarkably universal across various families of cuprates as well as at various doping levels. However, once again, this phenomenology has been found to break down in some more recent papers where the ARPES resolution has been increased (Plumb \etal (2010), Vishik \etal (2010), and Anzar \etal (2010)). Near the Fermi energy there is a new velocity scale which is different from that observed by Zhou \etal and is not universal.

\begin{figure}
        \begin{center}
         \vspace*{-0.5 cm}%
                \leavevmode
                \includegraphics[origin=c, angle=0, width=10cm, clip]{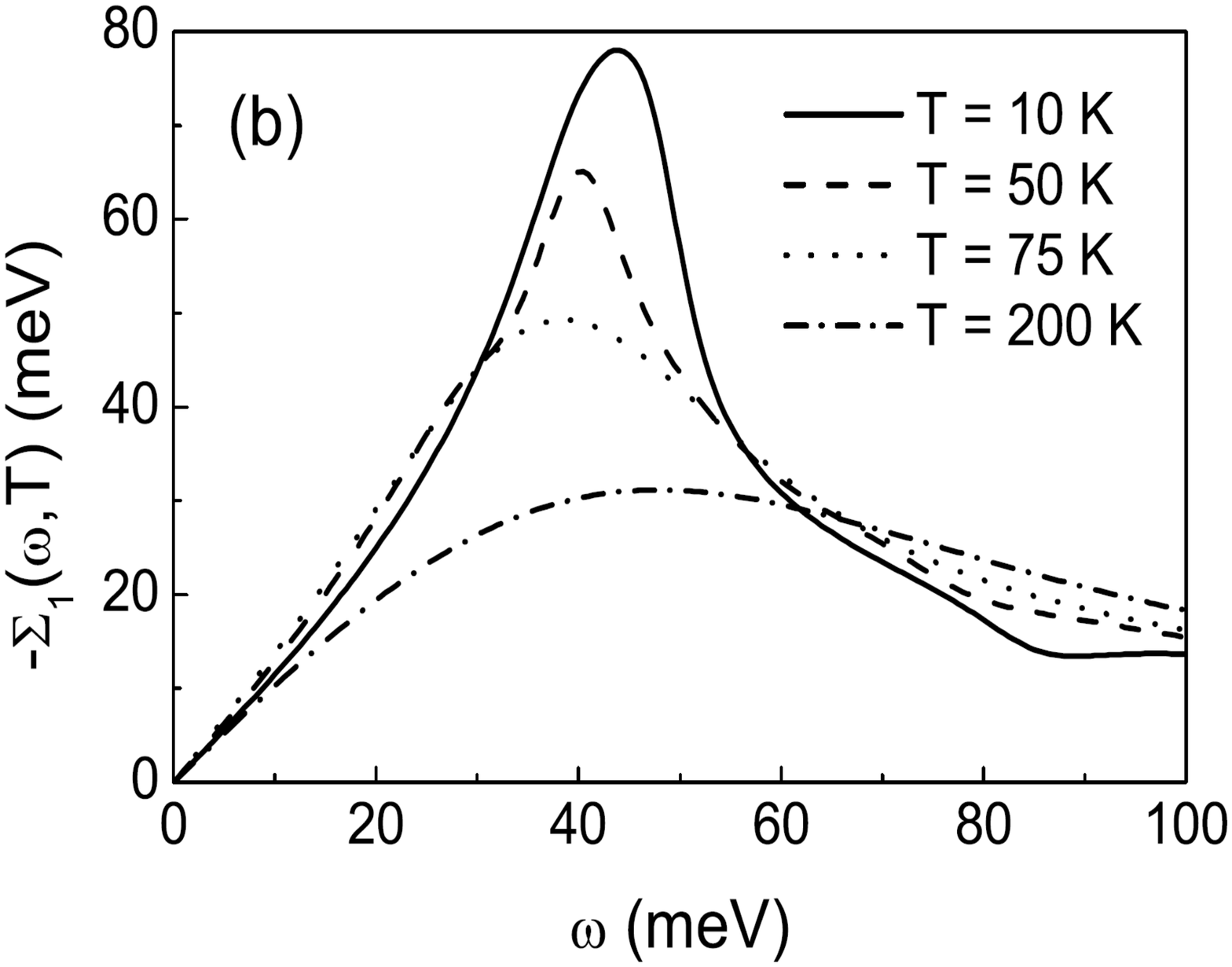}
         \vspace*{-1.1 cm}%
        \end{center}
\caption{Model calculation of the temperature dependence of the real part of the quasi particle self-energy, $\Sigma_1(\omega)$ for a Lorentzian model $I^2\chi(\omega)$ with central frequency $\omega_L=36.4$ meV, width $\Gamma = 3$ meV and $\lambda = 1.2$, giving $T_c = 54.5$ K (from Schachinger and Carbotte (2009)). There is a small shift to lower frequencies in the superconducting state due to the closing of the superconducting gap.  In the normal state there is a shift to higher frequencies with temperature. The underlying bosonic function is temperature independent.}
\label{fgARPES8}
\end{figure}

Schachinger and Carbotte (2009) have provided numerical simulations that help in understanding how boson structure encoded in the
electron-boson spectral density $\alpha^2F(\omega)$ presents itself in the self energy. These results are based on numerical solutions of the generalized Eliashberg equations for a d-wave superconductor with an electron-boson kernel taken to be a Lorentzian form centred on a specific frequency $\omega_L $ meV and of width $\Gamma$. For simplicity, the same kernel is assumed in both the gap and the renormalization channel. The authors find that while for low values of $\Gamma$ there is good correspondence between the position of the peak in the real part of the self energy and the sum of the value of $\omega_L$ plus the superconducting gap, as $\Gamma$ increases this is no longer true and the shifts in the position of the peak  on entering the superconducting state can be much less than the gap value. A similar situation holds when the temperature variations are considered. 
Pertinent theoretical results are summarized in figure \ref{fgARPES8}. The results are for $\omega_L = 36.4 $ meV and width $\Gamma = 3$ meV, $\lambda = 1.2$ and $T_c = 54.5$ K and a gap $\Delta_0$ at 10 K equal to 12.6 meV. The figure shows how the real part of $\Sigma(\omega)$ varies with $\omega$ for four temperatures 10, 50, 75 and 200 K. We note that while the underlying $\alpha^2F(\omega)$ is not changed, temperature effect in the self energy smears out the corresponding structure in $\Sigma$ and also shifts the position of its peak. The interpretation of the temperature dependence of boson structures in the real part of the self energy requires some care. In a recent paper (Lee \etal 2008), the temperature variation of the nodal kink in optimally doped Bi$_2$Sr$_2$Y$_{0.08}$Cu$_2$O$_{8+\delta}$ was traced and a superconducting induced shift in the boson energy involved identified. The data was interpreted in terms of a two Einstein mode model at 36 and 70 meV respectively and a superconducting gap value of 37 meV at $T$ = 0. The prominent peak in their normal state data at $T$ = 104 K is associated with the 70 meV mode while that in the superconducting state at $T$ = 10 K is associated with the 36 meV mode. This interpretation however depends on the use of $\delta$-functions and does not consider possible changes brought about by extended spectra (Schachinger and Carbotte 2009a). If sufficiently broad such spectra do not necessarily show a significant shift due to the opening of a superconducting gap.

\begin{figure}
        \begin{center}
         \vspace*{-0.7 cm}%
                \leavevmode
                \includegraphics[origin=c, angle=0, width=15cm, clip]{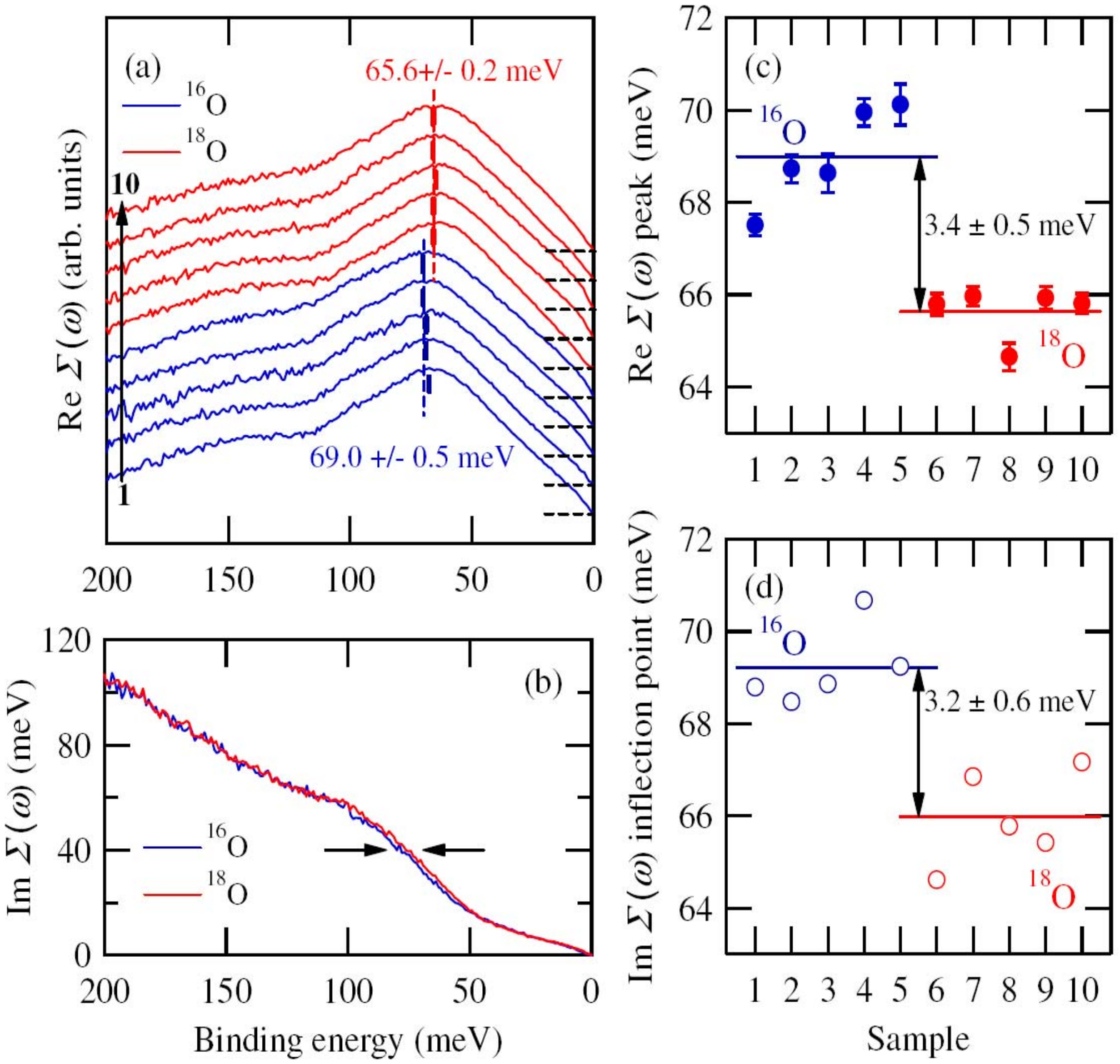}
         \vspace*{-1.1 cm}%
        \end{center}
\caption{(a) The effect of oxygen isotope substitution on the real part of the self-energy Re$\Sigma(\omega)$ from ARPES data of Iwasawa \etal (2008). Five samples of \BISCCO substituted with  $^{16}$O (blue lines) and $^{18}$O (red lines)  were measured along
the nodal direction. The curves are offset for clarity. (b) Imaginary part of the self-energy Im$\Sigma(\omega)$ determined from MDC full widths. (c), (d) Kink energy as a function of sample numbers both for $^{16}$O (blue line) and $^{18}$O (red line) from Re$\Sigma(\omega)$ and Im$\Sigma(\omega)$, respectively.}
\label{fgARPES10}
\end{figure}

One can get additional information of the possible origin of the nodal direction kink structure in the renormalized quasiparticle dispersion curves from isotope substitution effects. Oxygen $^{16}$O can be exchanged to $^{18}$O. Early experiments by Gweon \etal (2004) showed relatively small changes in the dispersion curves in the nodal direction at low energy but unexpectedly large (much larger than the 6 \% energy shift expected for lattice vibrations involving oxygen) at higher energies. Even larger shifts were found away from the nodes in MDC as well as EDC distributions (Gweon \etal 2006) with low energy photons and greatly improved spectral resolution compared with
conventional ARPES (Karalek \etal 2006). The expected isotope shift in boson frequency $\Delta\Omega/\Omega = (1-\sqrt{16/18})$ is a few meV at most.  Recent advances in ARPES resolution have made such measurements feasible.  Results on optimally doped \BISCCO from Iwasawa \etal (2008) are reproduced in figure \ref{fgARPES10}. These new results do not confirm the large changes at high energies found by Gweon et al. (2006). To extract the self energy from ARPES dispersion curves a bare dispersion is needed. Three different forms were tried by Iwasawa \etal and gave essentially the same results and are reproduced in figure \ref{fgARPES10} (a) for the real part of the self energy Re$\Sigma(\omega)$ and corresponding results for Im$\Sigma(\omega)$ are found in frame (b). Both sets of results give almost the same shift of $\approx$ 3.4 meV and 3.2 meV going from $^{16}$O to $^{18}$O as seen in (c) and (d) respectively. These shifts are in accord with an in-plane half-breathing phonon mode with $\Omega = 70$ meV as noted by Iwasawa \etal (2008). On the other hand, for $\Omega \approx 70$ meV the necessary structure of $\alpha^2F(\Omega)$ would have to be in the form of a distributed spectrum at this energy. Only in this case will the boson structure in Re$\Sigma(\omega)$ and $\alpha^2F(\Omega)$ be unshifted by the superconducting gap $\Delta$. We will return to our discussion of the isotope shift after we consider the inversion of ARPES data to recover a complete picture of the underlying electron-boson spectral density.

A difficultly in finding the boson spectral function lies in the fact that the experimentally determined quantity, the real part of the
quasiparticle self energy Re$\Sigma(\epsilon,{\bf k},\omega)$, is related to the boson spectral density of Eliashberg theory $\alpha_{\bf
k}^2F_{\bf k}(\Omega)$ through a convolution integral.  Such 'inverse' problems are inherently difficult and can result in ambiguous
solutions. A maximum entropy method (Jaynes 1957) can be used to invert this equation and retrieve an estimate of the electron-boson
spectral density. An early example of the application of this method to ARPES spectra was the work of Shi \etal (2004) who
recovered an electron-phonon function on the surface of Be and Zhou \etal (2005) and Yoshida \etal (2007) who used the same method in their study of electron self energy effects in \LSCO as a function of doping $x$. An analysis of nodal direction data in the normal state of underdoped samples yielded fine structures in the (23 - 29), (40 - 46), (58 - 63), and (75 - 85) meV energy ranges. Comparison of these results with inelastic neutron scattering data in the phonon spectrum leads these authors to support a phonon mechanism for superconductivity in this material. A related study on the heavily overdoped (Bi,Pb)$_2$Sr$_2$CuO$_{6+\delta}$ with a $T_c \approx 5$ K has also been interpreted in terms of coupling to multiple phonon modes with an electron-phonon mass enhancement of $\lambda = 0.42$ which covers the $\approx 70$ meV nodal kink in the dressed electronic dispersion curves as measured by ARPES (Zhao \etal 2010). Meevasana \etal (2006) have considered two different dopings of Bi$_2$Sr$_2$CuO$_6$, namely optimally doped (OP) Pb$_{0.55}$Bi$_{1.5}$Sr$_{1.6}$La$_{0.4}$CuO$_{6+\delta}$ with a $T_c$ = 35 K and an overdoped (OD) non-superconducting Pb$_{0.38}$Bi$_{1.74}$Sr$_{1.88}$CuO$_{6+\delta}$ ($T_c <$ 4 K) and invert their data of the quasiparticle self energy by a maximum entropy method (MEM) to recover the Eliashberg function $\alpha^2F(\omega)$ for the nodal cut at $T =$ 45 K in the OP sample and at $T =$ 8 K in the OD sample. The authors suggest that the peak in $\alpha^2F(\omega)$ in the range 70 - 90 meV involves planar oxygen motion.

\begin{figure}
        \begin{center}
          \vspace*{-1.0 cm}%
                \leavevmode
                \includegraphics[origin=c, angle=0, width=13cm, clip]{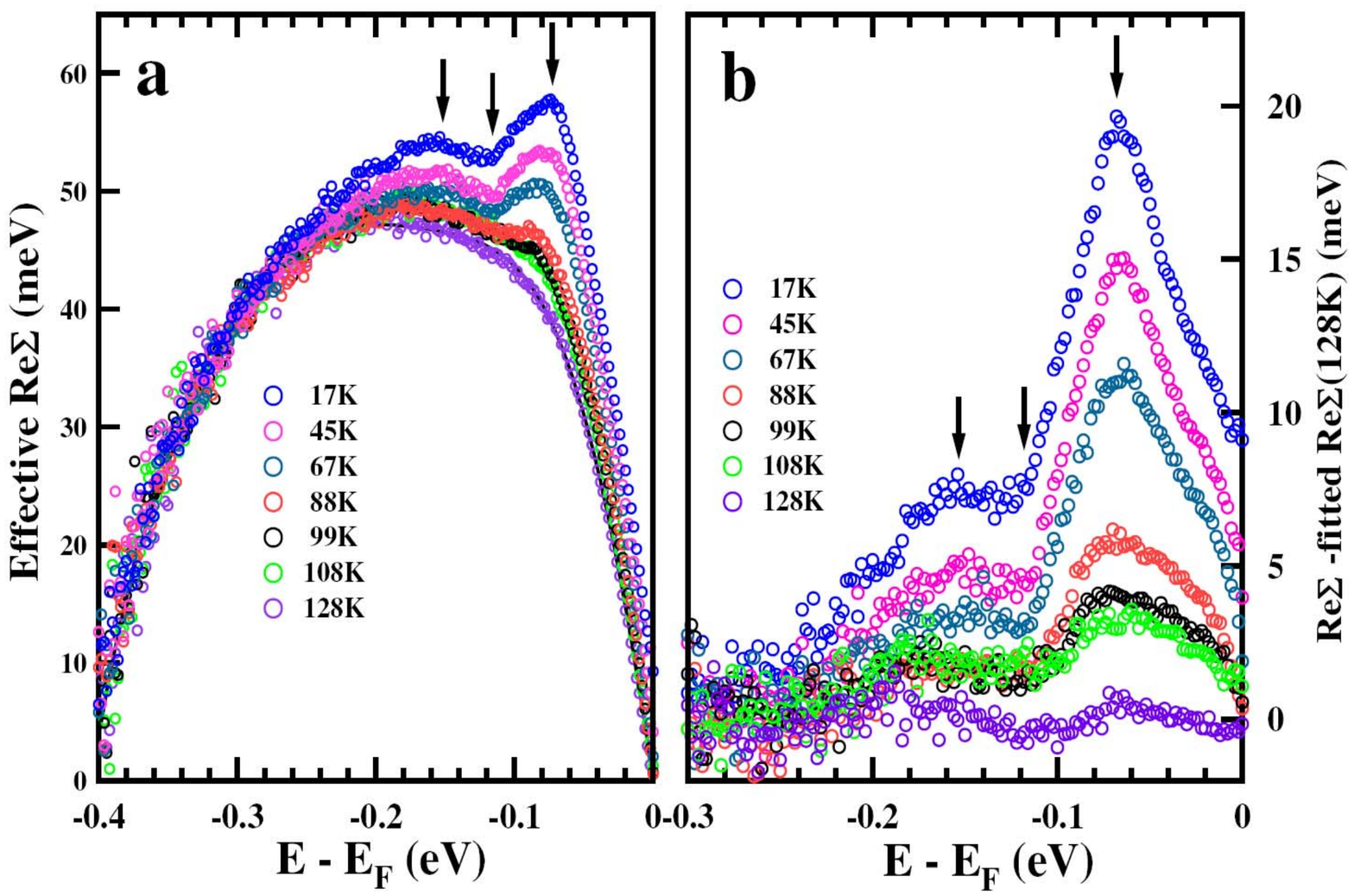}
         \vspace*{-2.0 cm}%
        \end{center}
\caption{(a) Temperature dependence of the effective real part of electron self-energy in \BISCCO from Zhang \etal (2008a). A straight line from $E_F$ to 0.4 eV has been assumed for the bare dispersion. (b) The difference between the measured self-energy in (a) and a polynomial fit to the 128 K spectrum. The arrows at 115 meV and 150 maV point to features in spectrum not previously observed.}
\label{fgARPES12}
\end{figure}

\begin{figure}
        \begin{center}
         \vspace*{-0.5 cm}%
                \leavevmode
                \includegraphics[origin=c, angle=0, width=15cm, clip]{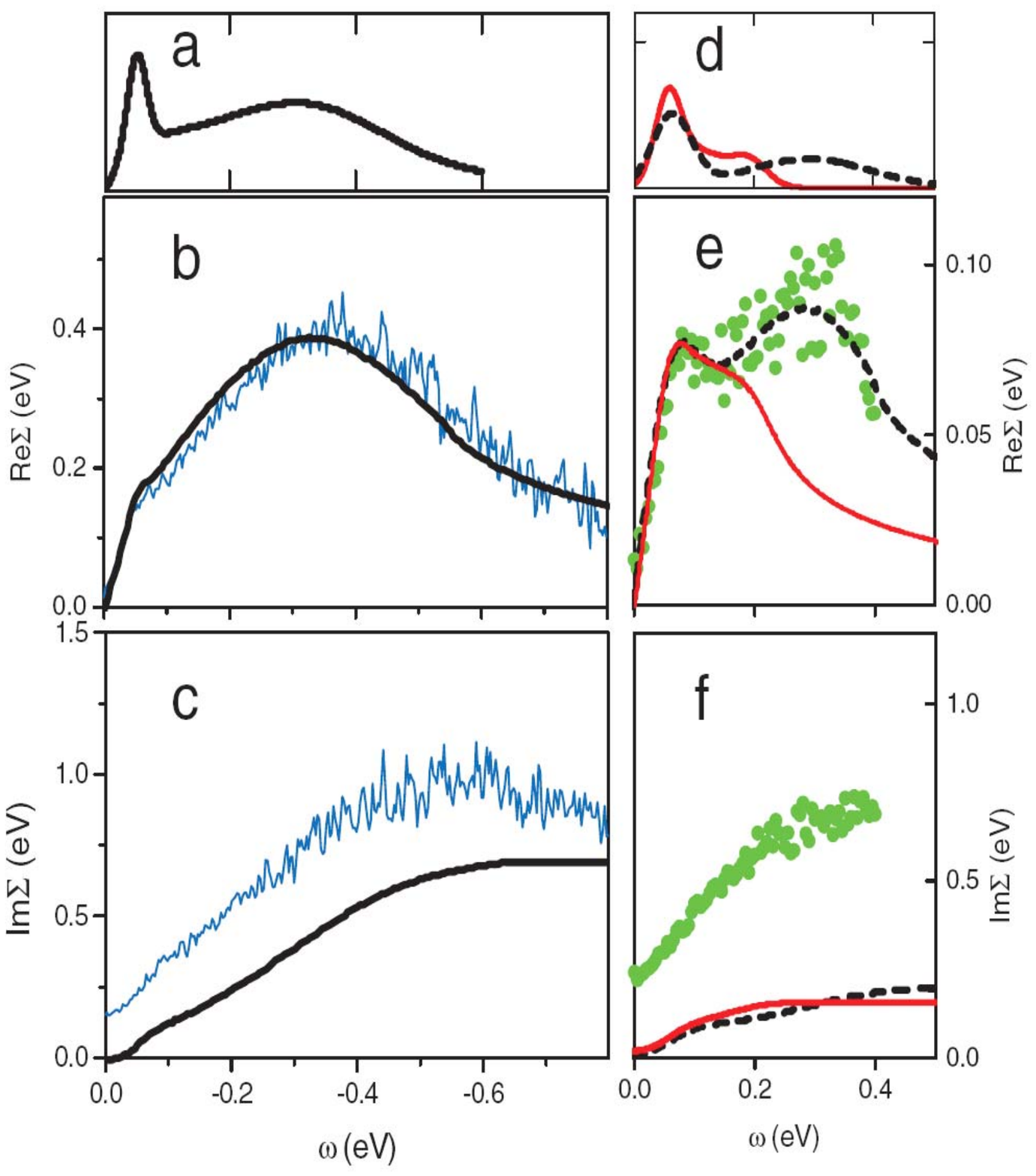}
         \vspace*{-1.2 cm}%
        \end{center}
\caption{ARPES self energy fits to models in \BISCCOa\  from Valla \etal (2007). a) A model bosonic spectrum used in fits to the data consisting of a peak and a high frequency background.  (b) Re$\Sigma$ measured in Bi 2212 (thin lines) and model Re$\Sigma$ obtained from the spectrum in (a) (bold line). (c) Corresponding Im$\Sigma$. (d)-(f) Same for \LBCO at $x=1/8$. Solid lines represent neutron scattering data from Tranquada \etal and the self-energies derived from them. Dashed lines represent the excitation spectrum and derived self-energies that better model the high-energy region of measured self-energies (green circles). (Near or nodal direction). Note the need for high energy spectral weight beyond the spin excitations revealed by neutron scattering.}
\label{fgARPES13}
\end{figure}

The above observation should be contrasted with results on optimally doped \BISCCO in which Zhang \etal (2008a) have identified in the real part of the quasiparticle self energy features at $\approx 115$ meV and $\approx 150$ meV in addition to the prominent feature at $\approx 70$ meV. Their results are reproduced in figure  \ref{fgARPES12}. The arrows (frame (a)) point to structures in the self energy which gradually smear as the temperature is increased. This is seen better in frame (b) which shows the difference of the data at a temperature $T$ and $T= 128$ K.  High energy features in ARPES have been discussed by many authors including Meevasana \etal (2007), Xie \etal (2007), Valla \etal (2007), Graf \etal (2007), Chang \etal (2008), Ikeda \etal (2009),   Pan \etal (2010), Inosov \etal (2007) and Zhang \etal (2008b). As an example, the high energy structure in the ARPES spectra of the cuprates has also been analyzed in Valla \etal (2007) for several doping levels in \BISCCO and \LBCOa. figure  \ref{fgARPES13} reproduces their results for one sample of \BISCCO and \LBCO at $x= 1/8$, the critical doping where, due to stripe formation, $T_c \rightarrow 0$ in this material. Frame (a) gives a model for the electron-boson spectral density which was chosen to reproduce the data for the real part of the self energy shown in (b). In (c) the corresponding imaginary part is shown (heavy lines) and compared with the data (thin lines). A similar comparison for the \LBCO sample is shown in (d) to (f). Here the solid red line is the spin fluctuation spectrum  determined from magnetic neutron scattering by Tranquada \etal (2004) and the black dashes are the modified spectrum used to get a better fit to the ARPES data (frame (e)). The authors conclude that this represents evidence for coupling to spin fluctuations. There have also been reports of additional low energy structures seen in ARPES spectra of optimally doped \BISCCO by Rameau \etal (2009) who report coupling to a zero momentum optical out-of-plane c-axis even-phonon mode with energy 8 meV and a $\lambda \approx 0.5$ to $0.4$. Recently several groups have reported new kink-like features in the ARPES dispersion curves at very low energies which depend on temperature and doping (Plumb \etal (2010), Vishnik \etal (2010) and Anzai \etal (2010)).

\begin{figure}
        \begin{center}
         \vspace*{-0.5 cm}%
                \leavevmode
                \includegraphics[origin=c, angle=0, width=15cm, clip]{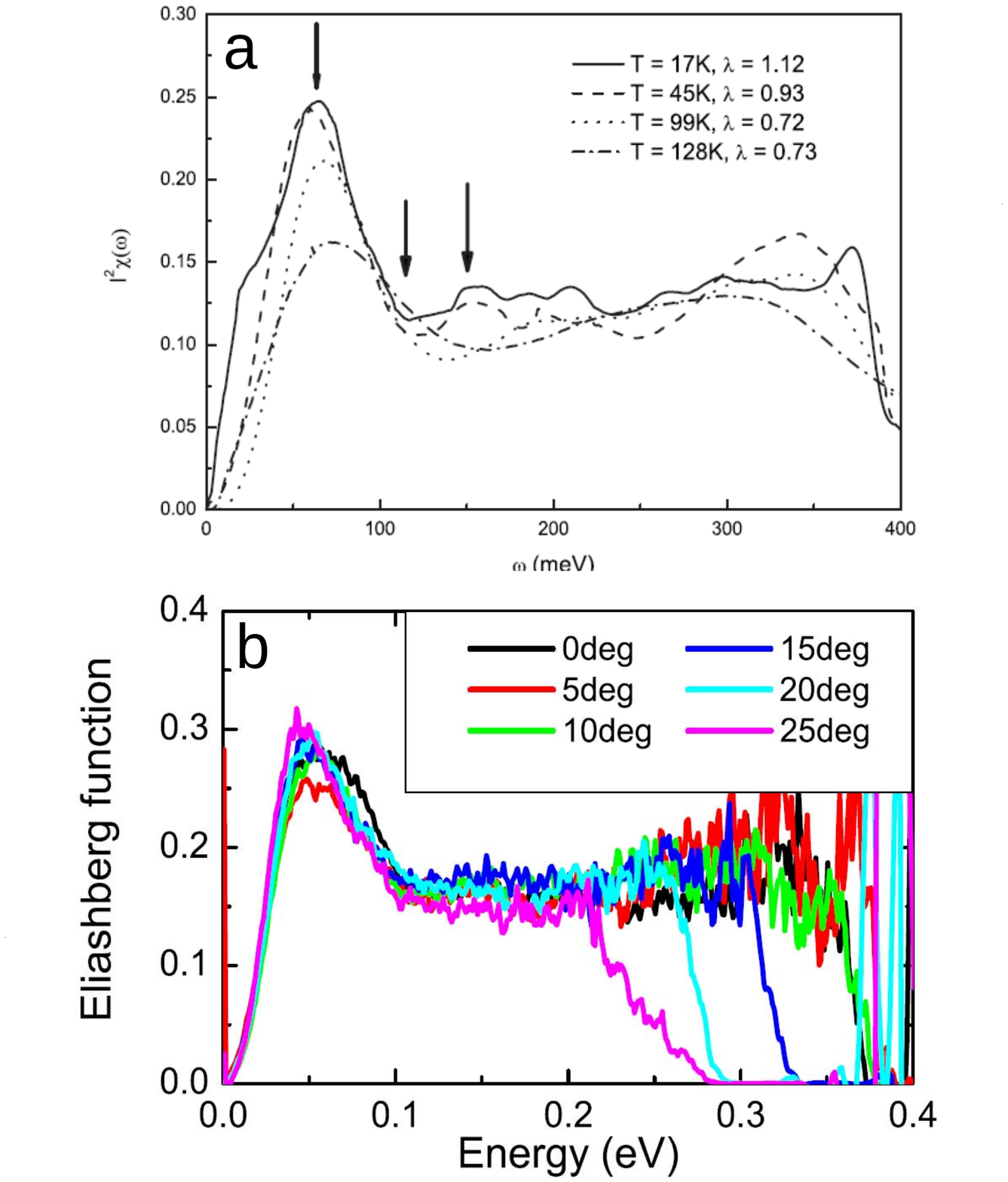}
         \vspace*{-1.0 cm}%
        \end{center}
\caption{{ Panel (a). The temperature dependence of }the electron-boson spectral density $I^2\chi(\omega)$ obtained by inversion by Schachinger and Carbotte (2008) of ARPES from Zhang \etal (2008a) along the nodal direction , $\lambda = 0.93$, superconducting state), and $T = 17$ K (solid line, $\lambda = 1.12$, superconducting state).  Panel (b) Results for  $\alpha^2F(\omega)$, obtained by inversion of ARPES data by Bok \etal (2010) for six different directions in k-space at $T =$ 107 K.}
\label{fgARPES14}
\end{figure}

The nodal direction high precision data of Zhang \etal (2008a) in optimally doped \BISCCO was inverted using the maximum entropy (MaxEnt)
technique by Schachinger and Carbotte (2008). They discuss finite band effects and obtain the results shown in panel (a) of figure  \ref{fgARPES14} for four temperatures, namely 17 K (solid), 45 K (long dashed), 99 K (dotted) and 128 K (dash dotted). In their work the MaxEnt technique is used to get an  estimate of the electron-boson spectral density using an approximate relationship between the self
energy and the bosonic spectral function. The resulting approximate electron-boson spectral density is then parameterized and used in the full Eliashberg equations with superconductivity included in d-wave symmetry and finally a least squares fit to the data is performed. The spectrum obtained evolves with temperature and the mass enhancement parameter $\lambda$ changes from 1.12 at 17 K to 0.73 at 128 K. The lowest $T$ case shows a peak around 65 meV which remains at 128 K but with reduced amplitude and its centre shifted to a higher frequency. A second component is a prominent background extending to $\approx 400$ meV. In the solid curve (at 17 K) we note a valley at $\approx 115$ meV and a second structure around 150 meV. A constant background for the spectral density is characteristic for the marginal Fermi liquid which displays quantum critical behaviour and was used early on to interpret ARPES spectra (Valla \etal 1999, 2000). This model however, is not consistent with a peak in the spectral density seen here around 65 meV. An explanation of this peak in terms of phonons has been advanced by Lanzara \etal (2001). The role of phonons in generating the nodal peak will be discussed further in the next section, but it is worth noting here that a recent study attempts to explain, within a phonon model, the evolution of the spectral function with temperature, an effect not normally expected for phonons. Meevasana \etal (2006) have made a comparative study by ARPES of an optimally doped and a strongly overdoped non-superconducting Bi$_2$Sr$_2$CuO$_6$ sample which is single-layered. They find a weakening of the self energy renormalization as well as a shift to higher energies with overdoping, an effect they related directly to changes in the coupling to c-axis phonons which results from increased metallicity (increased screening).

The data of Zhang \etal (2008a) were also analyzed by Bok \etal (2010). A maximum entropy inversion method was used and the results are
reproduced in panel (b) of figure  \ref{fgARPES14} for six directions, labeled from nodal at 0$^{\circ}$ to 25$^{\circ}$ in steps of 5$^{\circ}$. The results are very similar to those of Schachinger and Carbotte (2008) for the nodal direction. Off the node there is little change at energies below 200 meV. The main changes with angle along the Fermi surface are in the location of the high energy cutoff which starts around 400 meV at 0$^{\circ}$ and is reduced to roughly 250 meV at the largest angle considered, 25$^{\circ}$. All this provides evidence that the mechanism involved in the charge carrier self energy is not exclusively phonons and is in fact dominated by some higher energy electronic mechanism, possibly spin fluctuations. We return next to our discussion of the nodal direction isotope effect data of Iwasawa \etal (2008). Schachinger \etal (2009a) (see also Schachinger \etal 2010) have used the spectrum recovered from nodal direction ARPES data in \BISCCO which is displayed in panel (a) of figure \ref{fgARPES14} to calculate in the d-wave superconducting state the real part of the quasiparticle self energy Re$\Sigma(\omega)$ expected when the isotope O$^{16}$ is substituted by O$^{18}$. In this calculation they assume that only the peak in $\alpha_{\bf k}^2 F_{\bf k}(\omega)$ around $\omega \cong$ 60 meV seen in panel (a) of figure \ref{fgARPES14} is due to coupling to an oxygen phonon. The background below this peak which also extends to 400 meV is assumed to be due to some other interaction. Applying an isotope shift of $\sqrt{16/18}$ = 0.94 to this peak part only, they are able to explain the shift seen in the data of Iwasawa \etal (2008). They take this as evidence that in optimally doped \BISCCO phonons contribute $\sim$ 10 \% to the area under the nodal direction electron-boson spectral density.

\begin{figure}
        \begin{center}
         \vspace*{-0.5 cm}%
                \leavevmode
                \includegraphics[origin=c, angle=0, width=15cm, clip]{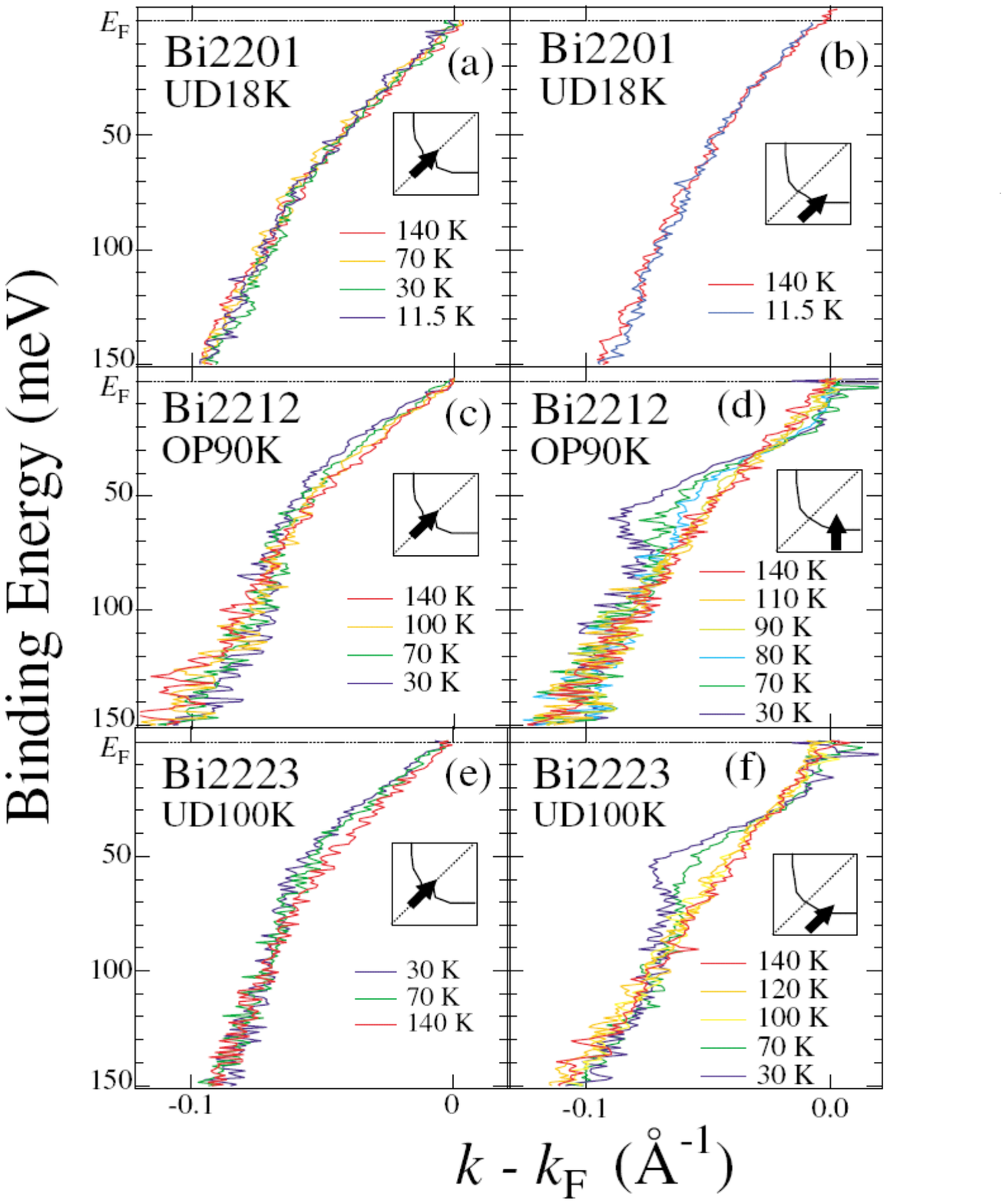}
         \vspace*{-1.2 cm}%
        \end{center}
\caption{ARPES dispersion of one, two and three layer superconductors from Sato \etal (2003). Left panels are Brlllouin zone cuts in the nodal direction, right panels closer to the antinode (the arrow in each inset shows the cut).  There is a systematic increase in the strength of the kink at 50 meV as one moves away from the node in all cases.  Also the kink is stronger as the number of layers increases.}
\label{fgARPES16}
\end{figure}

We return next to studies of boson structure away from the nodal direction towards the antinode. Studies across families of superconductors as well as away from the nodal direction can be helpful. One, two and three layer Bi$_2$Sr$_2$Ca$_{n-1}$Cu$_{n}$O$_{2n+4}$ ($n=1,2,3$) are compared in the work of Sato \etal (2003). Renormalized dispersion curves are shown in figure  \ref{fgARPES16}. Top to bottom are Bi2201 ($n=$1) UD (underdoped) $T_c = $18 K, Bi2212 ($n=$2) OP (optimally doped) $T_c = 90$ K and Bi2223 ($n=$3) UD (underdoped) $T_c = 100$ K. Left columns are nodal data at various temperatures while the right columns give results for a cut closer to the antinodal direction as shown in the inset. In contrast to the single layer case, for $n=2$ and $n=3$ there is a strong evolution of the dispersion curve kinks as one moves towards the antinode and lowers the temperature into the superconducting state. These features correlate with some of the features associated with the spin resonance. On the other hand for $n=1$ almost no temperature dependence is observed. This would seem to imply that in this case there is no coupling to the resonance mode which gives rise to the strong temperature dependence in the other two cases. The authors further conclude that the kink in the nodal direction may have a different origin from that in the antinodal which is possibly the spin resonance. Lee \etal (2009) have studied the Tl compounds and find a similar behaviour. For this family however a spin one resonance has been observed (He \etal 2002) in its single layer version. This leads Lee \etal (2009) to suggest that this disfavours strong coupling to the spin resonance in all cases and suggests that coupling to a c-axis phonon can provide a more natural explanation of the data. On the other hand Wei \etal (2008) have observed a peak-dip-hump structure in the antinodal direction in an optimally doped Bi$_2$Sr$_{1.6}$La$_{0.4}$CuO$_{6+\delta}$ sample with $T_c$ = 34 K. They find a peak-dip separation of only 19 meV in this system, much smaller than for multilayered samples. This scale is consistent with observed values of the spin excitations in single layered cuprates. This leads the authors to favour a spin origin for their observations and reject the motion that it may be due to a phonon. Certainly all oxygen-related phonons have energies above 35 meV.

\begin{figure}
        \begin{center}
         \vspace*{-1.5 cm}%
                \leavevmode
                \includegraphics[origin=c, angle=0, width=13cm, clip]{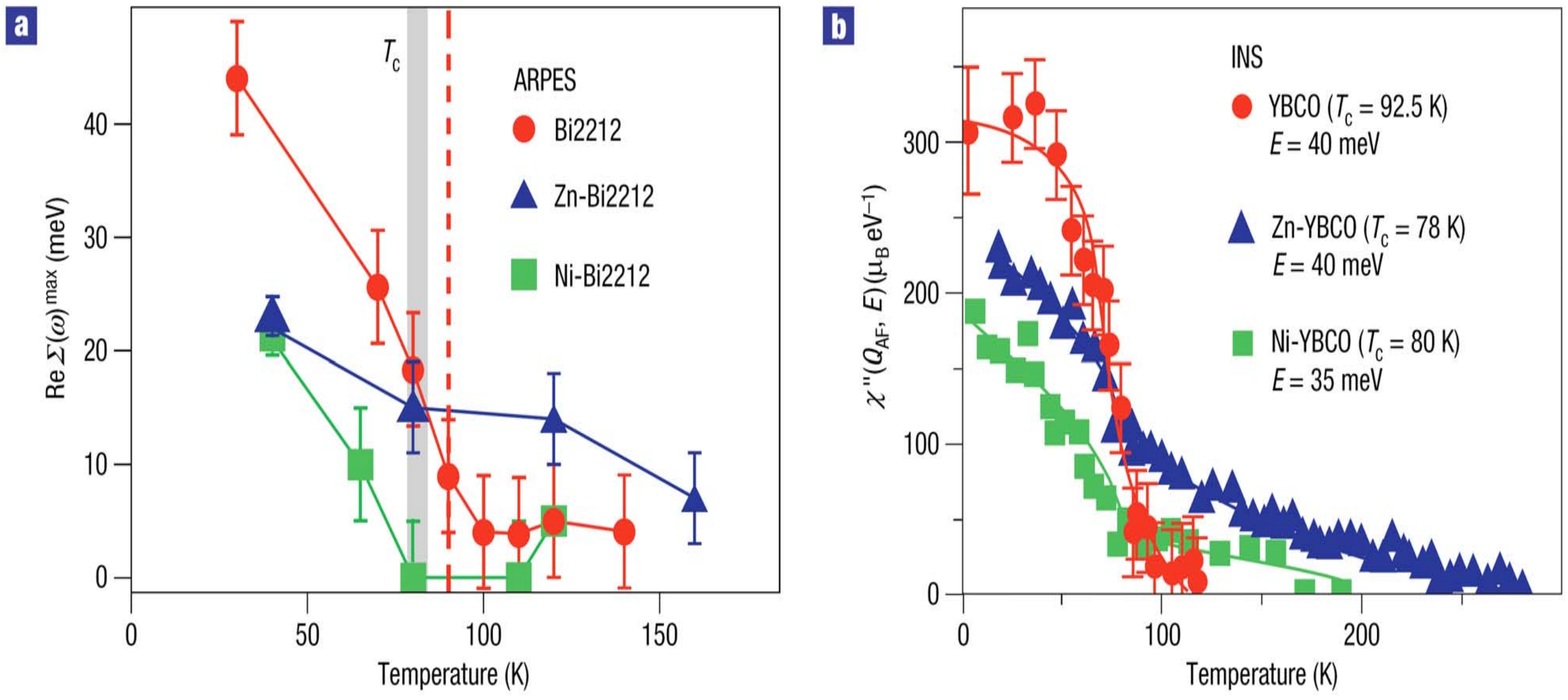}
         \vspace*{-2.3 cm}%
        \end{center}
\caption{The effect of Zn and Ni doping on the ARPES kink (left panels) and inelastic neutron scattering (right panels) from the work of Terashima \etal (2006). a) The temperature dependence of the amplitude of the maximum of the real part of the self energy in undoped,  Zn-substituted and Ni-substituted Bi2212 as determined by ARPES.  b) The temperature dependence of the imaginary part of the spin susceptibility at $(\pi,\pi)$, the antiferromagnetic wave vector, at the resonance energy measured by INS experiments for undoped, Zn-substituted and Ni-substituted YBCO (Sidis \etal (2000). Note that in both sets of experiments the resonance persists in the normal state for Zn doping but vanishes at $T_c$ with Ni doping.}
\label{fgARPES16b}
\end{figure}

Terashima \etal (2006) have studied the effect of substituting Ni or Zn for the in plane Cu in \BISCCOa\  (see also Zabolotnyy \etal 2006). Their pristine sample has a $T_c \sim$ 91 K dropping to 80 - 85 K for the  (0.5 - 1.0 \%)  Zn-Ni substituted samples. With impurities the off-diagonal kink is found to be noticeably weakened which is taken as an indication that the coupling to the boson involved is reduced. As a function of angle in the Brillouin zone the weakening is largest in the antinodal direction and is almost zero in the nodal direction. As figure  \ref{fgARPES16b} shows the antinodal kink survives as a function of temperature well above $T_c$ for the Zn-substituted sample (non-magnetic impurity) but not in the pristine and Ni substituted (magnetic impurity) case. This observation parallels what is seen in inelastic polarized neutron scattering experiments in YBa$_2$Cu$_3$O$_{7-\delta}$ for the effect of Zn and Ni on the spin resonance { shown in the right panel of figure  \ref{fgARPES16b}}. It is important to note that both Zn and Ni have a mass similar to that of Cu and would not be expected to change the lattice dynamics significantly.

Cuk \etal (2004) among others have studied boson structure evolution as the antinodal direction is approached. Their optimum
Bi$_2$Sr$_2$Ca$_{0.92}$Y$_{0.08}$Cu$_2$O$_{8+\delta}$ sample had a $T_c$ = 94 K. They find a kink in the normal (above $T_c$) state at $\omega \sim$ 40 meV for angles 22$^{\circ}$ to 27$^{\circ}$ off the antinodal direction. In the superconducting state at $T =$ 10 K a peak-dip-hump structure is seen with boson signature at 70 meV, a value which agrees with 40 meV shifted by the superconducting gap value. They also note that for an underdoped sample with $T_c =$ 85 K and a large superconducting gap the structure remains at $\sim$ 70 meV, while for a deeply overdoped sample ($T_c =$ 65 K or $\delta \cong$ 22 \%) the kink energy shifts to 40 meV consistent with a much smaller superconducting gap value. While {\color{b} some} of these observations are consistent with a spin resonance interpretation of the boson involved, some are not. The 40 meV kink remains in the normal state and is sharp in the superconducting state of the overdoped $T_c =$ 65 K sample. On the basis of this evidence and other facts, the authors suggest that the boson involved is the 40 meV B$_{1g}$ phonon involving out-of-plane motion of the in-plane oxygen. For more details see also Devereaux \etal (2004). Sandvik \etal (2004) have studied theoretically how the momentum dependence specific to coupling to a phonon or a spin fluctuation manifests itself in ARPES. They find only minimum qualitative changes between the two mechanisms and no easily recognized  qualitatively distinct signature which would unambiguously favour phonons or spin fluctuations.

Returning to the work of Borisenko et al. (2006) we note that in their experiment on YBCO they compared their ARPES spectra with spin susceptibility measured with inelastic neutron scattering on exactly the same samples. Using a simple model these authors found that the various "kinks, dips and humps" in the ARPES spectra across the whole Brillouin zone were caused by spin fluctuations. It should be noted that the several of the magnetic excitations, including a new high mode and continuum contributed to the spectra.

\section{Optical Properties}
One of the simplest measurements that can be made on a metallic system is the dc resistivity. Right from the beginning it was obvious that the high temperature superconductors were very unlike ordinary metals (Gurvitch and Fiory 1987 and Martin \etal 1990). Their dc resistivity had a linear temperature dependence with a zero temperature intercept at zero resistivity. In simple metals where the electron phonon interaction is the dominant scattering mechanism this intercept is approximately at $\theta_{Debye}/4$ and in a model system with an Einstein phonon at $\Omega_E$ one would expect  the intercept to be at $\Omega_E/4$. These early observations motivated the search for scattering models with a continuous spectrum of bosonic excitations without an energy scale such as the marginal fermi liquid theory (Varma \etal 1989) but studies of dc transport as a function of doping soon showed that the situation was much more complicated at doping levels away from optimal doping where this simple behaviour is seen: at low doping the pseduogap had a profound influence on dc transport while in the overdoped region a quadratic temperature dependence seemed to emerge, characteristic of electron-electron scattering. The optical conductivity offers a deeper insight into the influence of bosonic excitations on the transport of charge in a metallic system. A comprehensive review of the optical properties of high temperature superconductors can be found in Basov and Timusk (2005).
In a metal the standard formula for the current ${\bf J}$ in the presence of an electric field ${\bf E}$ (Ziman 1972) is given by:
\begin{equation}
{\bf J}={\bf {\sigma}} {\bf {\cdot}}{\bf E}={e^2 \over 4\pi^3}\int {dS_F \over v_{\bf k}} {\tau_{\bf k}{\bf v_k v_k {\bf \cdot E}}\over
1-i\tau_{\bf k}\omega}
\label{ziman}
\end{equation}
where ${\bf \sigma}$ is the conductivity tensor, ${\bf E}$ the applied field, $\tau_{\bf k}$ the lifetime of state ${\bf k}$ and ${\bf v_k}$ its velocity.  The integration is carried out over the Fermi surface. The  conductivity is proportional to the component of
the velocity in the direction of the applied field ${\bf E}$ averaged over the Fermi surface and it is also proportional to the factor
$\tau_{\bf k}$ which can vary over the Fermi surface independently.

In a dirty metal where impurity scattering dominates  $\tau$ is a constant and the classical Drude formula for the conductivity holds:

\begin{equation}
\sigma(\omega)= {1 \over {4\pi}}{{\omega_p^2} \over {1/\tau-i\omega}}
\end{equation}
The real part of $\sigma(\omega)$ is a Lorentzian at zero frequency with an oscillator strength $\omega_p^2/8$. For a spherical Fermi surface $\omega_p^2=4{\pi}ne^2/m_e$, where $n$ is the free--carrier density and $m_e$ is the electronic band mass. The imaginary part of $\sigma(\omega)$ is just the real part multiplied by $\omega\tau$.

If the electrons are scattered by bosonic fluctuations $\tau$ becomes frequency dependent, the extended Drude model or memory function
technique can be used (Mori 1965, G\"otze and W\"olfle 1972, Allen and Mikkelsen 1977, and Puchkov \etal 1996). The extended Drude formula is written as:

\begin{equation}
\sigma(\omega,T)=\sigma_1(\omega,T)+i\sigma_2(\omega,T)=
{1 \over {4\pi}} {{-i\omega_p^2}\over {1/\tau^{op}(\omega,T)-i\omega[1+\lambda^{op}(\omega,T)]}}
\label{OPTconduc}
\end{equation}
where $1/\tau^{op}(\omega,T)$ describes the frequency-dependent optical scattering rate and $\lambda^{op}(\omega,T)$ is the optical mass enhancement factor. By analogy to the case of quasiparticles, an optical self energy, $\Sigma^{op}(\omega)$ can be defined with minus twice its imaginary part given by $1/\tau^{op}(\omega)$ and the real party related to $\lambda^{op}(\omega)$.

One can solve for $1/\tau^{op}(\omega)$ and $1+\lambda^{op}(\omega)$ in terms of the experimentally determined optical conductivity to find

\begin{equation}
\frac{1}{\tau^{op}(\omega)}={\omega_p^2 \over {4\pi}} Re\Big{[}{1 \over \sigma(\omega)}\Big{]}.
\label{tausig}
\end{equation}
The dc resistivity is the zero frequency limit $\rho_{dc}(T)=1/\sigma_{dc}(T)=m_e/[\tau(T)ne^2]$ since $\sigma(\omega)$ is real in the zero frequency limit. The mass enhancement factor $\lambda(\omega)$ is given as the imaginary part of $1/\sigma(\omega)$:

\begin{equation}
1+\lambda^{op}(\omega)=-{\omega_p^2 \over {4\pi}} {1 \over \omega} Im\Big{[}{1\over \sigma(\omega)}\Big{]}.
\end{equation}
The total plasma frequency $\omega_p^2$ can be found from the sum rule $\int_0^{\infty}\sigma_1(\omega)d\omega=\omega_p^2/8$.
Allen (1971) and Shulga \etal (1991) give the following expression for $1/\tau(\omega,T)$:
\begin{equation}
{1 \over \tau^{op}(\omega,T=0)}=\frac{2 \pi}{\omega} \int_{0}^{\omega}d\Omega \: \alpha_{tr}^2(\Omega)F(\Omega)\:(\omega-\Omega) + {1 \over
\tau_{imp}},
\label{allen}
\end{equation}
\begin{eqnarray}
{1 \over \tau^{op}(\omega,T)}&=&{\pi \over \omega} \int_0^{\infty} d\Omega\:\:
\alpha_{tr}^2(\Omega)F(\Omega)\Big{[}\:2{\omega}\:{\coth}\Big{(}{\Omega \over {2T}}\Big{)}- (\omega+\Omega){\coth}\Big{(}{{\omega+\Omega}
\over {2T}}\Big{)}\nonumber \\ &+&(\omega-\Omega){\coth}\Big{(}{{\omega-\Omega} \over {2T}}\Big{)}\Big{]} +
{1 \over \tau_{imp}}.
\label{Shulga}
\end{eqnarray}
Here $\alpha_{tr}^2(\Omega)F(\Omega)$ is a weighted phonon density of states appropriate to transport and $T$ is the temperature measured in frequency units. Transport and quasiparticle electron-boson spectral density differ through a factor that weighs more strongly  back scattering in transport as compared with quasiparticle scattering but we will not emphasize these differences here. The last
term in equation \ref{Shulga} represents impurity scattering.

The simple approximations of Allen (1971) and Shulga \etal (1991) can be justified  through the use of the Kubo formula approach where the optical conductivity is related to the current-current correlation function which involves the overlap of two Green's functions, one displaced in energy  from the other by the photon energy $\Omega$. In the most general formulation vertex corrections need also to be considered. To include inelastic scattering by bosons, an energy dependent self energy needs to be included in the formalism written in real frequency axis formulation. An alternative is to stay on the imaginary Matsubara frequency axis and use analytical continuation using Pad\'{e} approximant (Nicol \etal 1991). For numerical implementation on the real axis an efficient formula within  Eliashberg theory including an s-wave superconducting gap was given by Marsiglio \etal (1988). To go from the Green's function to the frequency dependent conductivity $\sigma(\omega,T)$ the necessary formula was given by Lee \etal (1989). This was evaluated numerically by Akis \etal (1991) who provided results for the real part of the conductivity including s-wave superconductivity, and Marsiglio \etal (1996) who considered the imaginary part. The calculations can also be generalized to the marginal Fermi liquid (Nicol and Carbotte 1991). The necessary generalization to d-wave superconductivity was given by Carbotte \etal (1995) and by Jiang \etal (1996a). To consider a gap with d-wave symmetry, it was first necessary to include the possibility of different spectral densities $\alpha^2F(\Omega)$ entering the gap and the renormalization channels. In the simplest possible approximation the same shape $\alpha^2F(\Omega)$ is retained but its magnitude is allowed to be different in each channel with a parameter giving the ratio of d- to s- wave projections of the underlying spectral density (Carbotte and Jiang 1993, 1994). These equations have since been applied to the calculation of other related properties including phonon self energies (Marsiglio \etal 1992), penetration depth (Arberg \etal 1993, 1994 and Schachinger \etal 1997), microwave conductivity (Schachinger \etal 1998a), thermal conductivity (Schachinger \etal 1998b) and Raman scattering (Jiang and Carbotte 1996b). A comparison of quasiparticle self energy effects and optical signature was given by Schachinger \etal (2003) and by Carbotte \etal (2005). Most formulations as described above involve the infinite band approximations with a constant density of states. There are corrections coming from finite band effects both in the normal state properties (Mitrovi\'{c} and Carbotte 1983a) and the superconducting ones (Mitrovi\'{c} and Carbotte 1983b) and more specifically on the optical conductivity (Knigavko and Carbotte 2005, 2006, Knigavko \etal 2004, Cappelluti and Pietronero 2003, and Do\v{g}an and Marsiglio 2003).

In the infinite band approximation with a constant density of electronic states the conductivity at temperature $T$ and frequency $\omega$ is given by:
\begin{eqnarray}
\sigma(T,\omega)={\omega_p^2 \over {4\pi}}{ i \over \omega}\int_{-\infty}^{+\infty}{{f(\nu) - f(\nu+\omega)} \over {\omega +
\Sigma^*(\nu)-\Sigma(\nu+\omega)}} d\nu
\label{MoreExact}
\end{eqnarray}
where $\omega_p$ is the plasma frequency, $f$ the Fermi function and $\Sigma(\omega)$ the complex quasiparticle self energy. At $T=0$ this reduces to:

\begin{eqnarray}
\sigma(T=0,\omega)={\omega_p^2 \over {4\pi}} { i \over \omega}\int_{0}^{\omega} {{1} \over {\omega +i/\tau^{op}_{imp}-
\Sigma^*(\nu)-\Sigma(\omega-\nu)}} d\nu
\end{eqnarray}
where we have made the elastic impurity contribution to the imaginary part of the quasiparticle self energy explicit with
$1/\tau^{op}_{imp}\equiv 2/\tau^{qp}_{imp}$ with $1/\tau^{qp}_{imp}$ the quasiparticle elastic scattering rate. In deriving these equations it is assumed that the anisotropy in the electron-boson scattering rate can be neglected and the self energy depends only on frequency and not on the direction or magnitude of the momentum. For many purposes this is perhaps a reasonable first approximation but it needs to be kept in mind that even in simple metals such as aluminum the electron-phonon interaction does exhibit anisotropy. (see Leung \etal 1976)

\begin{figure}
        \begin{center}
         \vspace*{-0.7 cm}%
                \leavevmode
                \includegraphics[origin=c, angle=0, width=15 cm, clip]{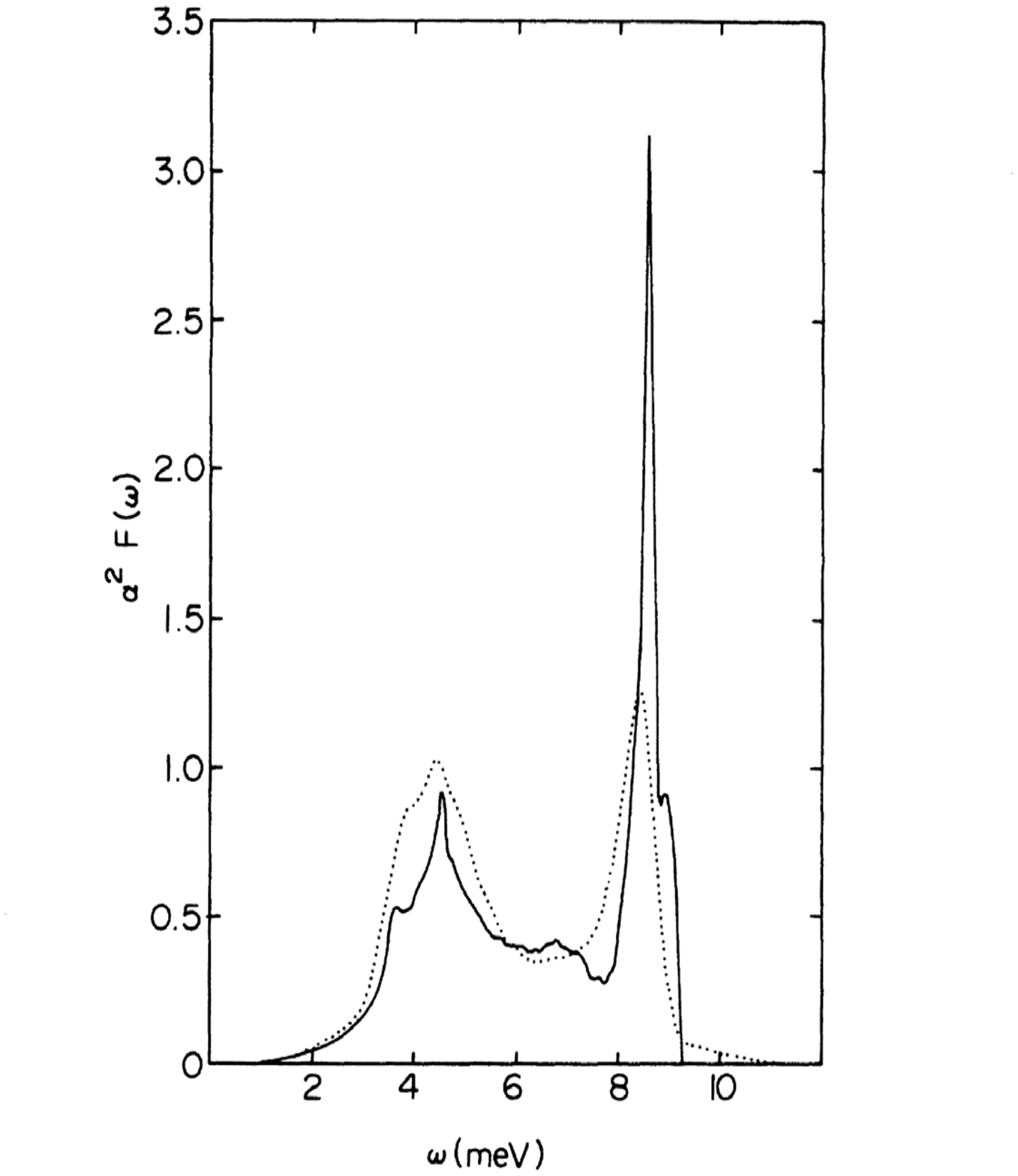}
         \vspace*{-1.1 cm}%
        \end{center}
\caption{Bosonic spectral function $\alpha^2F(\Omega)$ from tunnelling data of Pb from McMillan and Rowell (1965) (dotted curve) compared with a calculated $\alpha^2F(\Omega)$ (solid curve) based on a four plane wave electronic wave function (Tomlinson and Carbotte 1976). The phonons were modelled by a Born von Karman model based on neutron scattering.}
\label{fgOptics4}
\end{figure}

A multiple plane wave calculation of the electron-phonon interaction by Leung \etal (1976) shows that the directional spectral density
$\alpha^2F(\phi,\theta,\Omega)$ vs. $\Omega$ varies considerably with the initial state position $(\phi,\theta)$ on the Fermi surface. Note that
\begin{equation}
\alpha^2({\bf k},\omega)F({\bf k},\Omega) \equiv {V  \over {8\pi^3 \hbar}}\int_{S_F}{{dS_{\bf k'}} \over {\hbar v_{\bf k'}}
}\sum_{\lambda}|g_{{\bf k' k}\lambda}|^2 \delta(\Omega-\Omega_{\lambda}({\bf k'}-{\bf k}))
\end{equation}
where $dS_{\bf k}$ is a Fermi surface element, $v_{\bf k}$ the Fermi velocity, $V$ the volume, $\hbar$ Planck's constant, $g_{{\bf k'
k}\lambda}$ the electron-phonon matrix element for scattering from initial state ${\bf k}$ on the Fermi surface to all final states ${\bf k'}$ again on the Fermi surface emitting a phonon of energy $\Omega_{\lambda}({\bf k'} - {\bf k})$ with $\lambda$ a branch index. The mass enhancement factor $\lambda_{\bf k}$ associated with the directional $\alpha^2({\bf k},\Omega)F({\bf k},\Omega) $ and given by $2\int
\alpha^2({\bf k},\Omega)F({\bf k},\Omega)d\Omega/\Omega$ displays considerable anisotropy, of order 20 \%. The
superconducting critical temperature of such a material is determined not by the directional $\alpha^2({\bf k},\Omega)F({\bf k},\Omega)$ but by its Fermi surface average
\begin{equation}
\alpha^2F(\Omega)={{\int_{S_F} {{dS_{\bf k}} \over {\hbar v_{\bf k}}}\alpha^2({\bf k},\Omega)F({\bf k},\Omega)}   \over   {\int_{S_F}  {{dS_{\bf k}} \over
{\hbar v_{\bf k}}}}}
\end{equation}
over all the initial states of the Fermi surface. Anisotropy can increase the value of the critical temperature but this is usually a very small effect. Impurities wash out anisotropy and conventional metals are often in the isotropic limit. For example, in figure  \ref{fgOptics4} we show the results of multiple plane wave calculations of $\alpha^2F(\Omega)$ for Pb done by Tomlinson and Carbotte (1976). The solid curve shows the result of the numerical calculation while the dotted curve given for comparison are the experimental results obtained from superconducting tunnelling experiments by McMillan and Rowell (1965). It is this average function which is most relevant to discussions of the mechanism of superconductivity. In this regard the experimentally measured optical conductivity, like the quasiparticle density of states of tunnelling, is an average property over all the electrons. Angle resolved photoemission is momentum specific and yields a directionally specific spectral density which needs to be averaged over directions before it can be made the basis for a discussion of the magnitude of the critical temperature. This same average function determines the many properties of the superconducting state in real materials which differ from BCS predictions (Carbotte \etal1986, Marsiglio \etal 1987, 1992, Mitrovi\'{c} \etal 1980 and Schachinger \etal 1980). Carbotte \etal (2005) have given a discussion of the relationship between optics and quasiparticle self energy. While these are not the same they have some commonality.

In modern discussions of the optical conductivity it has become standard practice to represent the results of experiments as well as
calculations based on the Kubo formula in the extended Drude form, equation \ref{OPTconduc}. A very useful but approximate formula for $1/\tau^{op}(\omega)$ was given by Allen (1971) and extended to finite temperatures by Shulga \etal (1991) as equation \ref{allen} and equation \ref{Shulga}, respectively. The Shulga formula, equation \ref{Shulga}, was originally derived using ordinary perturbation theory without explicit reference to many body theory and in particular to Eliashberg theory. Others (Marsiglio \etal 1998 and Marsiglio 1999) have given more analytical arguments that justify equation \ref{Shulga}. Extensions to energy dependent densities of states have been provided by Sharapov \etal (2005) based on previous work of Mitrovi\'{c} \etal (1985).

\begin{figure}
        \begin{center}
         \vspace*{-0.7 cm}%
                \leavevmode
                \includegraphics[origin=c, angle=0, width=12 cm, clip]{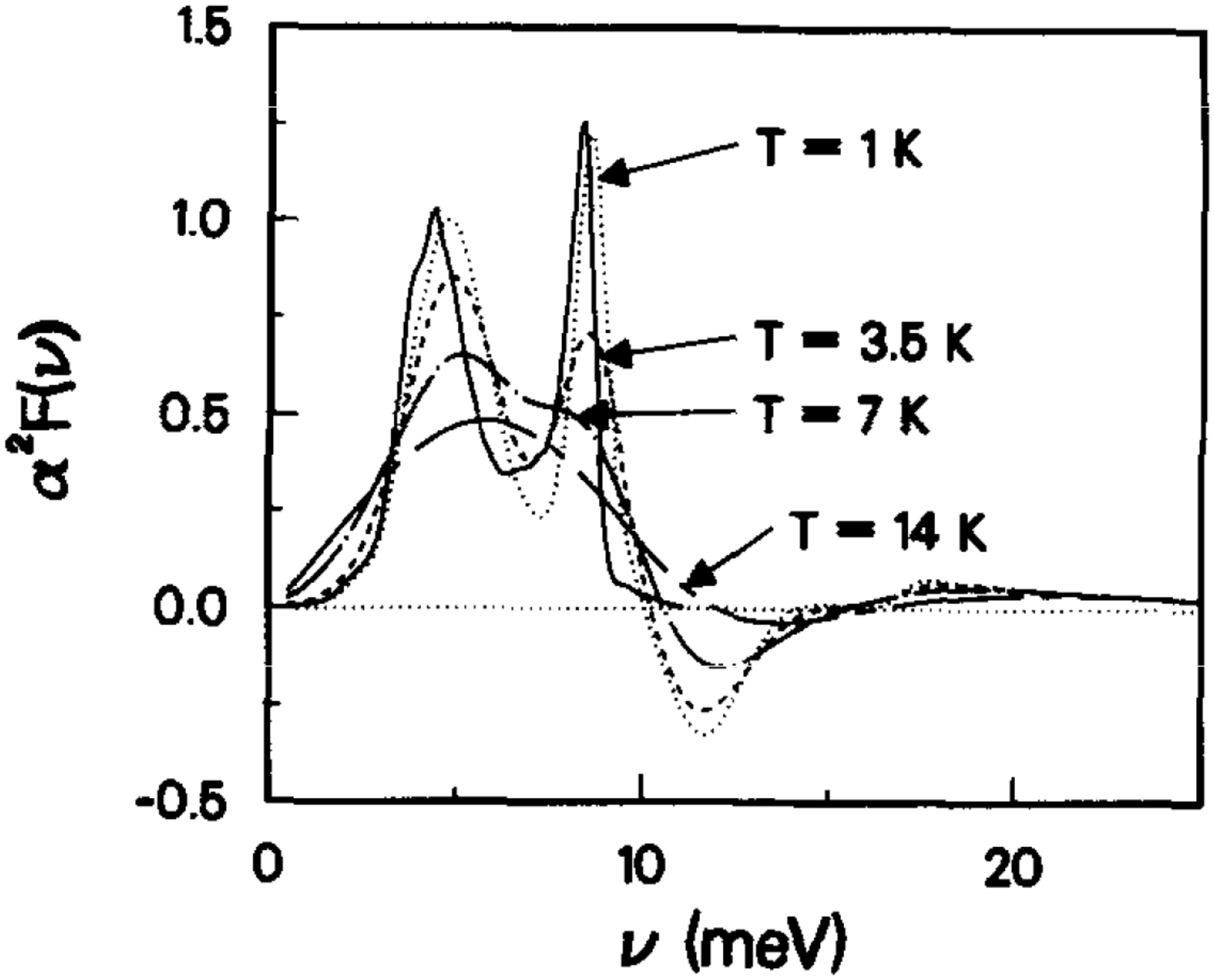}
         \vspace*{-1.2 cm}%
        \end{center}
\caption{The spectral function $\alpha^2F(\Omega)$ calculated from the optical conductivity with the use of equation \ref{second_derivative}. The optical conductivity  is calculated from a model $\alpha^2F(\Omega)$ based on tunnelling data for Pb (solid curve).   At low temperatures the spectral function is retrieved accurately but as the temperature is raised, the fine details are lost, but the overall energy scale
remains. From Marsiglio \etal (1998).}
\label{fgOptics5}
\end{figure}

In as much as the approximate equation \ref{allen} is applicable one immediately finds that
\begin{equation}
\alpha^2(\Omega)F(\Omega) \cong W(\omega) \equiv {1 \over 2\pi}{d^2 \over d \omega^2}\Bigg[\omega{1 \over
\tau^{op}(\omega)}\Bigg].
\label{second_derivative}
\end{equation}

We can write the equation above in terms of the optical conductivity as follows using equation \ref{tausig}:
\begin{equation}
W(\omega) \cong  {\omega_p^2 \over 8\pi^2}{d^2 \over d \omega^2}\Big{\{}\omega {\rm Re}\Big{[}{1 \over
\sigma(\omega)}\Big{]}\Big{\}}.
\label{a2F}
\end{equation}

The validity and limitations of equation \ref{second_derivative} can be tested against complete numerical calculations of the conductivity based on the Kubo formula and Eliashberg equations. Results are shown in figure  \ref{fgOptics5} from Marsiglio \etal (1998). The solid curve is the $\alpha^2F(\Omega)$ vs. $\Omega$ obtained for Pb by McMillan and Rowell (1965) from tunnelling data. This function is then used as input for the normal state numerical calculation of the optical conductivity. The derivative indicated by equation \ref{second_derivative} is carried out on the numeric data for the conductivity and $W(\Omega)$ (denoted $\alpha^2F(\Omega)$ in the figure) computed at $T=1 $ K (the dotted curve) agrees remarkably well with the input $\alpha^2F(\Omega)$ and shows that in this case at least, the optical conductivity can provide information on the underlying electron-boson spectral density in the system. Note the negative tails at higher energies which appear in $W(\Omega)$ but have to be ignored as they cannot be part of the spectral density, which by definition is positive definite. But in the phonon energy range, the second derivative $W(\Omega)$ and $\alpha^2F(\Omega)$ are very close to each other in shape and in absolute value. In this regard $W(\Omega)$ is a dimensionless function. Of course, while equation (\ref{second_derivative}) was established only at $T=0$ K and is approximate, it can be generalized to finite temperatures but for now we point out that a simple second derivative of the optical scattering rate starts to deviate more seriously from the spectral density we wish to know as we raise the temperature. At larger $T=14 $ K the curve for $W(\Omega)$  shows a single peak which is much broader than the two peak structure of the input function. Roughly speaking there is a thermal smearing which makes the simple correspondence
between $W(\Omega)$ and $\alpha^2F(\Omega)$ less accurate and eventually, at higher temperatures, completely break down.

\begin{figure}
        \begin{center}
         \vspace*{-0.5 cm}%
                \leavevmode
                \includegraphics[origin=c, angle=0, width=12cm, clip]{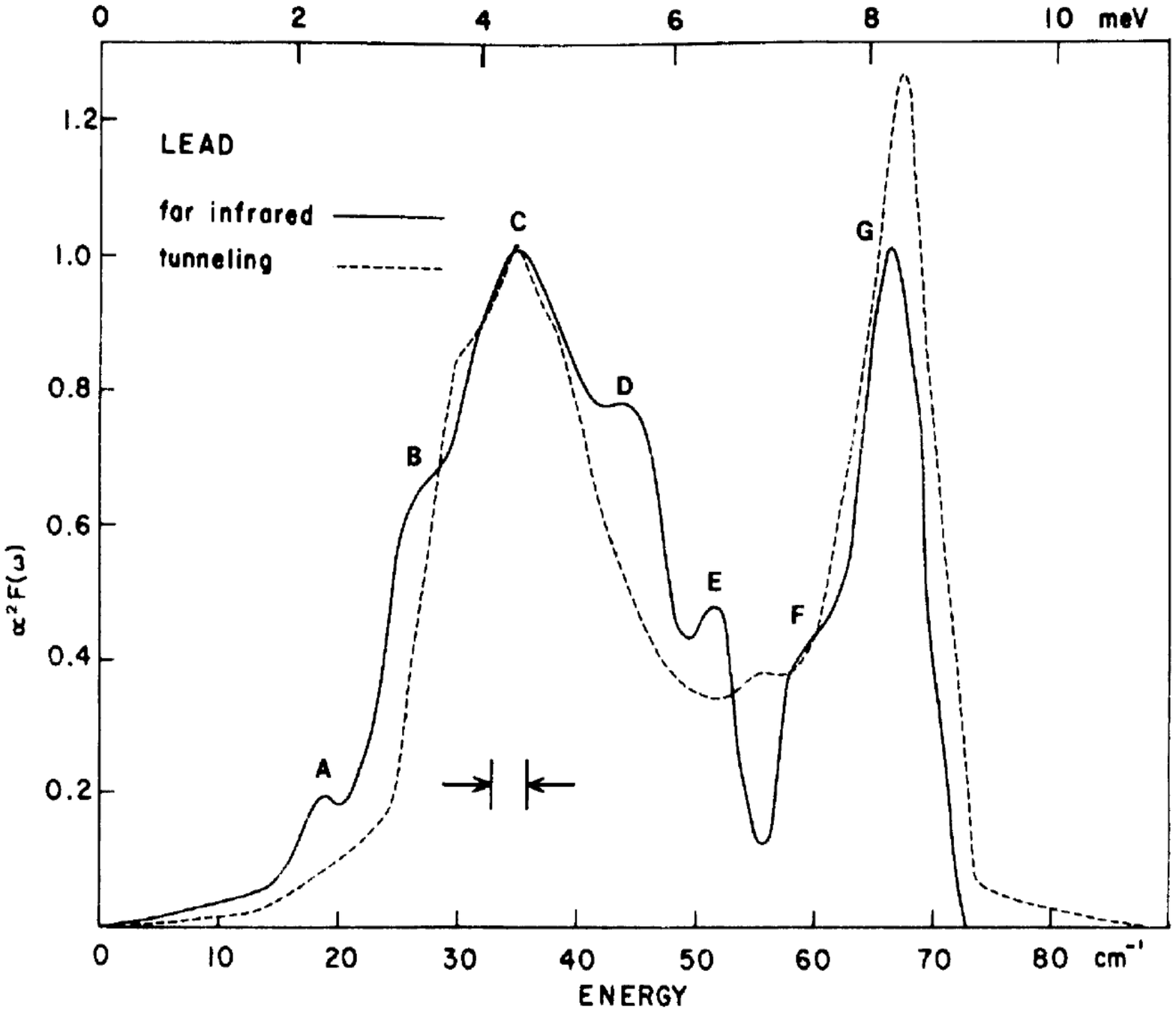}
         \vspace*{-1.1 cm}%
        \end{center}
\caption{Phonon density of states from the optical absorption (solid line) of Pb compared with tunnelling density of states (dashed line) in the same material from Farnworth and Timusk (1976). The curves have been normalized at the point C.  The peaks B, D and F are in good agreement with features in the neutron density-of-states of  lead. The peak A at $\omega=2\Delta$ is associated with phonon lifetime effects.}
\label{fgOptics6}
\end{figure}

One can ask, can one obtain experimental data of sufficient accuracy to be able to extract from it a detailed and accurate image of
$\alpha^2F(\Omega)$? The answer is yes as demonstrated by Farworth and Timusk (1976).  In figure  \ref{fgOptics6} their optical results (solid curve) are compared with the dotted curve obtained from tunnelling. The agreement is very good and shows quite explicitly that the optical conductivity can yield the electron-phonon spectral density in the case of Pb. We note that the optical data seem to contain more details than does the tunnelling data and show in particular an image of phonon lifetime effects at twice the gap (structure labelled A).

\begin{figure}
        \begin{center}
         \vspace*{-0.5 cm}%
                \leavevmode
                \includegraphics[origin=c, angle=0, width=12cm, clip]{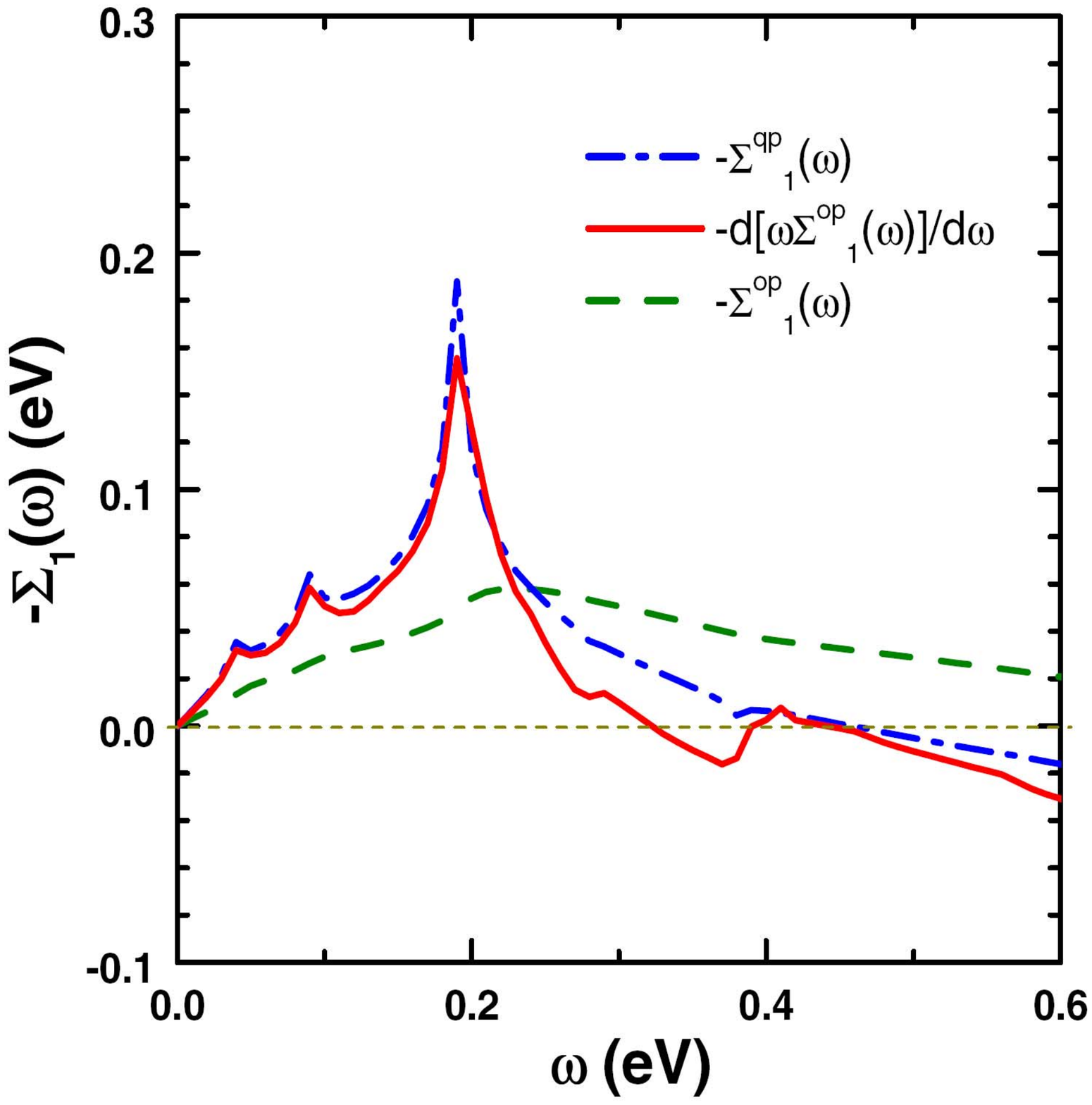}
         \vspace*{-1.1 cm}%
        \end{center}
\caption{Calculation by Hwang \etal (2007a) that shows that an image of the quasiparticle self-energy $\Sigma^{qp}$ can be extracted from optical self energy   via the derivative of $\omega\Sigma^{op}$ (solid red curve). It agrees well in magnitude and detailed structure with the exact theoretical curve for $\Sigma^{qp}$ (dash-dotted blue). The optical self-energy itself (dashed green) is almost structureless and does not agree with the $\Sigma^{qp}$ curve.}
\label{fgOptics7}
\end{figure}

In isotropic systems at low temperature our results show that there exists a close relationship between the quasiparticle self energy
$\Sigma^{qp}(\omega)$ and the corresponding optical self energy $\Sigma^{op}(\omega)$ defined in terms of the generalized Drude formula equation \ref{OPTconduc}. As we have verified that equation \ref{second_derivative} holds with $1/\tau^{op}$ given by equation \ref{allen} we see that $d/d\omega[\omega/\tau^{op}(\omega)] \cong \pi \int_0^{\omega} \alpha^2 F(\Omega) d\Omega$ which is exactly equal to twice the quasiparticle scattering rate in the same approximation. Carbotte \etal (2005) provided a finite temperature generalization and Hwang \etal (2007a) have considered the case when the density of electronic states is not constant as in finite bands. One can define an intermediate quantity $1/\tau^{op-qp}(\omega)$ directly through the measured conductivity scattering rate:
\begin{equation}
{1 \over \tau^{op-qp}(\omega)} \equiv {d \over d\omega }\Big{[}{\omega \over {\tau^{op}(\omega})}\Big{]}
\label{op-qp}
\end{equation}
and take this to provide a first estimate of the momentum averaged quasiparticle scattering rate from which the real part of
$\Sigma^{qp}(\omega)$ follows by Kramers-Kronig transformation.  Complete numerical results based on the Eliashberg equation  are presented in figure \ref{fgOptics7} for a three-delta function model of $\alpha^2F(\Omega)$ with peaks at 0.04 eV, 0.09 eV, and 0.19 eV. The green dashed curve shows the results of a numeric calculation of the Kubo formula for the conductivity from which $\Sigma_1^{op}(\omega)$ is extracted. We see little boson structure in this curve. The red curve however is extracted from the green as $-d[\omega\Sigma_1^{op}(\omega)]/d\omega$ and is to be compared with the blue dash-dotted curve for the quasiparticle self energy calculated for the same spectral density within the Eliashberg theory. The agreement with the solid red curve is good and shows that both methods are able to give the same results. Of course, this only holds if anisotropy can be neglected.

\begin{figure}
        \begin{center}
         \vspace*{-0.5 cm}%
                \leavevmode
                \includegraphics[origin=c, angle=0, width=15cm, clip]{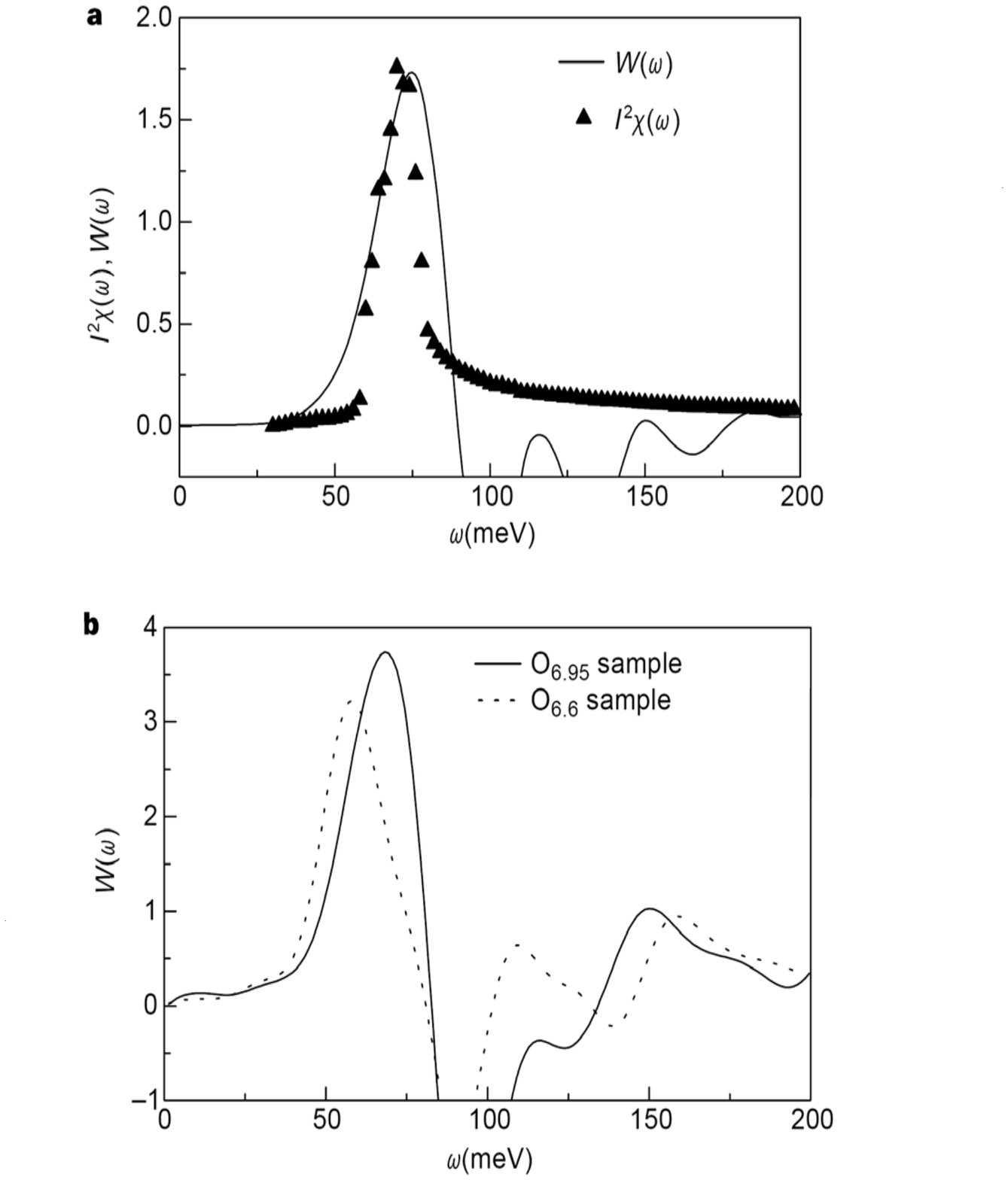}
         \vspace*{-1.1 cm}%
        \end{center}
\caption{Inversion of the superconducting state optical conductivity for YBCO from Carbotte \etal (1999). {\bf a}  Comparison of a model calculation of the bosonic spectral function  based on measured neutron spin susceptibility (triangles) with the derivative of the optical conductivity $W(\omega)$ (equation \ref{a2F}).   b) $W(\omega)$ derived from the experimental data for the conductivity of an optimally doped YBa$_2$Cu$_3$O$_{6.95}$ single crystal with a-axis polarization (solid curve) and for an underdoped, untwinned YBa$_2$Cu$_3$O$_{6.6}$ single crystal (dashed line).}
\label{fgOptics8}
\end{figure}

\begin{figure}
        \begin{center}
         \vspace*{-0.5 cm}%
                \leavevmode
                \includegraphics[origin=c, angle=0, width=15cm, clip]{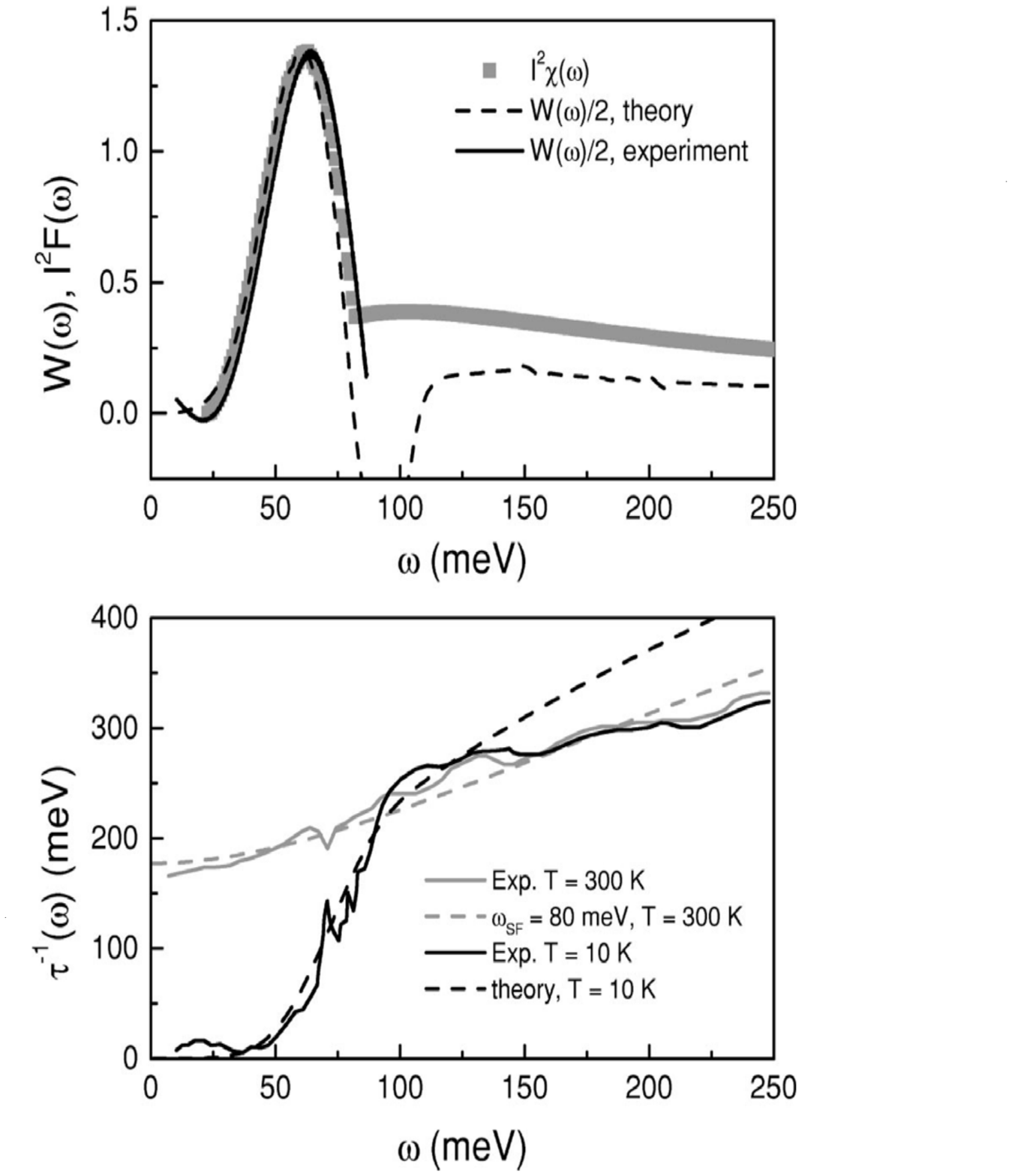}
         \vspace*{-1.1 cm}%
        \end{center}
\caption{The top frame gives the model of Schachinger and Carbotte (2000) for the spin-fluctuation spectral density (displaced by the theoretical gap $\Delta_0 = 24$ meV) for YBa$_2$Cu$_4$O$_8$ (Y124) in the superconducting state at $T= 10$ K (gray solid squares). The dashed line is $W(\omega)/2$ obtained from the calculated conductivity. The bottom frame shows two sets of optical scattering rates and theoretical fits to these. The solid lines are experimental and the dashed lines are the theoretical results. The gray lines are for the normal state at $T = 300$ K and the black ones are for the superconducting state at $T = 10$ K.}
\label{fgOptics9}
\end{figure}
Thomas \etal (1988) analyzed their reflectivity data in \YBCO with $T_c = 92$ K to recover a rough estimate of the underlying spectral
density. Within the analysis they equate the optical scattering rate with its quasiparticle equivalence and recover a spectrum extending from 30 meV to 70 meV with a mass enhancement parameter of the order of 9. A similar calculation was
performed by Collins \etal (1989) who find a peak at about 33 meV and broad background that extends to rather high energies with a coupling constant $\lambda$ of about 2 to 3. Carbotte \etal (1999) took a somewhat different approach taking guidance from measured spin polarized neutron scattering of Bourges \etal (1999a). A spin resonance peak at 41 meV is observed around $(\pi,\pi)$ in momentum space. This peak occurs only below $T_c$. They used their Eliashberg analysis generalized to include approximately a superconducting transition to a d-wave gap to perform two calculations. In the first they use a smooth spectrum without the low temperature resonance peak to model an $\alpha^2F(\Omega)$ function which they denote by $I^2\chi(\Omega)$ to better reflect its inspiration from spin fluctuation theory and adjust its strength to get the measured $T_c$. The coupling between electrons and spin fluctuations is then kept fixed but the shape of the spectrum is readjusted to include the measured spin resonance peak and the conductivity at 5 K in the superconducting state is calculated. In figure  \ref{fgOptics8} (top frame) we show the model $I^2\chi(\Omega)$ (solid triangles) used as well as the second derivative of the real part of the inverse of the conductivity which  according to equation (\ref{a2F}) serves as the definition of $W(\Omega)$  in the superconducting state. On theoretical grounds there is no reason why this derivative should agree with the input
$I^2\chi(\Omega)$. But it was found numerically that the resonance peak in $I^2\chi(\Omega)$ agrees well with $W(\omega)$ provided its frequencies are displaced by the superconducting gap value. In the Eliashberg calculations the gap was 28 meV so that the resonance plus gap frequency was 69 meV. Processing the experimental data in the same way gives the solid curve in the upper frame of figure  \ref{fgOptics8} which bears considerable resemblance to the theoretical curve and gives strong evidence for coupling to the spin resonance in the superconducting state at 5 K. In the lower frame of figure  \ref{fgOptics8} the results for $W(\omega)$  in YBa$_2$Cu$_3$O$_{6.95}$ (solid curve) are compared with results for underdoped sample of YBa$_2$Cu$_3$O$_{6.6}$ (dashed curve) which shows that the resonance seen in optics moves to lower frequency with underdoping.  In later work Schachinger \etal (2000) further applied their method to other systems. From the second derivative of the measured optical conductivity in the superconducting state they conclude that YBa$_2$Cu$_4$O$_8$ (Y124) ($T_c = 82$ K), \TL2201 ($T_c = 90$ K), and \BISCCO ($T_c = 90$ K) should have spin one resonances at 38, 43 and 46 meV respectively while overdoped \TL2201 ($T_c = 23$ K) should have none. After their work, a resonance in \TL2201 with ($T_c = 90$ K) was indeed measured (He \etal 2002) at 47meV but with a rather small resolution limited width. In figure  \ref{fgOptics9} we show results obtained by Schachinger \etal (2000) for the case of Y124 with ($T_c=82$ K). What is shown in the lower frame are the experimental results of Puchkov \etal (1996) at 300 K (light solid) in the normal state for the optical scattering rate which is fit (light dashed line) in Eliashberg calculations of the conductivity with input spectral density given by a simple spin fluctuation form (MMP):
\begin{equation}
I^2\chi(\omega)=I_s {\omega \over {\omega^2+\omega_{sf}^2}}
\label{mmp}
\end{equation}
cut off at $\omega_c \sim 300$ to 400 meV (Millis \etal 1990) and $\omega_{sf} = 80$ meV referred to as an MMP form. The heavy solid black curve are the experimental results at $T = 10$ K in the superconducting state. These experimental results were then processed to get the second derivative $W(\omega)$ and only its resonant piece which peaked at 38 meV (resonance) + 24 meV (gap) = 62 meV was retained and used to modify the MMP form found at $T_c$ by replacing its value below 38 meV by the resonance peak which was interpreted as a spin resonance contribution. The result was the heavy dashed curve for the calculated superconducting state optical scattering rate of figure  \ref{fgOptics9} (lower frame) which agreed well with the data (heavy solid curve). The second derivative curve obtained from the theoretical data is shown as the dashed curve in the upper frame which agrees well with the experimental data (solid curve) in the resonance energy region up to 62 meV and with the ($T = 10$ K) input spectral density (solid squares) for $I^2\chi(\omega)$. It is important to realize that no reference to neutron scattering data is made in this comparison which proceeds entirely on the basis of charge dynamic data based on optics.

\begin{figure}
        \begin{center}
         \vspace*{-0.7 cm}%
                \leavevmode
                \includegraphics[origin=c, angle=0, width=15cm, clip]{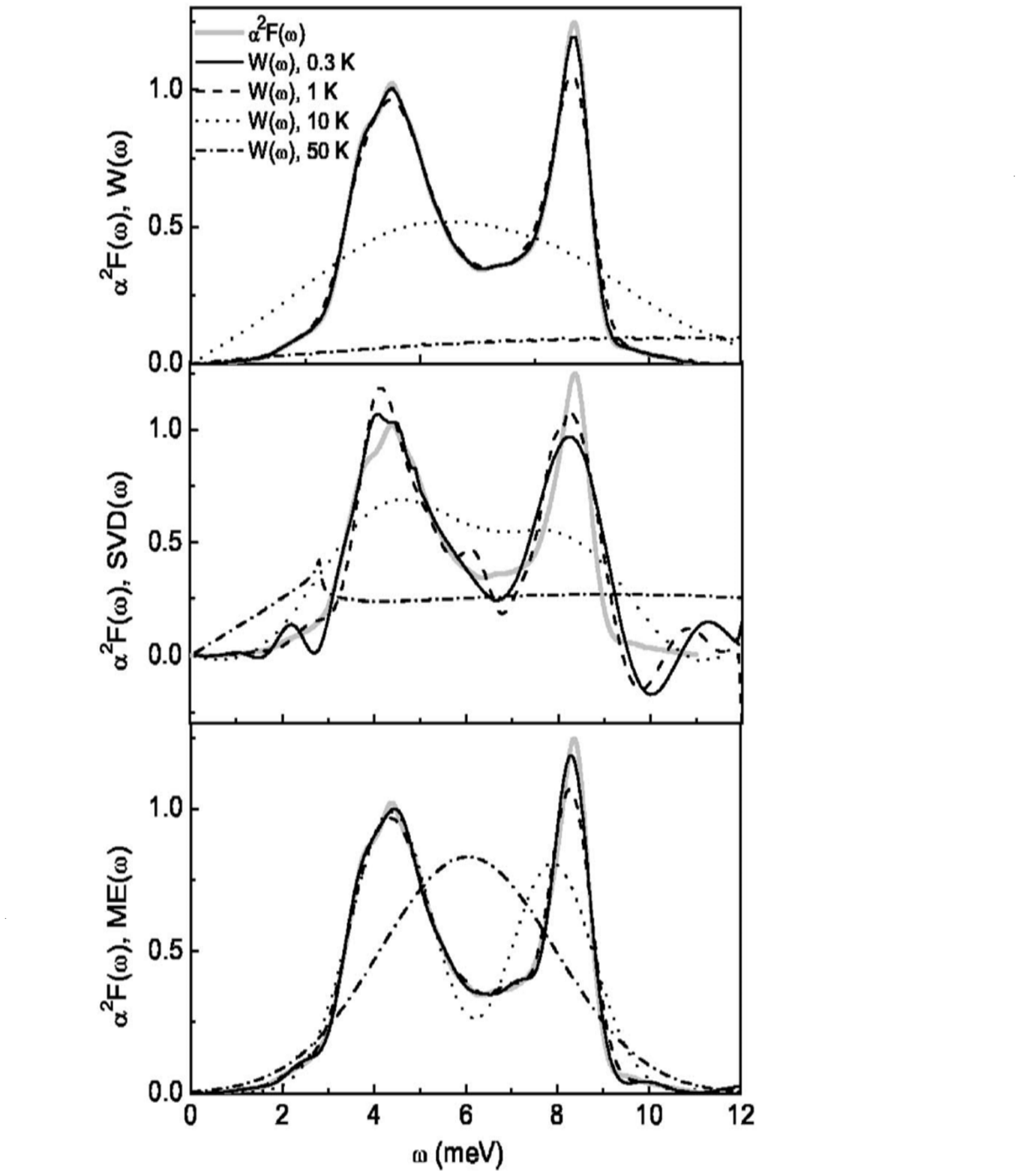}
         \vspace*{-1.1 cm}%
        \end{center}
\caption{Inversion of finite temperature, normal state $1 / \tau^{op}(\omega,T)$ data computer generated using kernel equation \ref{Shulga}. The solid lines correspond to the temperature $T = 0.3$ K, dashed lines to 1 K, dotted lines to 10 K, and dash-dotted lines to 50 K. The grey solid line represents the $\alpha^2(\Omega)F(\Omega)$ spectral function applied to calculate the optical scattering rate data. Top frame: Second derivative method. Center frame: SVD method. Bottom frame: MaxEnt method. From Schachinger \etal (2006)}
\label{fgOptics10}
\end{figure}
\begin{figure}
        \begin{center}
         \vspace*{-0.5 cm}%
                \leavevmode
                  \includegraphics[origin=c, angle=0, width=15cm, clip]{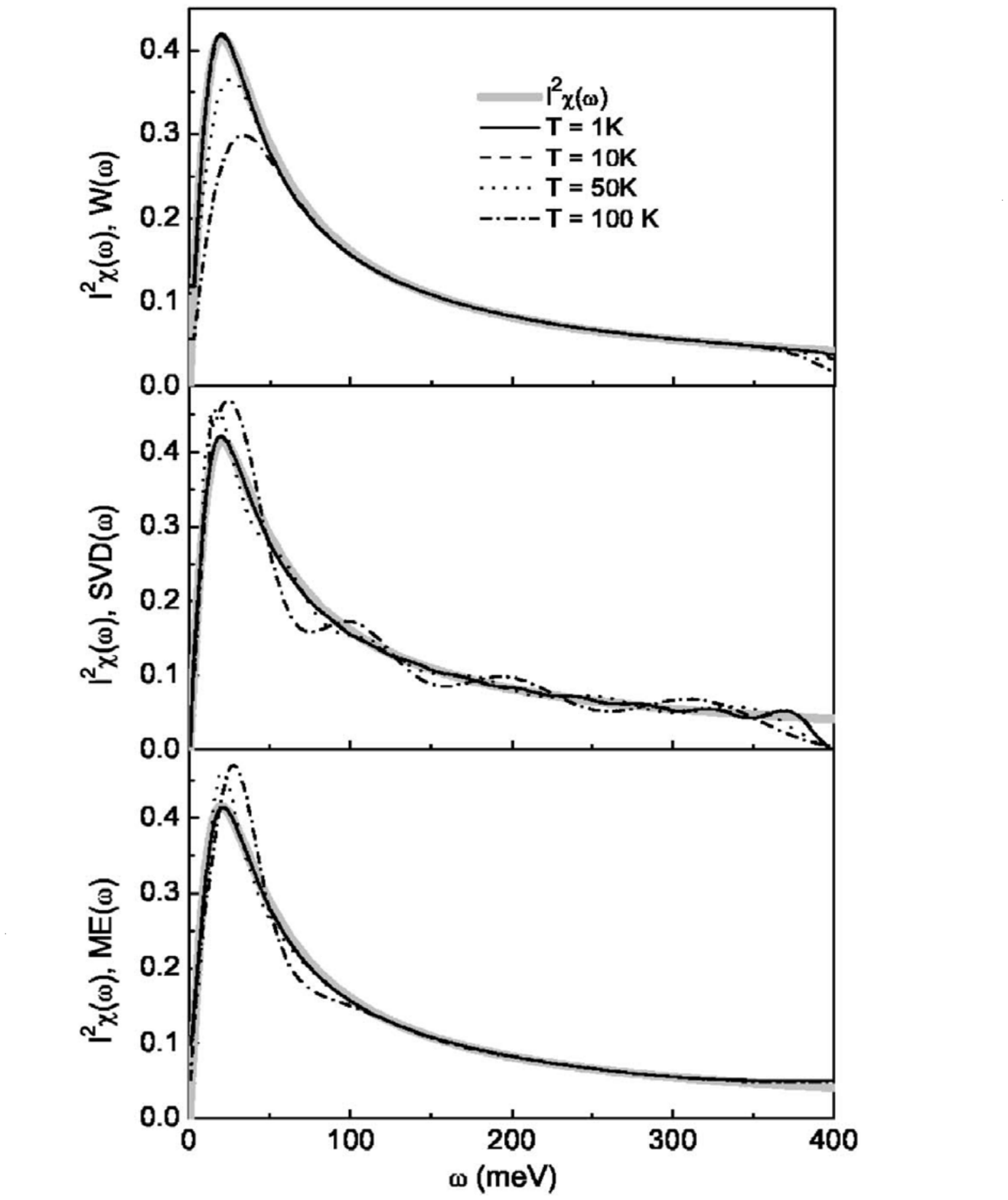}
         \vspace*{-1.1 cm}%
        \end{center}
\caption{Same as figure  \ref{fgOptics10} but now with $I^2\chi(\Omega)$ given by equation \ref{mmp} with $I^2 = 20$ meV and $\omega_{sf} = 20$ meV has been used and the input optical scattering time is taken from the optical calculation calculated in Eliashberg theory without simplications.}
\label{fgOptics11}
\end{figure}

While the second derivative method has been quite useful in the analysis of the optical data, in view of its obvious limitations,  it seems worth while to look at the possibility of having an alternate method which does not suffer from unwanted and unphysical
negative tails in $I^2\chi(\omega)$, although it should be noted that Abanov et al. (2001) have made use of this negative tail to extract the superconducting gap. One can start with a model problem namely the inversion of the equation
\begin{equation}
{1 \over \tau^{op}}(\omega,T)= \int_0^{\infty}d\Omega\:I^2(\Omega)\chi(\Omega)\:K(\omega, \Omega, T)
\label{tau1}
\end{equation}
with the kernel $K(\omega,\Omega,T)$ given by the thermal factor in the square brackets of equation equation \ref{Shulga} (times $\pi/\omega$) to recover the spectral density $\alpha^2(\Omega)F(\Omega)$ from a finite temperature optical scattering rate $1 / \tau^{op}(\omega)$. Dordevic \etal (2005) have considered the so-called singular value decomposition (SVD) method and Schachinger \etal (2006) considered in addition a maximum entropy scheme and offer comparisons between the two methods. In figure  \ref{fgOptics10} we reproduce a comparison from this last reference. In each frame the input $\alpha^2(\Omega)F(\Omega)$ is given as the shaded curve while solid; the dashed, dotted and the dash dotted curves give results of the inversion of equation \ref{tau1} for $T=$ 0.3 K, 1 K, 10 K, and 50 K respectively, with computer data for $1/\tau^{op}(\omega,T)$ calculated from (approximate) equation \ref{Shulga}. The top frame uses the second derivative method, the middle is for SVD and the bottom frame the maximum entropy case, denoted by ME$(\omega)$. All show thermal smearing as $T$ is increased with the double peak moving towards a single peak structure. The SVD curve shows considerable unwanted oscillations and produces negative values for the spectral density. This is not allowed in the maximum entropy case which is a better choice. In figure  \ref{fgOptics11} we show additional inversion results but with two differences. First a spectrum, more appropriate for the normal state when a spin fluctuation mechanism applies with $I^2\chi(\Omega)$ given by equation \ref{mmp}. The second difference is that now  the scattering rate on the left hand side of equation \ref{tau1} is based on results obtained from full numerical solutions of the Eliashberg equations and the Kubo formula. As before, the shaded curve is the input $I^2\chi(\Omega)$ and the others the results of the inversion at different temperatures. Note that the SVD solutions (middle frame) have extra wiggles, which are not part of the input $I^2\chi(\Omega)$, and these are absent in the top and bottom frames. While the results in the top frame show a temperature smearing of the peak around $\omega_{sf}$ and a small shift in this peak towards higher energies not part of the input data, the differences are not large. The same is true for the bottom frame where a maximum entropy method is used. As previously commented, it is the entire function $I^2\chi(\Omega)$ which determines the size of the critical temperature in a d-wave Eliashberg superconductor so that the small differences just mentioned as $T$ is increased are not of major concern, at least for this purpose. We will return to the issue later however when we look at the experimental results.

\begin{figure}
        \begin{center}
          \vspace*{-0.5 cm}%
                \leavevmode
                  \includegraphics[origin=c, angle=0, width=15cm, clip]{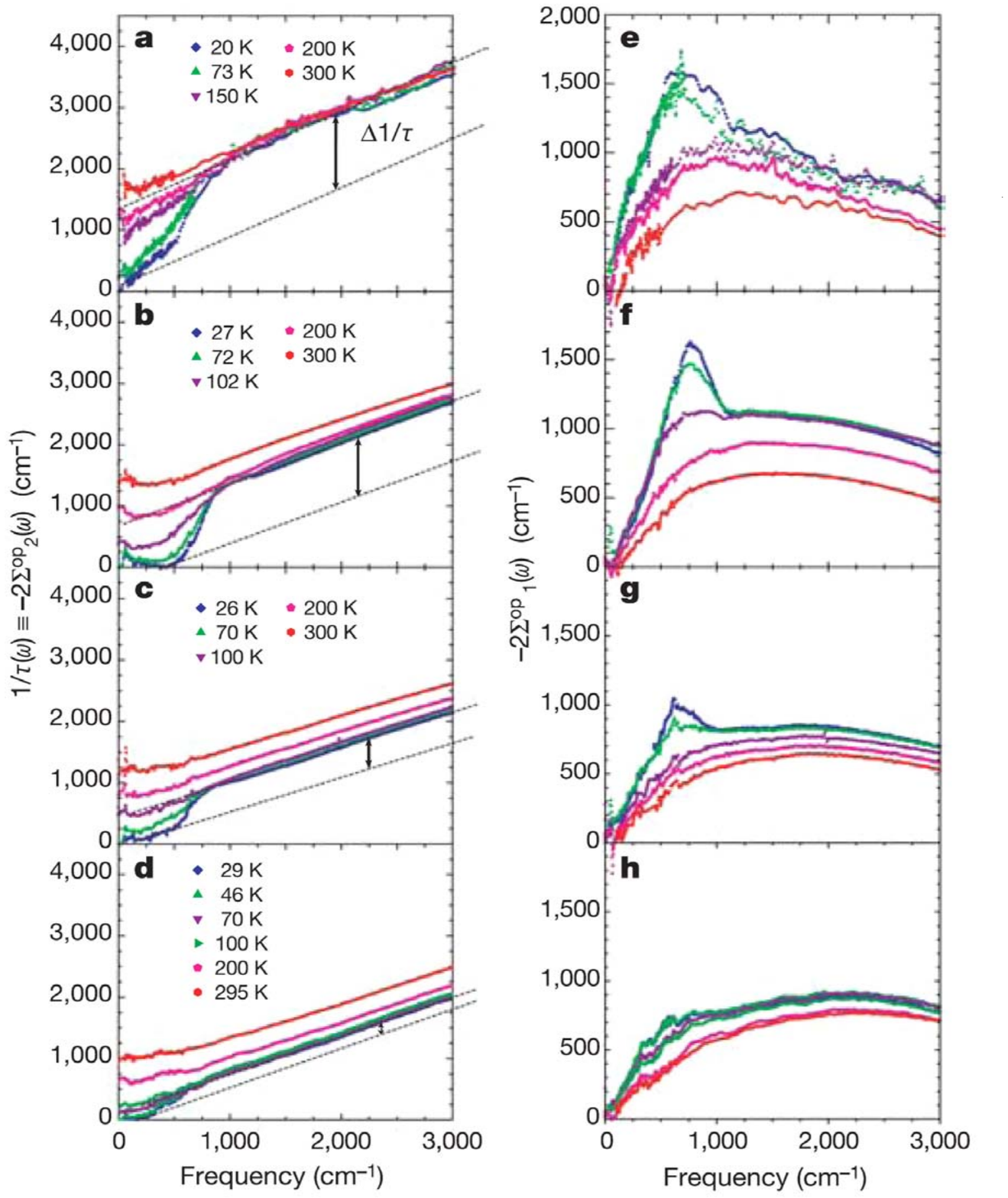}
         \vspace*{-1.0 cm}%
        \end{center}
\caption{The optical single-particle self-energy of \BISCCO from Hwang \etal (2004). {\bf a-d}, The doping and temperature dependent optical scattering rate, $1/\tau(\omega)$ at four doping levels. {\bf a}, $T_c = 67$ K (underdoped); {\bf b}, 96 K (optimally doped); {\bf c}, 82 K (overdoped); {\bf d}, 60 K (overdoped). {\bf e-h}, The real part of the optical self-energy. The slope of $-2\Sigma_1^{op}(\omega)$ near $\omega = 0$ is proportional to the mass enhancement factor, and also decreases as the doping increases, consistent with other studies. We note the weakening of the feature at 700 \cm in both sets of curves as the doping level increases.}
\label{fgOptics12}
\end{figure}

\begin{figure}
        \begin{center}
         \vspace*{-0.5 cm}%
                \leavevmode
                  \includegraphics[origin=c, angle=0, width=15cm, clip]{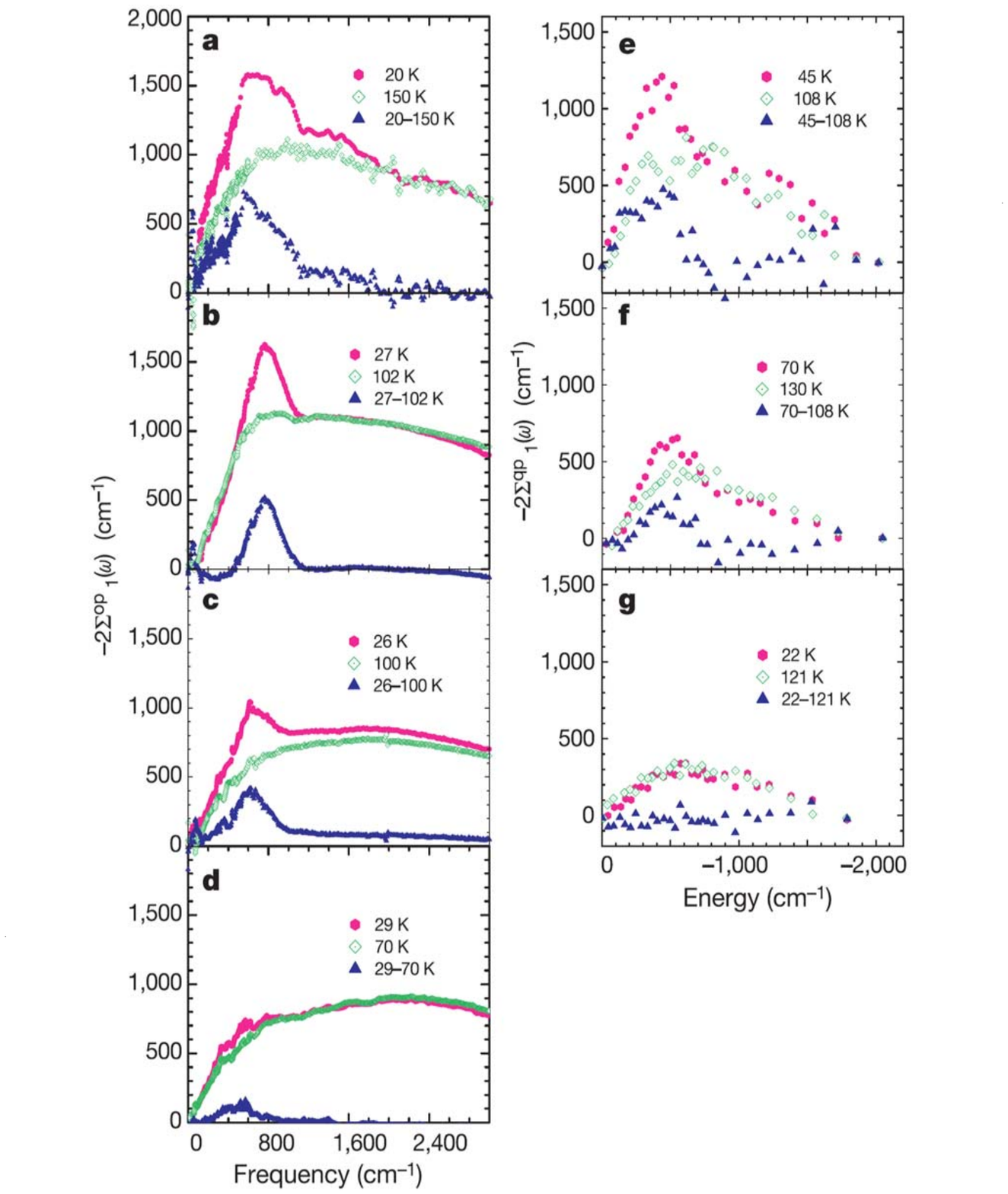}
         \vspace*{-1.1 cm}%
        \end{center}
\caption{Comparison of the self-energy measured with infrared and angle-resolved photoemission for Bi-2212. {\bf a-d}. The real part of the optical self-energies in the normal and superconducting phases derived from the scattering rates of figure  \ref{fgOptics13}. {\bf e-g} The real part of the self-energies, from the ARPES measurements of Johnson \etal (2001). We note the close correlation of the two sets of data. The parallel evolution of the spectroscopic features such as the peak at 400 \cm with temperature and doping in the two sets of experiments is striking (allowing for the higher noise level and lower resolution of the photoemission data). From Hwang \etal (2004)}
\label{fgOptics13}
\end{figure}
We have seen that the function $I^2\chi(\Omega)$ can be found by the inversion of the optical scattering rate, but it is the real part of the optical self energy that  is most closely related to the renormalized quasiparticle dispersion curves measured in ARPES. We begin with a discussion of data on $\Sigma_1^{op}(\omega)$ with an eye at a qualitative comparison with ARPES. In figure  \ref{fgOptics12} we show results for $1/\tau^{op}(\omega)$ or $-2\Sigma_2^{op}(\omega)$, in the left column and the corresponding real part $-2\Sigma_1^{op}(\omega)$  in the right column from  Hwang \etal (2004). From top to bottom we go from underdoped to overdoped samples of \BISCCOa, as indicated in the caption and various temperatures above and below $T_c$. Starting with $-2\Sigma_1^{op}(\omega)$ we see a remarkable resemblance with the ARPES measurement of Johnson \etal (2001) given in the previous ARPES section (figure  \ref{fgARPES3}). A more critical comparison is made in figure  \ref{fgOptics13} which we reproduce here from Hwang \etal (2004). As was done in the ARPES comparison we subtract from the data at low temperature in the superconducting state, the value of $-2\Sigma_1^{op}(\omega)$ just above $T_c$ in the normal state. The difference, shown as blue triangles, in the left frame shows that there is an extra peak in the superconducting as compared to the normal state. The peak varies in energy with doping, becoming weaker as one goes from underdoped
to overdoped. This agrees with the data of Johnson \etal (2001) also shown for ease of reference in the right hand column. The temperature and doping dependence of the peak seen in optical characteristics can be taken as further support for the interpretation given to this peak by Johnson \etal (2001), as related to spin fluctuations. We point out however that this peaks is far too weak to completely saturate the optical self energy and consequently corresponds to only a small part of the total $I^2\chi(\Omega)$ which, it should be noted, is the quantity most relevant to the glue which causes the superconductivity. We can sharpen this argument by looking not at the self energy or scattering rates themselves but at the underlying spectra $I^2\chi(\Omega)$ obtained through maximum entropy inversion of the optical data.

\begin{figure}
        \begin{center}
         \vspace*{-0.5 cm}%
                \leavevmode
                  \includegraphics[origin=c, angle=0, width=15cm, clip]{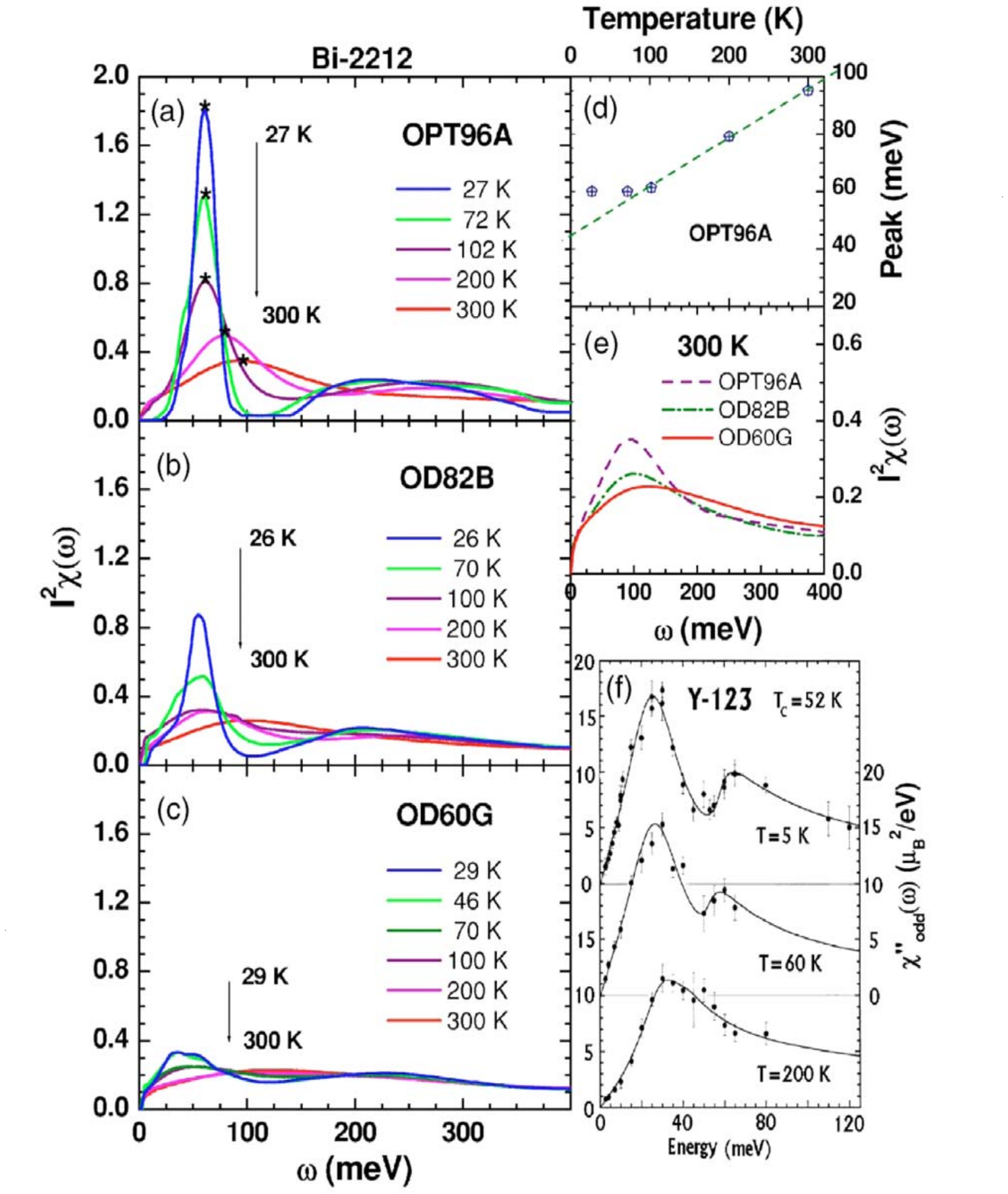}
         \vspace*{-1.1 cm}%
        \end{center}
\caption{The bosonic spectral density $I^2\chi(\Omega)$ of \BISCCO as determined by a maximum entropy inversion of the optical scattering rate at three doping levels (OPT96A $T_c=96$ K, OD82B $T_c=82$ K, and OD60G $T_c=60$ K) from Hwang \etal (2007b). At room temperature all samples exhibit a broad continuum background shown in panel (e). On lowering the temperature, the broad background peak in the 300 K spectrum evolves into low-energy peak with a deep valley above it. The spectral weight gained in the peak is roughly balanced by the spectral weight loss in the valley in all three samples. With increasing doping the intensity of the peak and intensity lost in the valley are significantly weakened (see figure \ref{fgOptics15} b). Panel (d) shows the temperature dependence of the frequency of the peak maximum in OPT96A marked by a star. Panel (f) shows the local magnetic susceptibility of underdoped \YBCO from neutron scattering. We note a qualitatively similar temperature evolution of panel (a).}
\label{fgOptics14}
\end{figure}

\begin{figure}
        \begin{center}
         \vspace*{-0.5 cm}%
                \leavevmode
                  \includegraphics[origin=c, angle=0, width=15cm, clip]{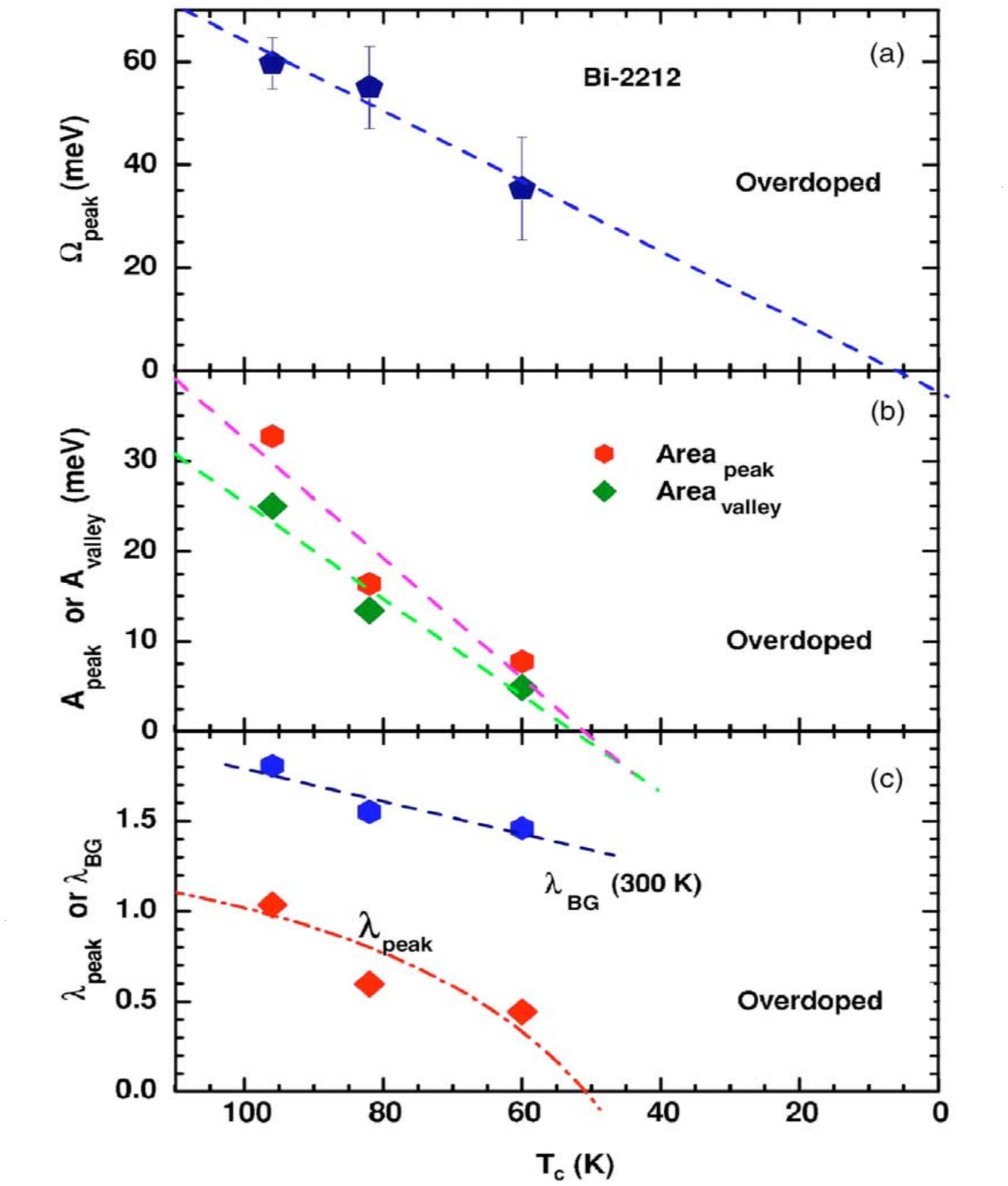}
         \vspace*{-1.1 cm}%
        \end{center}
\caption{$T_c$ dependent properties of the peak and the valley in the bosonic spectral function of figure  \ref{fgOptics14} from Hwang \etal (2007b). The central peak frequency is proportional to $T_c$, i.e., $\omega_{peak} = 8.0 k_B T_c$. The peak and valley are closely connected, have a very similar $T_c$ dependence, and vanish at the same $T_c = 50$ K. (c) The coupling constant $\lambda_{peak}$ and $\lambda_{BG}$ (background). $\lambda_{peak}$ vanishes at the same $T_c$ as the peak and valley.}
\label{fgOptics15}
\end{figure}

This is shown in figure  \ref{fgOptics14} for three samples of \BISCCO OPT96A OD82B and OD60G, from top to bottom, left hand column, at various temperatures reproduced from Hwang \etal (2007b). As we expect from our previous discussion of the corresponding real part of the self energy, a peak is seen in  $I^2\chi(\Omega)$ around 60 meV in the OPT96A sample. This peak is reduced in height as $T$ increases and shifts to higher energies. Also, as the doping increases, the peak loses amplitude and shifts somewhat in frequency. In the overdoped sample OD60G it has almost disappeared. More importantly, the overall magnitude of the spectral density is reduced in the overdoped regime as compared with underdoped. These trends are quantified in figure  \ref{fgOptics15}, where we show the energy of the peak position as a function of critical temperature value in the top frame. For our samples $\Omega_{peak}\cong 8.0 k_BT_c$. Noting that above the peak there is a valley which is particularly prominent at optimum doping (top frame of figure  \ref{fgOptics14}, we can also define an area associated with it as we can for the peak. As temperature or doping is increased this valley gets progressively filled in as the peak size reduces and the peak energy moves to higher values. These trends with temperature are qualitatively captured in the phenomenological analysis of Prelo\^sek  and Sega (2006) of the temperature evolution of the magnetic collective mode which they describe using a memory function approach.  They also contrast their results with those obtained in a more conventional formalism based on the random phase approximation for the spin susceptibility.
In the middle frame of  figure  \ref{fgOptics15} we show the peak area (red hexagons) as well as the area of the valley (green diamonds). Both quantities drop towards zero, as $T_c$ is  reduced. Similarly the authors define a value of the mass enhancement $\lambda$ associated with the background (blue hexagons) which extends to very high values of energy as well as a mass enhancement associated with the peak (red squares). While the $\lambda$ associated with the peaks is rapidly dropping towards a zero value for $T_c \sim 50$ K, the background $\lambda$ is not. It does decrease with decreasing value of $T_c$ but remains significantly above unity  for $T_c \sim 50$ K. It is this $\lambda$ which in the spin fluctuation formulation controls the size of $T_c$ and not the value of $\lambda$ associated with the peaks seen only in the superconducting state at low temperatures. Returning to the right hand column of figure  \ref{fgOptics14}, we show in frame (d) the temperature dependence found for the peak position in the OPT96A sample.  It moves from 60 meV in the superconducting state at 27 K to nearly 100 meV at room temperature. At the same time the shape of the spectrum has changed from a sharp peak, around 60 meV followed by a valley which is then followed by a second very broad peak at 200 - 300 meV, to a simple background which looks much more like the MMP spectrum of equation \ref{mmp} of spin fluctuation theory. Below $T_c$, in the superconducting state, the evolution of the peak looks more like a simple reduction of the area under it, while above $T_c$ it also shows a tendency to move. The $T =$ 300 K spectra for each of the three samples are compared in frame (e). Each look like MMP forms and are all comparable in size. The temperature evolution just outlined for Bi-2212 is very similar to that for the odd component of the spin susceptibility seen in a sample of \YBCO with $T_c= 52$ K reproduced in frame (f) from the inelastic neutron scattering (Bourges \etal 1997, 1999b and Fong \etal 2000).

\begin{figure}
        \begin{center}
         \vspace*{-0.5 cm}%
                \leavevmode
                \includegraphics[origin=c, angle=0, width=15cm, clip]{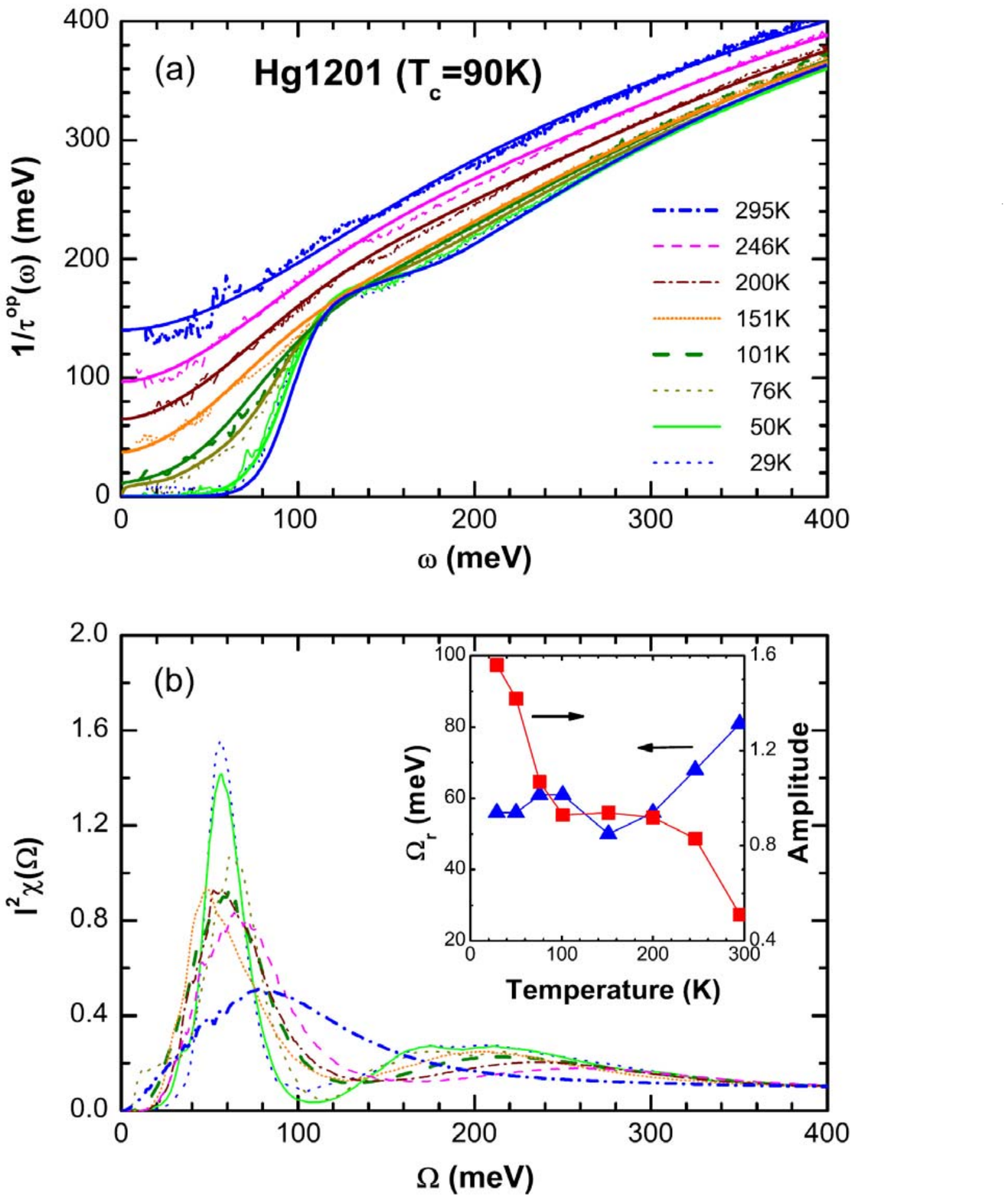}
         \vspace*{-1.2 cm}%
        \end{center}
\caption{Top frame, the optical scattering rate $1/\tau^{op}(T,\omega)$ for Hg1201 vs $\omega$ for 8 temperatures (light curves). The wider curves are our maximum entropy reconstructions. Bottom frame, the electron-boson spectral function $I^2\chi(\Omega)$ vs $\Omega$. The inset gives the peak position (blue triangles) left scales as a function of temperature and the red squares give the corresponding peak amplitude. From Yang \etal (2009).
}
\label{fgOptics17}
\end{figure}

We remind the reader that the superconductivity state spectra of figure  \ref{fgOptics14} were obtained using a maximum entropy inversion
technique based on a generalization to the superconducting state of our equations \ref{Shulga} and \ref{tau1} as developed by Schachinger
\etal (2006). At zero temperature the appropriate kernel which replaced the kernel $K(\omega, \Omega,T)$ in equation \ref{tau1} is
\begin{equation}
K_{sc}(\omega, \Omega)={{2\pi} \over {\omega}}\Big{\langle}(\omega - \Omega)\theta(\omega + 2 \Delta_0(\phi)-\Omega) {{\rm
E}\Big{(}\sqrt{1-{{4\Delta_0^2(\phi)} \over {(\omega-\Omega)^2}}}\Big{)}}\Big{\rangle}
\label{skernel}
\end{equation}
where the bracket $\langle \  \rangle$ indicates an integration over angles $\phi$, $\Delta_0(\phi)$ is the d-wave superconducting gap and ${\rm E}(\omega)$ is the elliptic integral. This form is used to get a first estimate of $I^2\chi(\Omega)$ in the superconducting state. In a final calculation, the spectral density obtained is further modified slightly through a least square fit to the experimental scattering rates of the full Kubo formula results obtained from the d-wave Eliashberg equations. These are given in Schachinger \etal (2006) where the procedure is further described.
A more recent application of the MaxEnt inversion is
reproduced in figure  \ref{fgOptics17}  for the one-layer mercury compound  Hg1201 from Yang \etal (2009). The top frame shows the experimentally measured scattering rate at a series of temperatures while the bottom frame displays the bosonic spectral function for the eight values of temperature as labeled in the top frame. At low temperature in the superconducting state we see a prominent low energy peak around 56 meV, which is followed by a valley and then by a second broad region of spectral weight extending to high energies. Subsequent to the optical work, Yu \etal (2010) found a spin resonance in inelastic neutron scattering studies at $\Omega_R$ = 56 meV in excellent agreement with the above data. This surprisingly high frequency for the neutron resonance was confirmed subsequently by van Heumen \etal (2009). The energy of the lower peak $\Omega_r$ while almost constant in the superconducting state, begins to move to higher energy as $T$ rises towards room temperature as seen in this inset to the frame (solid blue triangles). At the same time the amplitude of this peak drops by a factor of 2 in height by the time $T_c$ is reached (solid red squares). Also the two peak structure evolves into a single peak MMP like spectrum (equation \ref{mmp}) at room temperature.

\begin{figure}
        \begin{center}
         \vspace*{-0.5 cm}%
                \leavevmode
                \includegraphics[origin=c, angle=0, width=15cm, clip]{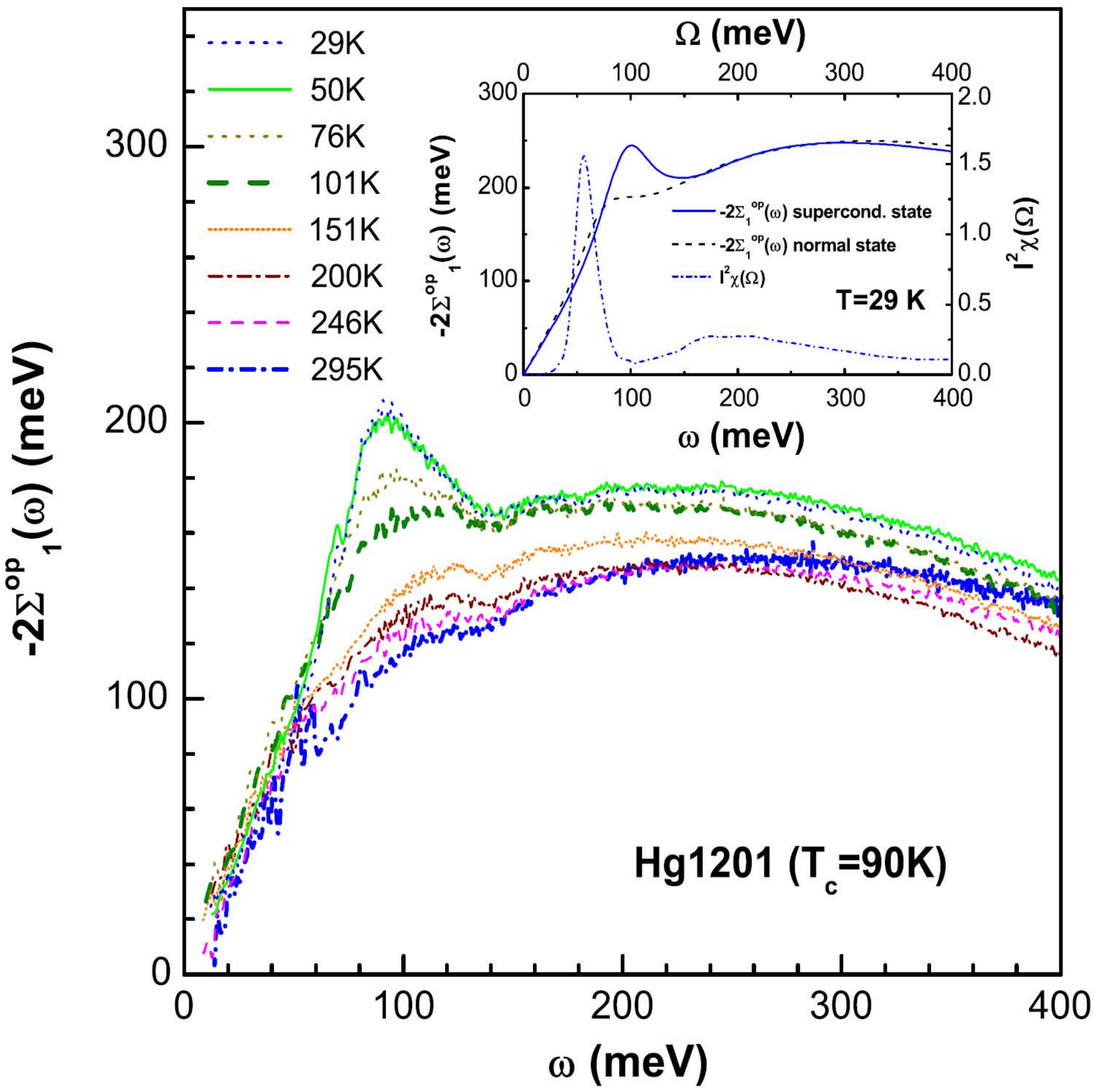}
         \vspace*{-1.2 cm}%
        \end{center}
\caption{The optical self-energy of Hg-1201 from Yang \etal (2009). Inset, theoretical results based on numerical solutions of the generalized Eliashberg equations. (Solid blue superconducting and dashed black normal state.) The electron-boson exchange spectral density used is shown as the dash-dotted blue curve. The superconducting gap value is $\Delta =$ 22.4 meV.}
\label{fgOptics18}
\end{figure}

\begin{figure}
        \begin{center}
         \vspace*{-1.0 cm}%
                \leavevmode
                \includegraphics[origin=c, angle=0, width=13cm, clip]{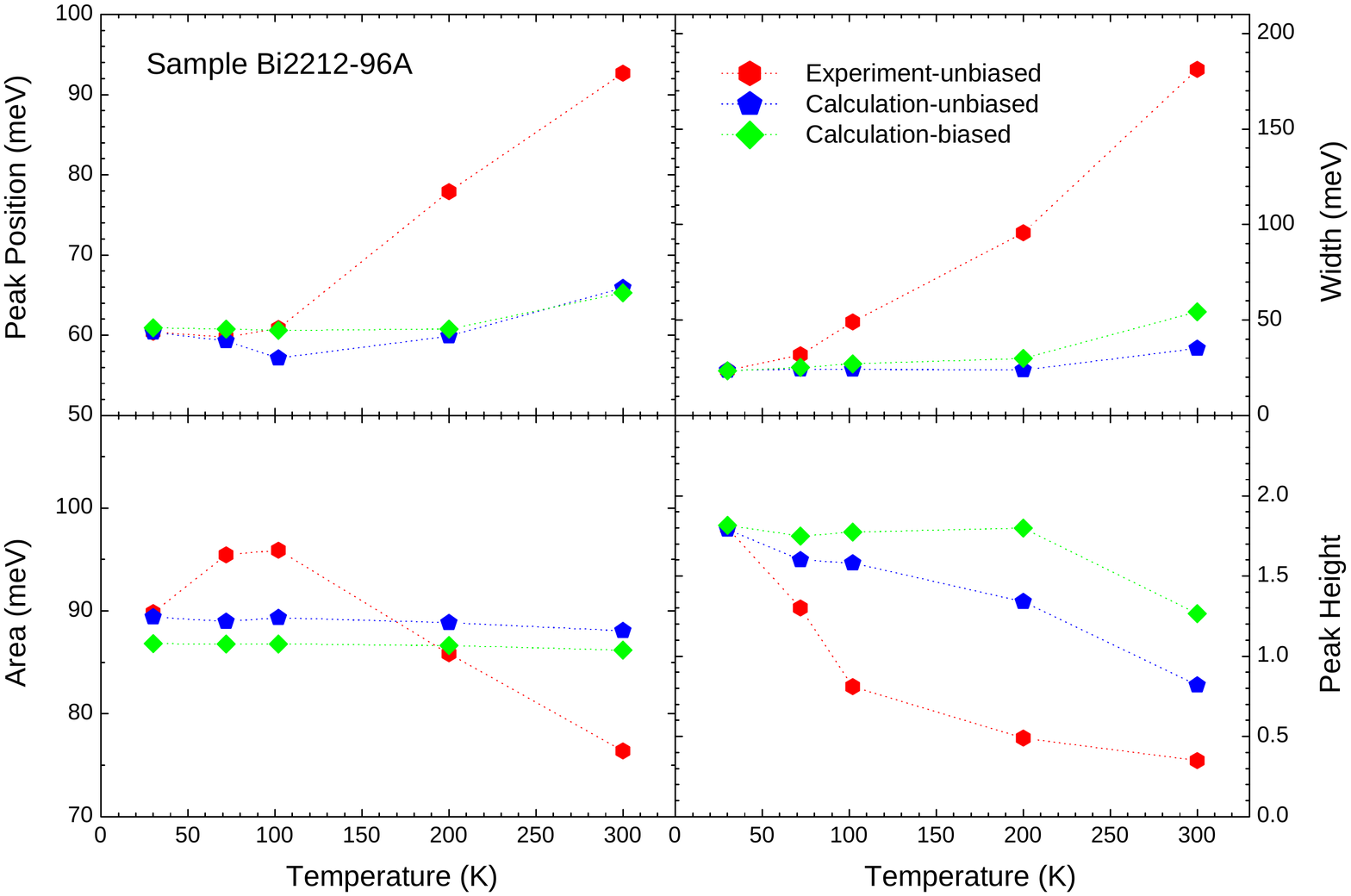}
         \vspace*{-1.1 cm}%
        \end{center}
\caption{Comparison of fitting parameters obtained from the actual measured optical data and the simulated optical scattering rate calculated from the Kubo formula and full Eliashberg equations. (see in the text)}
\label{fgOptics16a}
\end{figure}

The optical scattering rates on which these data are based are shown in the top frame of figure \ref{fgOptics17} where the maximum entropy fits are also shown. In the superconducting state these results are based on the Kubo formula for the conductivity with full numerical solutions of the generalized Eliashberg equation for a d-wave superconductor. The value of the low temperature superconducting gap returned from the solution of the Eliashberg equations was equal to 22.4 meV for a gap to $T_c$ ratio $2\Delta_0/k_B T_c$ = 5.8. In figure  \ref{fgOptics18} we show further results from the work of Yang \etal (2009) where they plot the real part of the optical self energy $-2\Sigma_1^{op}(\omega)$ in meV vs. energy $\omega$ in meV. Note the peak in this function at low temperature in the superconducting state. This peak is mainly attributed to the effect of the superconducting gap as illustrated in the inset of the figure where we show the input spectrum $I^2\chi(\Omega)$ (blue dot dashed curve), the optical self energy $-2\Sigma_1^{op}(\omega)$ (solid blue) and the same quantity in the normal state (dashed black curve). The second curve does not have the peak around 100 meV seen in the superconductivity case which falls, as we expect, at an energy closer to $\Omega_r +2\Delta_0 \cong 101$ meV.

\begin{figure}
        \begin{center}
         \vspace*{-1.0 cm}%
                \leavevmode
                \includegraphics[origin=c, angle=0, width=15cm, clip]{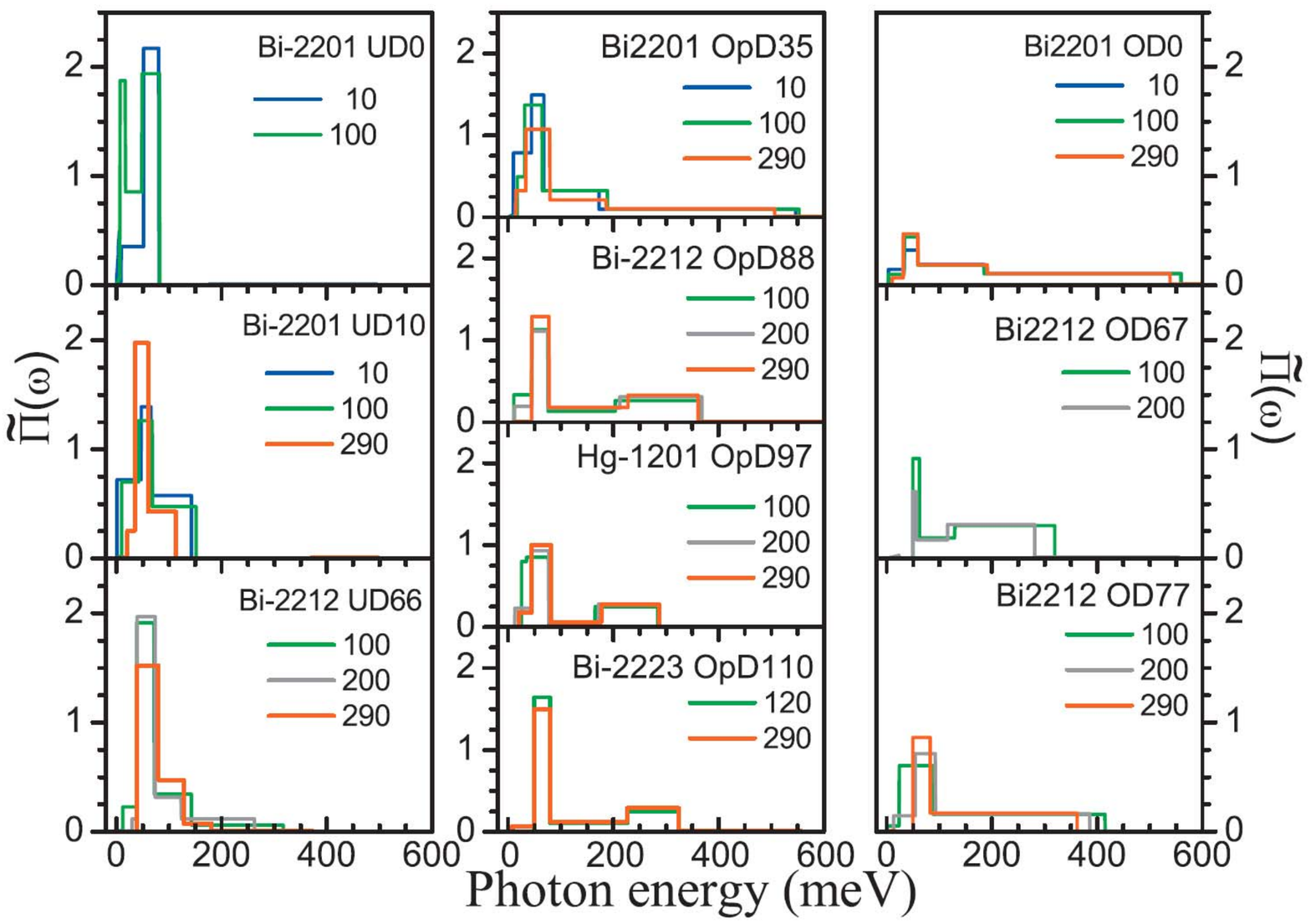}
         \vspace*{-1.5 cm}%
        \end{center}
\caption{Electron-boson coupling function for different cuprate superconductors from van Heumen \etal (2009). Left panels, underdoped
materials, middle panels, optimally doped and right panels, overdoped.}
\label{fgOptics16}
\end{figure}

In the maximum entropy inversions that we have described so far an unbiased mode was used at each temperature. It is instructive to use instead a biased mode in which the starting spectral density for inversion at temperature $T$ is taken to be the converged answer obtained for the previous next lowest temperature. While this usually provides equally good fits to the scattering rate data, it gives solutions for $I^2\chi(\Omega)$ which show considerably less temperature dependencies and are closer to those obtained on the basis of equation \ref{MoreExact} with a least squares fit of a histogram as in figure  \ref{fgOptics16}. Part of, but not all, the temperature smearing found in figure  \ref{fgOptics14} can be traced to the method of inversion itself. This is illustrated in figure  \ref{fgOptics16a} from unpublished work of Hwang \etal (2011) where we show results of two numerical simulations labeled calculated biased (green diamond) and unbiased (blue pentagon) and which we compare with the original inversion (red hexagon) of the data in Bi2212 OPT96A of figure  \ref{fgOptics14}. What is done is that the $I^2\chi(\Omega)$ recovered at $T \cong$ 30 K is retained at all temperatures and the conductivity calculated from the Kubo formula and full Eliashberg equations. The temperature and frequency dependent optical scattering rate so obtained are then inverted using an unbiased and also a biased technique. This leads respectively to the blue pentagon and green diamond results for peak position, width, height and area (clockwise from top left) shown in figure  \ref{fgOptics16a}. It is clear that biased or unbiased results are quite similar with greatest differences seen for peak height. Since $I^2\chi(\Omega)$ was never changed with temperature, the significant temperature dependence seen for the width and height of the peak around 60 meV in the spectral density used, is due to the maximum entropy inversion itself. It is not in the input spectral density. Nevertheless, when these results are compared with the red hexagon symbols of figure  \ref{fgOptics16a} which represent inversion based on the actual optical data found for Bi2212 OPT96A, we see that the real data shows much more evolution with temperature than do the numerical simulations. Thus, even though maximum entropy inversions do incorporate some extrinsic temperature shift towards higher energies, much of the temperature evolution found in the data represents a real shift in the spectral function itself.

\begin{figure}
        \begin{center}
         \vspace*{-0.5 cm}%
                \leavevmode
                \includegraphics[origin=c, angle=0, width=15cm, clip]{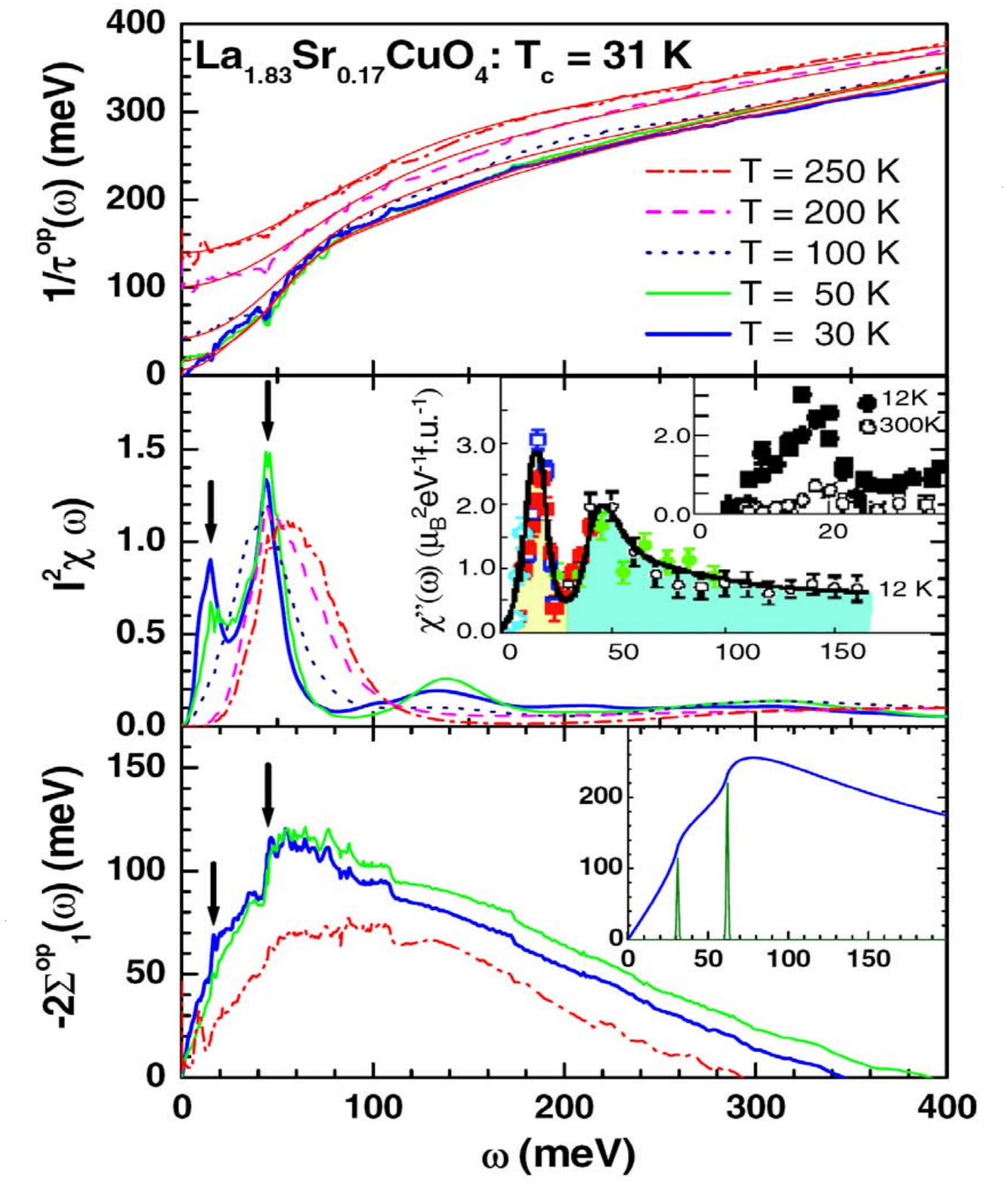}
         \vspace*{-1.1 cm}%
        \end{center}
\caption{Top panel optical scattering rate of \LSCO from Hwang \etal (2008). The heavier lines are the experimental data and the lighter solid lines are the fits to the data. Middle panel: The electron-boson spectral density obtained from Eliashberg inversion from the data in the top panel. The inset shows the data of Vignolle \etal (2007) for a closely related sample of La$_{2-x}$Sr$_x$CuO$_4$ with $x = 0.16$ and $T_c = 38.5$ K. In the inset of the inset the data with solid points are at 12 K and a comparison with 300 K data below 40 meV is also given (open points). Bottom panel: The real part of the optical self-energy  $\sigma_1(\omega)$ for \LSCO. The arrows show the positions of the sharp peaks found in the spectral density at low temperature, $\omega = 15$ and  44 meV. Note the sharp rise in the self-energy at these frequencies. In the inset a simulated self-energy using a mode with two Einstein modes is shown.}
\label{fgOptics20}
\end{figure}

A closer comparison between optics and the spin susceptibility measured in inelastic neutron scattering can be made in \LSCO as reported by Hwang \etal (2008). As we have already described, the electron-boson spectral density is a function associated with all the transitions of the electrons at or around the Fermi energy from an occupied state to all the possible final states, again on and around the Fermi energy to the unoccupied states through the exchange of a boson to which momentum has been transferred. These transitions provide an average over all the bosons defined in the Brillouin zone and as such the resulting spectral density should retain an identifiable similarity in shape to the local i.e. momentum averaged spin susceptibility. Vignolle \etal (2007) have given results for the local spin susceptibility in \LSCO and find a peak centred around 18 meV and another near 40 to 70 meV as well as smaller features extending as high as 150 meV. Optical data is also available in La$_{1.83}$Sr$_{0.17}$cuO$_4$ (Gao \etal 1993), close in doping to the sample investigated by neutron scattering La$_{1.84}$Sr$_{0.16}$CuO$_4$ which has a $T_c = 38.5$ K, and for which neutron scattering results are reproduced in the inset of the middle frame of figure  \ref{fgOptics20}. The results for $I^2\chi(\Omega)$ obtained for optics appear in the main frame of the middle panel of this same figure for several temperatures identified in the top frame. In the top frame, we also show the data for the optical scattering as well as the theoretical results obtained from the Kubo formula and Eliashberg equations with the $I^2\chi(\Omega)$ retrieved from a maximum entropy fit (middle frame). First, we note that the $I^2\chi(\Omega)$ at 30 and 50 K do not differ much and both show distinct peaks emphasized by the solid black arrows at the energy seen in the neutron data. As the temperature is increased the $I^2\chi(\Omega)$ obtained from optics evolve towards a single peak structure. The low energy peak at 15 meV is now gone as is the sharp peak at 44 meV which is replaced by a single much broader peak around 60 meV. This evolution in temperature is in qualitative agreement with the neutron data shown in the inset at 300 K (open circles) where we see the peak at 15 meV seen at 12 K (solid squares) has completely disappeared. This striking resemblance of the optically derived $I^2\chi(\Omega)$ and the local susceptibility is taken as evidence that the spin fluctuations play an important role in the many body renormalizations seen in \LSCO. In the lower frame of figure  \ref{fgOptics20} we show results for the real part of the optical self energy in which the two peak structure at 15 and 44 meV are clearly seen in the quantity without analysis. This is expected from theory as shown in the inset of the lower frame of the figure leaving little doubt that there are peaks at these energies in the electron boson spectral density $I^2\chi(\Omega)$.

A somewhat different approach to inversion of normal state optical data has been presented by van Heumen \etal (2009). They work with the
equation \ref{MoreExact} for the conductivity in the normal state at finite temperature and use a least square fit to data based on a histogram form for the spectral density $\alpha^2 F(\Omega)$. In this way they obtain the results produced in figure  \ref{fgOptics16}. Several different families of cuprates as labeled in the figure are considered and in each case at several temperatures. In particular, we focus on their data for \BISCCO OpD88 which might be compared with the data from Hwang \etal (2007b) shown in the left hand upper frame (a) of figure  \ref{fgOptics14} for a comparable material. Both spectra have a large peak around 50 - 60 meV followed by a valley and then a second peak providing a long high energy tail or background. At 100 K the peak-height is of order 1 in both sets of results. The main difference between the two spectra is the evolution with a temperature between 100 and 300 K. While Hwang \etal (2007b) find a clear evolution towards a broad spectrum with a single peak that has moved towards 100 meV, van Heumen \etal (2009) find much less change with temperature and note that the peak at 50 - 60 meV remains largely unchanged and is robust. As can be seen in figure  \ref{fgOptics12} both the optical scattering rate (left hand column) and corresponding real part of the optical self energy (right hand column) become rather smooth and unstructured with increasing temperature, and consequently an accurate determination of the underlying spectral density
becomes harder. As we saw in figure  \ref{fgOptics10}, temperature tends to smear out some of the features of the recovered spectral density whether one uses a second derivative technique, singular value decomposition, or maximum entropy. Another point of agreement between the data in figures  \ref{fgOptics14} and \ref{fgOptics16} on Bi-2212 is that in both sets of data the mass enhancement factor associated with the peak is reduced with increasing doping while the background is more constant.
Returning to the middle column in figure \ref{fgOptics16}, we note that in Hg1201 OpD07 van Heumen \etal (2009) find that $\alpha^2 F(\Omega)$ has very much the same shape as for Bi2212 OpD88, a large peak around 50 - 60 meV, a dip, followed by a second step which provides a background extending to higher energies of 300 meV. This agrees qualitatively with the findings of Yang \etal (2009).

\begin{figure}
        \begin{center}
         \vspace*{-0.5 cm}%
                \leavevmode
                \includegraphics[origin=c, angle=0, width=12cm, clip]{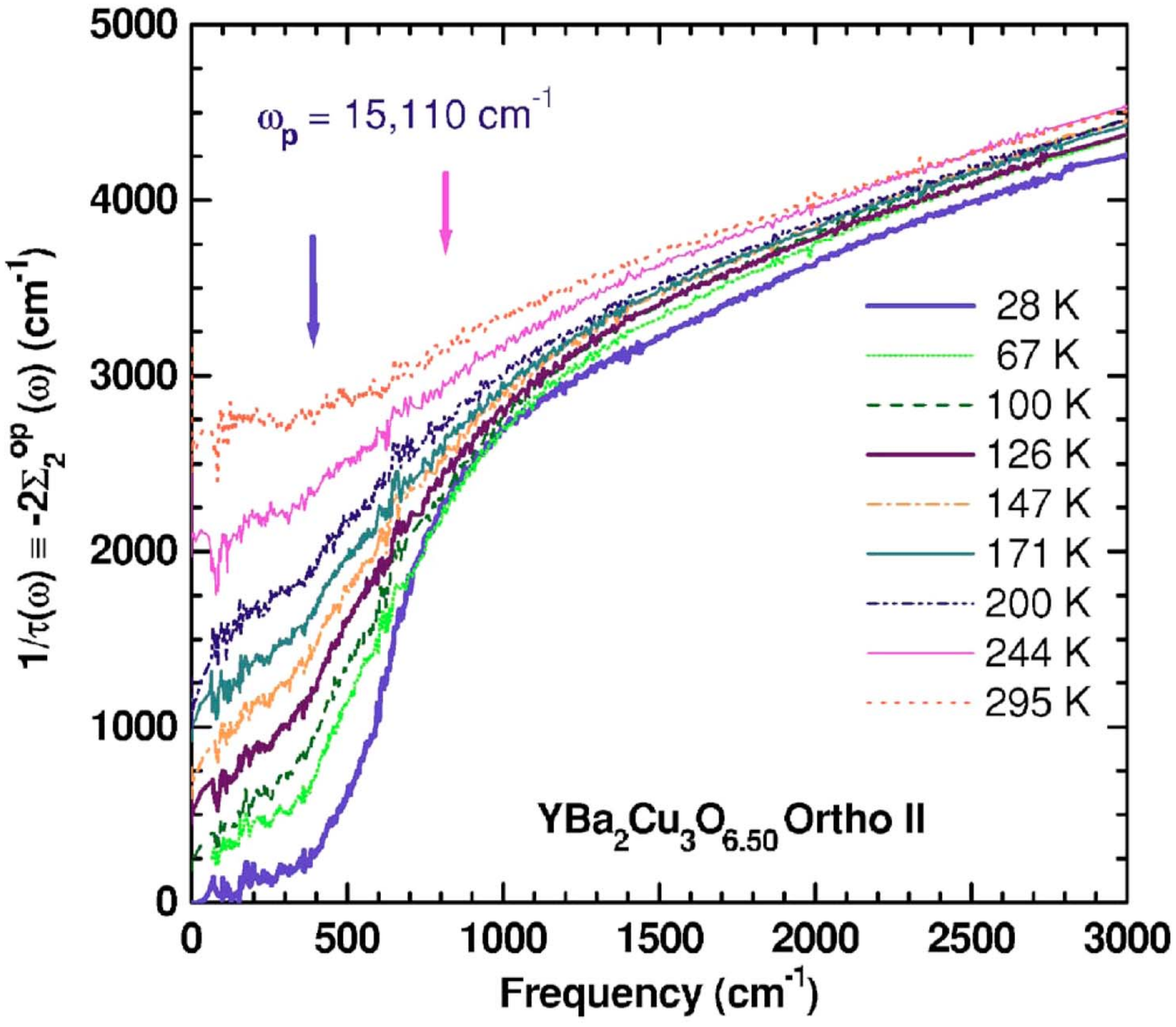}
         \vspace*{-1.1 cm}%
        \end{center}
\caption{The optical scattering rate of Ortho-II \YBCO from Hwang \etal (2006). Two onsets, denoted by arrows dominate the scattering.}
\label{fgOptics21}
\end{figure}

In the work described so far on optics no attempt was made to account explicitly for the opening of a pseudogap which is known to exist in underdoped cuprates (Timusk and Statt 1999 and Norman \etal 2005). In particular equation \ref{MoreExact} holds only under the assumption of an infinite band approximation with a constant bare band density of states. Formula \ref{tau1} with kernel $K(\omega,\Omega,T)$ of equation \ref{Shulga}, while it accounts for finite temperature effects, does not include the possibility of an energy
dependent dressed electronic density of states. The study of effects of an energy dependent density of states $\tilde{N}(\epsilon)$ on normal and superconducting state properties has a long history, in particular in relation to the A15 compounds as described by Mitrovi\'{c} \etal (1983a,1983b). More recent work has been aimed at understanding the influence of these finite band effects on the quasiparticle self energy (Do\v{g}an and Marsiglio 2003 and Cappelluti and Pietronero 2003) and optical properties (Knigavko \etal 2004, 2005, 2006). Following the work of Allen (1971, 1976), Mitrovi\'{c} \etal (1985) derived an approximate formula which is a generalization of equation \ref{allen} to include energy dependence in the effective renormalized electronic density of state defined by $\tilde{N}(\epsilon)$. Following this work Sharapov and Carbotte (2005) generalized it further by including both temperature dependence, as in the work of Shulga \etal (1991), and energy dependent electronic density of state, as in the work of Mitrovi\'{c} \etal (1985) to yield:
\begin{eqnarray}
\frac{1}{\tau^{op}(\omega,T)}&=&\frac{\pi}{\omega} \int_0^{\infty} d\Omega \alpha_{tr}^2(\Omega)F(\Omega)\int_{-\infty}^{+\infty}dz
[\tilde{N}(z-\Omega)+\tilde{N}(-z+\Omega)]\nonumber \\&\times&[n_B(\Omega) + 1 - f(z-\Omega)][f(z-\omega) - f(z+\omega)]
\label{sharapov}
\end{eqnarray}
where $\tilde{N}$ is the fully renormalized frequency dependent density of states. For a constant density of states this formula reduces to equation \ref{Shulga}  and at zero temperature it simplifies to:
\begin{equation}
{1 \over \tau^{op}(\omega,T=0)} = {2\pi \over \omega} \int_0^{\omega} d\Omega \alpha_{tr}^2(\Omega)F(\Omega)\int_{0}^{\omega-\Omega}dz {1
\over 2} [\tilde{N}(z)+\tilde{N}(-z)]
\label{sharapov0}
\end{equation}
which is the equation given by Mitrovi\'{c} and Fiorucci (1985). Consider scattering by a single boson mode $\Omega_E$, for $\omega<\Omega_E$ the scattering rate is zero and for $\omega>\Omega_E$ is given by:
\begin{equation}
{1 \over \tau^{op}(\omega,T=0)}={2\pi \over \omega} \int_0^{\omega- \Omega_E} dz \tilde{N}(z)
\label{sharapov1}
\end{equation}
where for simplicity we have taken a model where $\tilde{N}(z)$ is symmetric about the Fermi energy. It follows directly from equation
\ref{sharapov1} that a full gap $\Delta_{pg}$ in $\tilde{N}(z)$ for example, will look like the case of a constant density of states but with a boson mode set at $\Omega_E+\Delta_{pg}$. Thus the analysis of boson structure needs to account for pseudogap effects in the underdoped cuprates. Hwang \etal (2006) have considered the case of the a-axis conductivity of detwinned Ortho-II YBa$_2$Cu$_3$O$_{6.50}$. Their results for the optical scattering rate at a number of temperatures are shown in figure  \ref{fgOptics21}. The two vertical arrows at 400 \cm\ and 850 \cm\ indicate regions where the scattering rate undergoes a faster than average increase with frequency. To analyze these data Hwang \etal (2006) use the approximate formula equation \ref{sharapov} and allow for the existence of a pseudogap of size 350 \cm\  estimated from the c-axis infrared conductivity (Homes \etal 1993) and tunnelling (Kugler \etal 2001). For the electron boson spectral density they take a Gaussian peak and an MMP (equation \ref{mmp}) background. The position of the Gaussian and its width are taken from known neutron scattering data of Stock \etal (2005) for the local ($q$ integrated) spin susceptibility. The parameters of the MMP form are obtained from a fit to the 295 K optical data without the Gaussian peak.

\begin{figure}
        \begin{center}
         \vspace*{-0.5 cm}%
                \leavevmode
                \includegraphics[origin=c, angle=0, width=12cm, clip]{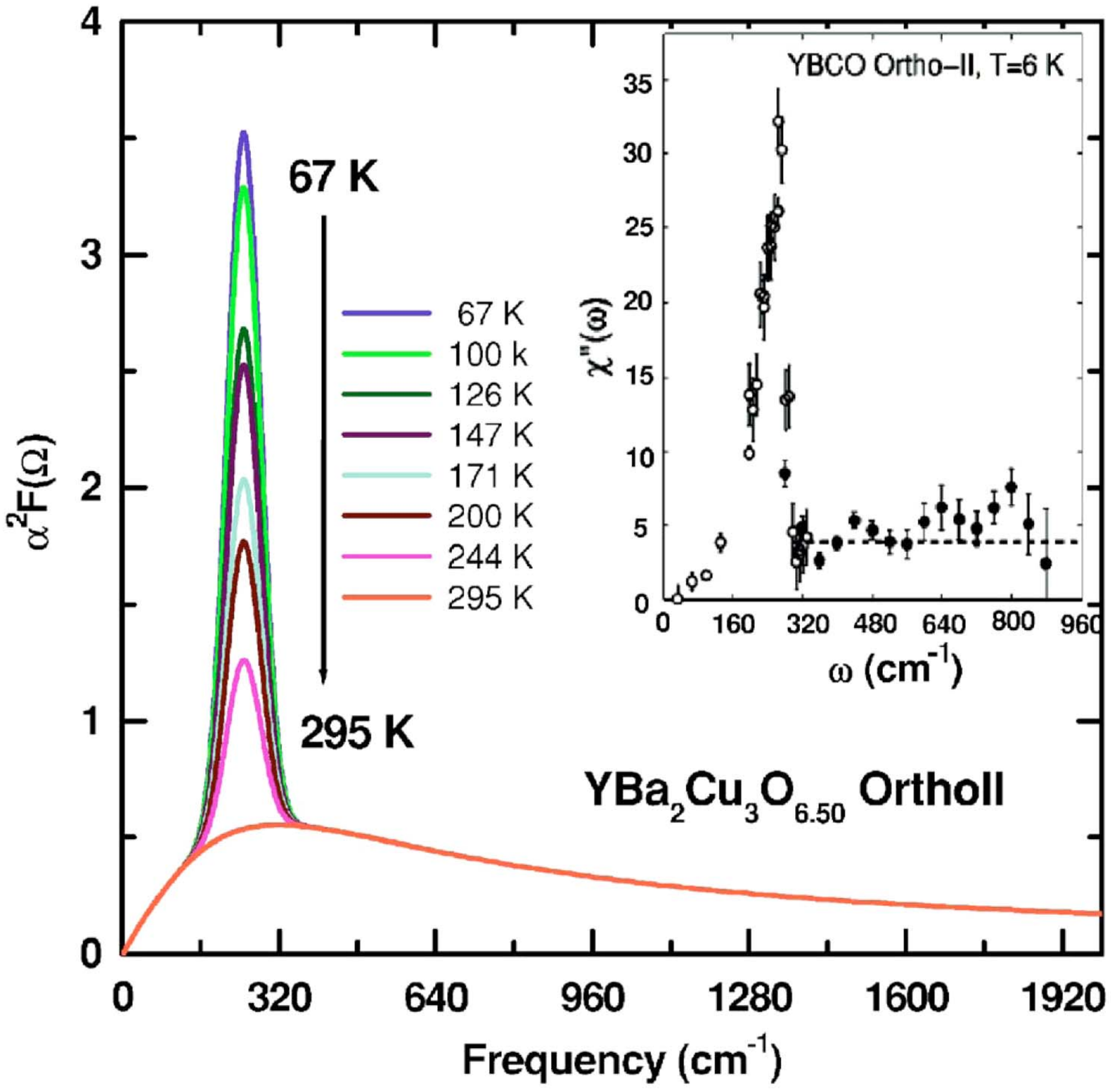}
         \vspace*{-1.1 cm}%
        \end{center}
\caption{The bosonic spectral function $\alpha^2 F(\Omega)$ obtained from the least square fits to the scattering rate data shown in figure \ref{fgOptics21}. The inset shows the $q$ integrated spin susceptibility determined by neutron scattering by Stock \etal (2005).}
\label{fgOptics22}
\end{figure}

\begin{figure}
        \begin{center}
         \vspace*{-0.5 cm}%
                \leavevmode
                \includegraphics[origin=c, angle=0, width=15cm, clip]{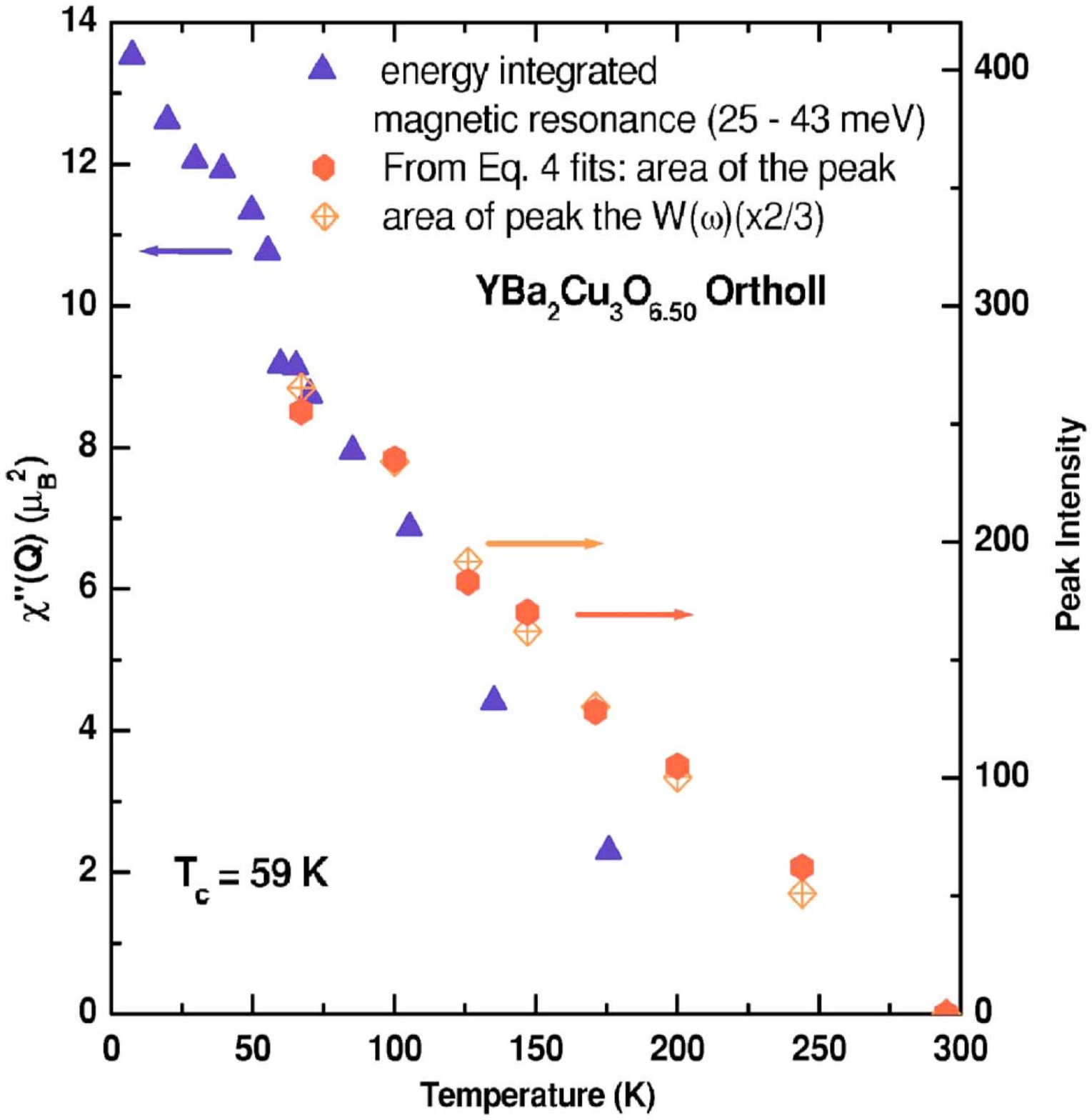}
         \vspace*{-1.1 cm}%
        \end{center}
\caption{Temperature dependence of the amplitude of the sharp bosonic mode in Ortho-II YBCO. The open diamonds with the cross and the closed hexagons are from fit a scattering rate to the model of the optical data of Hwang \etal (2006), including a sharp mode and a background. The upright triangles show the energy integrated amplitude of the neutron mode Stock \etal (2005).}
\label{fgOptics23}
\end{figure}

The remaining parameters of the model are determined through a least squares fit to the data as are the height of the Gaussian and the depth of the pseudogap depression below the Fermi energy. Reasonable values are obtained for both these parameters with the depth ranging from 40 \% of the value at low temperature to only 17 \% at room temperature. In figure \ref{fgOptics22} we show the $I^2\chi(\Omega)$ obtained in this way for the eight values of temperature labeled. In the inset we show the inelastic neutron scattering results for $T =$ 6 K in which the model for the Gaussian part of $I^2\chi(\omega)$ is based. In figure  \ref{fgOptics23} we show results for the area under the Gaussian peak obtained from the least square fit as function of temperature. These values compare with the energy integrated amplitude of the neutron data (upright triangles) and an area estimated from the peak in the second derivative function $W(\omega)$ obtained directly from data on $\sigma_1(\omega)$ according to equation \ref{second_derivative} open diamonds with crosses. The agreement between these points is good and is taken as an indication that the optical data of Ortho-II YBa$_2$Cu$_3$O$_{6.50}$ is consistent with the neutron data and the existence of a sharp peak in the local spin susceptibility at 248 \cm. This also serves to show that the pseudogap should be accounted for in studies of boson structure.

\begin{figure}
        \begin{center}
         \vspace*{-0.7 cm}%
                \leavevmode
                \includegraphics[origin=c, angle=0, width=15cm, clip]{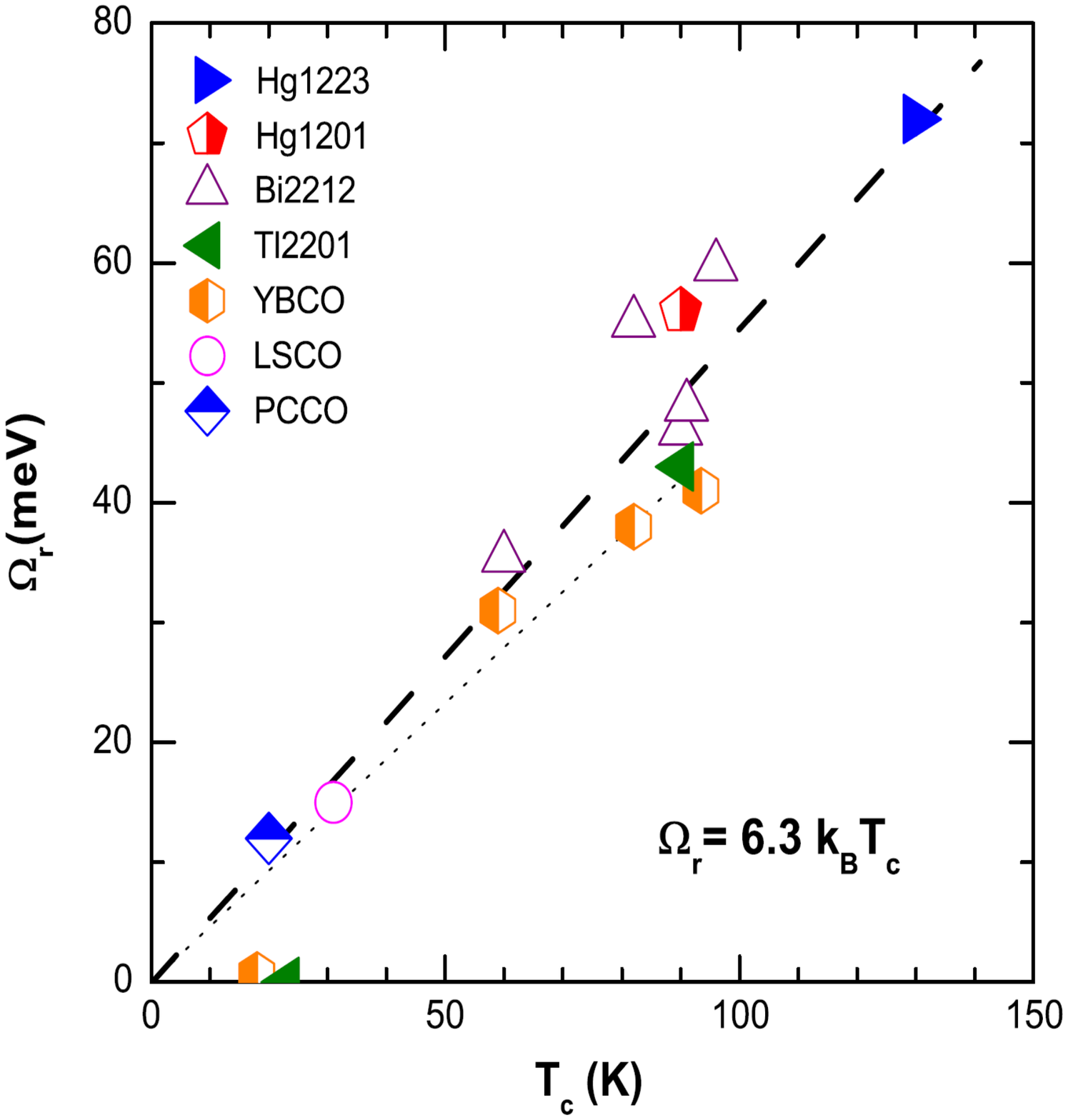}
         \vspace*{-1.1 cm}%
        \end{center}
\caption{The optical resonance frequency $\Omega_r$ as a function of $T_c$ the superconducting transition temperature from Yang \etal (2009). The points refer to various superconducting cuprate families at different doping levels. A linear relationship is found to hold over nearly an order of magnitude in $T_c$.}
\label{fgOptics19}
\end{figure}

In figure  \ref{fgOptics19} we summarize the results obtained from optical measurements for the peak energy $\Omega_r$ as a function of critical temperature. The black dashed line is a guide to the eye and corresponds to $\Omega_r = 6.3 k_B T_c$. This value is close to that obtained for the spin-one resonance in inelastic neutron scattering which is $\Omega_{res} = 5.4 k_B T_c$ and is shown as the dotted curve in the figure. A universal relationship between magnetic resonance and superconducting gap is also stressed in a recent paper by Yu \etal (2009).

Wang \etal (2002) have used optics to study the effect of a complete substitution of  $^{16}$O with $^{18}$O in an underdoped sample of \YBCO with $T_c=67.6$ K for $^{16}$O and $T_c=66.7$ K for $^{18}$O. Analysis of the shift in energy of the reflectance shoulder around 400 to 500 \cm\  leads these authors to conclude that the feature cannot be due to a copper-oxygen phonon stretching mode and that it is largely electronic in origin. The observation is consistent with the evidence presented in this review that phonon effects are small and can perhaps only be seen with probes that emphasize  the nodal direction where the magnetic contribution is smaller. This  would make such effects unlikely to show up in optics which is a momentum averaged probe.
\section{Tunnelling}

\begin{figure}
        \begin{center}
         \vspace*{-0.5 cm}%
                \leavevmode
                \includegraphics[origin=c, angle=0, width=15cm, clip]{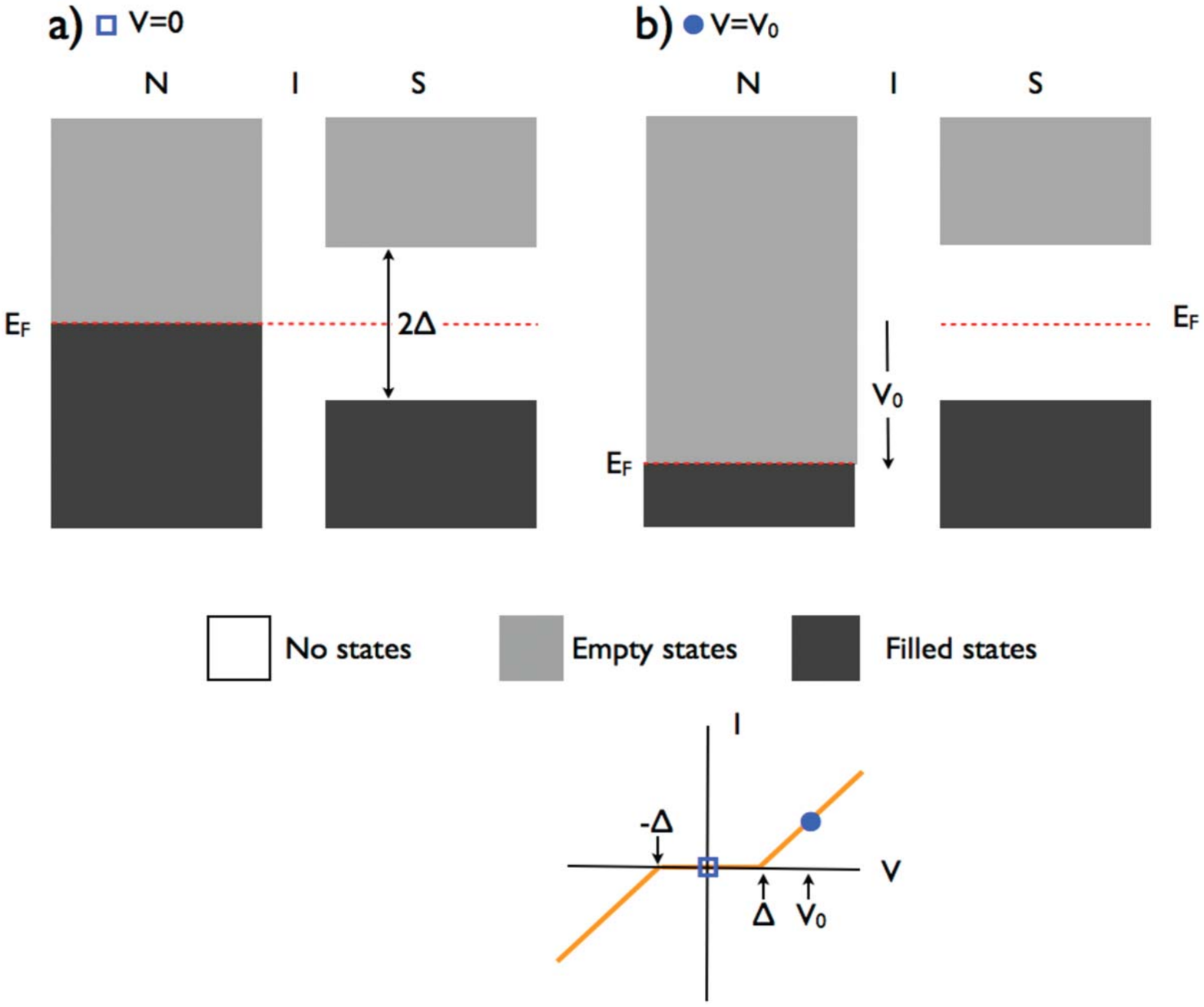}
         \vspace*{-1.0 cm}%
        \end{center}
\caption{ When a normal metal (N) and a superconductor (S) are separated by an insulating or vacuum barrier (I) no current can flow as shown in panel a) since filled states in the metal are faced with either states that are filled in the superconductor or no states in the gap.  When a bias V$_0$ is applied, filled states in the superconductor face empty states in the metal and electrons can flow from the superconductor into these states provided the bias exceeds the gap value, $V_0 > \Delta$.  As the bias increases the number of states rises linearly if the joint density of states is constant.}
\label{fgTunnelling0}
\end{figure}

Tunnelling spectroscopy was used in BCS superconductors soon after its discovery by Giaever in 1960.  It showed that there was a gap at the Fermi level when the superconducting state, as shown in figure  \ref{fgTunnelling0},  was formed and after some refinements subtle changes in the tunnelling conductance at higher energies were used to yield the spectrum of excitations responsible for superconducting pairing (Giaever 1960, McMillan and Rowell 1965, Scalapino 1969, Carbotte 1990).  In SIN (superconductor-insulator-normal metal) tunnelling an oxide layer separates the normal metal and a superconductor shown in figure  \ref{fgTunnelling0}a.  In STM (scanning tunnelling microscopy) the normal metal is a sharp tip and a vacuum space separates the tip and the surface of the sample. With the tip and the sample at the same voltage, their Fermi levels line up and no current can flow in either direction since the filled states on the left are on the same level with filled states on the right or no states in the gap.  When a bias is applied, such that $V>\Delta$ shown in    figure  \ref{fgTunnelling0}b electrons can flow from the filled states in the superconductor to the empty states in the metal.   When a bias of opposite polarity is applied the tunnelling current will flow in the opposite direction provided $V<\Delta$.  The inset shows the current as a function of bias. In this simple example, with a constant density of states the current will be proportional to the amount of overlap between the filled and the empty states and will rise linearly with bias.  In a more complicated situation the current will be proportional to the integral of the joint density of states of the two sides of the junction. The conductance defined as $dI/dV$ at zero temperature will be proportional to the density of states $N(\omega)$ where
\begin{equation}
N(\omega)=\sum_{{\bf k}, BZ}A({\bf k}, \omega).
\label{density}
\end{equation}
While there are issues associated with the precise form of the tunnelling matrix element, the current-voltage characteristic of the junction is related to the carrier spectral density averaged over all momenta ${\bf k}$ in the Brillouin zone as well as appropriate thermal factors.

\begin{figure}
        \begin{center}
         \vspace*{-0.5 cm}%
                \leavevmode
                \includegraphics[origin=c, angle=0, width=15cm, clip]{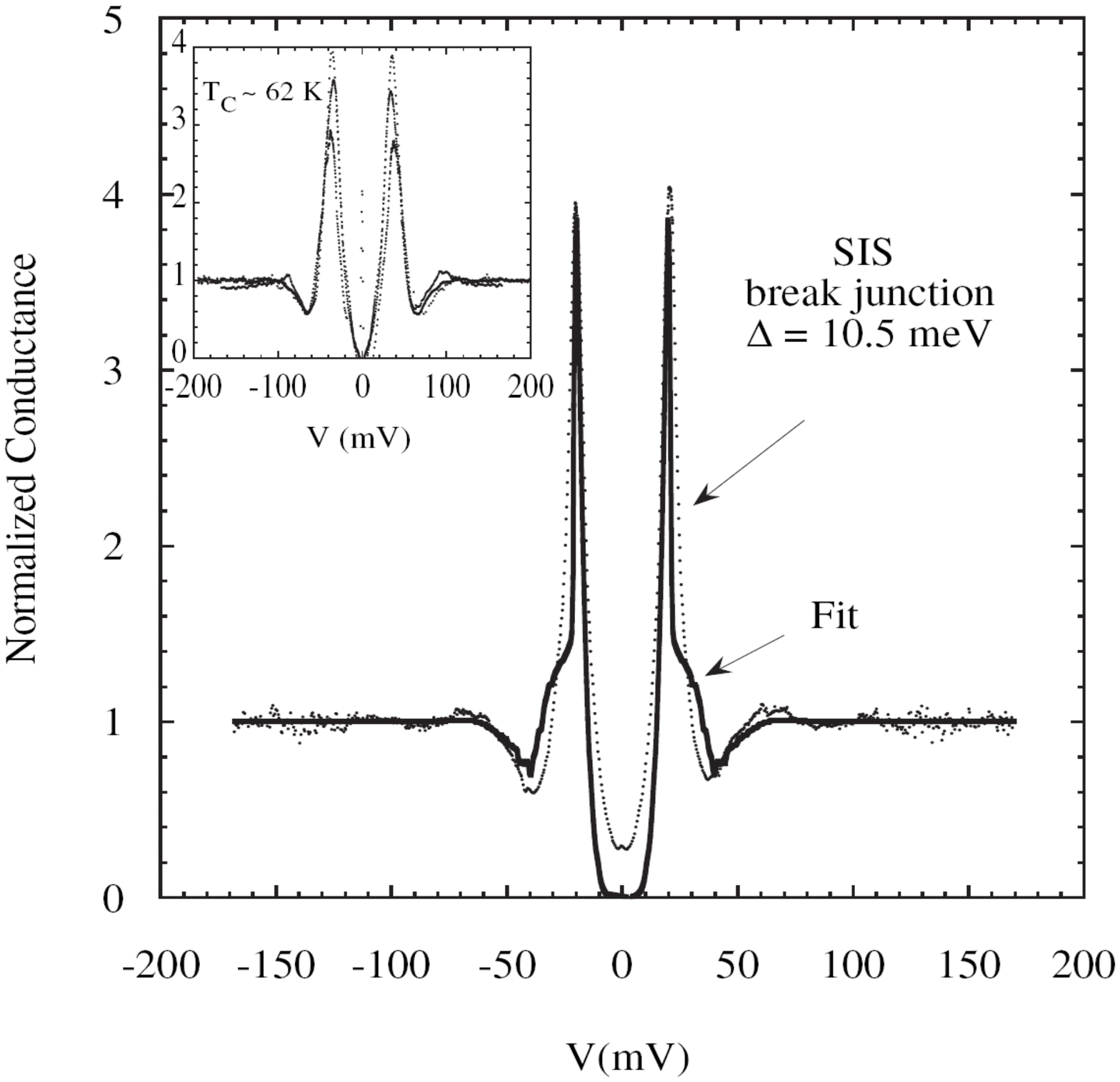}
         \vspace*{-1.2 cm}%
        \end{center}
\caption{SIS break-junction tunnelling conductance from Zasadzinski \etal (2006) (dots) and d-wave Eliashberg fit (solid line) for a junction of overdoped Bi-2212 with $\Delta = 10.5$ meV. The inset shows series of junctions from the same study with $T_c= 60$ K.}
\label{fgTunnelling1}
\end{figure}

\begin{figure}
        \begin{center}
         \vspace*{-0.8 cm}%
                \leavevmode
                \includegraphics[origin=c, angle=0, width=10cm, clip]{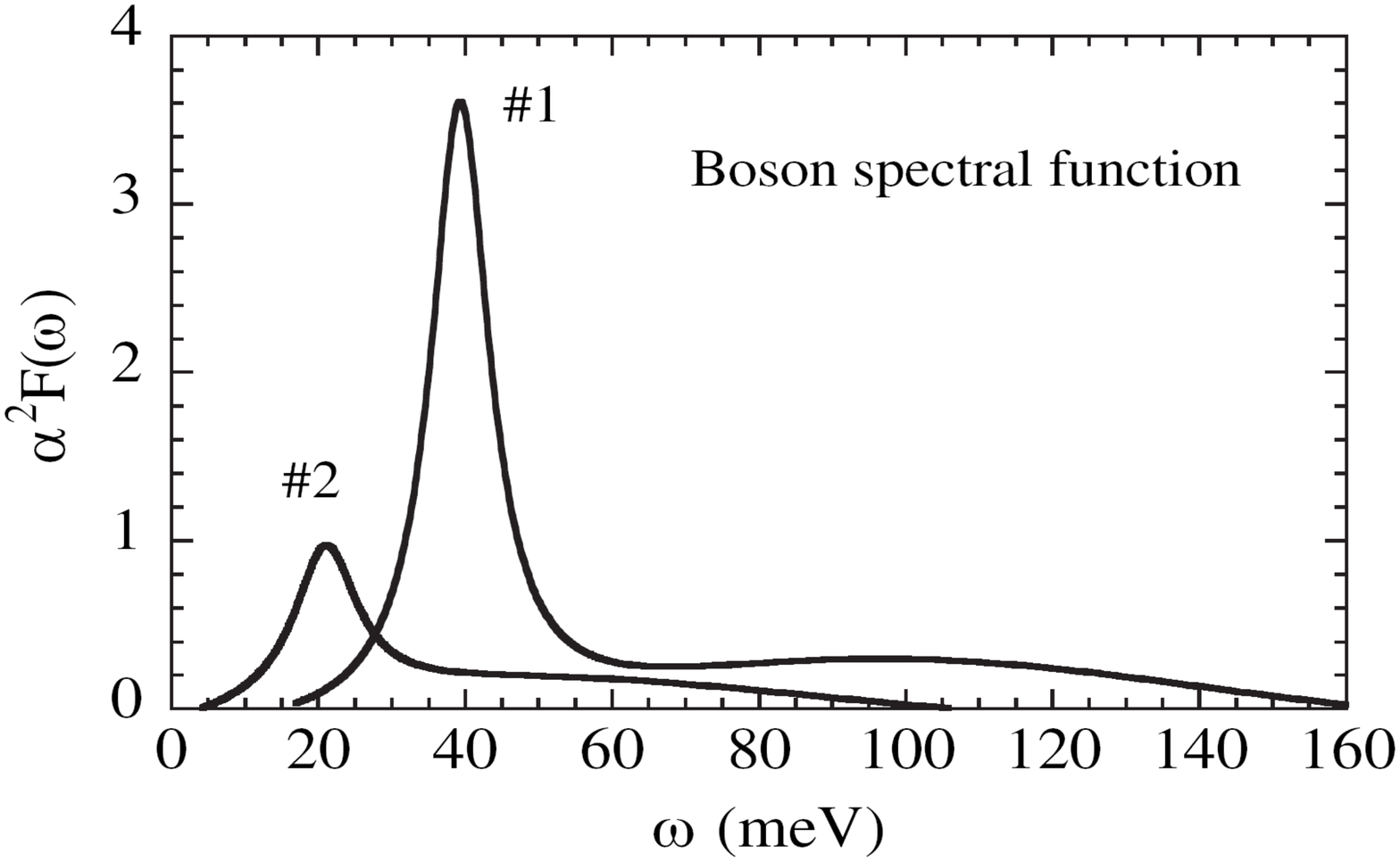}
         \vspace*{-1.7 cm}%
        \end{center}
\caption{Electron-boson functions $\alpha^2F(\Omega)$, which result from the strong-coupling fits to the near optimal SIN data of figure
\ref{fgTunnelling1} (inset) and as well as an overdoped SIS data from the same study, labeled as \#1 and \#2, respectively. The extracted
bosonic spectral function consists of a peak and a background in both cases with the peak weakening and shifting to lower frequencies with
overdoping.}
\label{fgTunnelling2}
\end{figure}

In scanning tunnelling spectroscopy (STS) it is the local density of states at an atomically resolved site that is measured and involves a superconductor-insulator-normal metal (S-I-N) junction. In break junction experiments, the junction is formed by mechanically breaking a
crystal and then allowing the resulting crack to close to
form a thin well-defined insulating barrier. In this case, it is the spatial average of  equation (\ref{density}) which is directly involved and a superconducting-insulating-superconducting (S-I-S) junction is formed. Zasadzinski \etal (2001,2003, 2006) and Ozyuzer \etal (2000) have considered both STM and break junctions of overdoped samples of \BISCCO with a gap of 10.5 meV and critical temperatures of 56 K as well as nearly optimally doped samples with a gap of 28 meV. After appropriate normalizations by a state-conserving background, representative of a normal state, the conductance of the highly overdoped sample is deduced and shown in the main frame of figure  \ref{fgTunnelling1} (dots). The solid curve is a d-wave Eliashberg fit to the data with a gap $\Delta = 10.5$ meV. The inset
shows results for three other break junctions with gaps in the range 17 - 19 meV and a $T_c$ of the order of 60 K. The spectral density
$\alpha^2F(\Omega)$ which provides the fit (solid black curves) is shown in figure  \ref{fgTunnelling2}, also shown for comparison is the
$\alpha^2F(\Omega)$ (\# 1) obtained for the nearly optimally doped sample with $\Delta = 28$ meV. Both spectra exhibit the same characteristic shape with a well-defined peak at $\approx 20$ and 40 meV respectively for \#1 and \#2 and a broad tail beyond extending to higher energies.

A comparison with spectra obtained from optics and shown in figure  \ref{fgOptics14} reveals  the same general trends. For optimally doped Bi-2212 with $T_c = 96$ K the peak in the spectral density is closer to 60 meV, but Schachinger \etal (2000) found that in a second Bi-2212 sample with a lower $T_c$ of 90 K from the optical data of Puchkov \etal (1996), the peak was instead at 43 meV. The same value was found by  Schachinger \etal (2006) based on data in
Bi-2212 by Tu \etal (2002).
To improve the quality of the sample the material used in figure  \ref{fgOptics14} had some Y doping and some of the differences in the recovered spectra could be related to sample dependence. Another reason for differences between optical and tunnelling spectra are certainly related to different relationships between the underlying spectral densities and the experimental data. The transport
spectral densities weigh backward scattering more strongly than forward scattering as these deplete the current more efficiently while
quasiparticle lifetime weighs both processes equally. It should be noted however, that a very different picture emerges in the tunnelling work of Shim \etal (2008) on films of La$_{1.84}$Sr$_{0.16}$CuO$_{4}$. These authors identify 11 minima in their second derivative data for current vs voltage and find that these match precisely the published Raman data on phonons.

\begin{figure}
        \begin{center}
         \vspace*{-0.5 cm}%
                \leavevmode
                \includegraphics[origin=c, angle=0, width=15cm, clip]{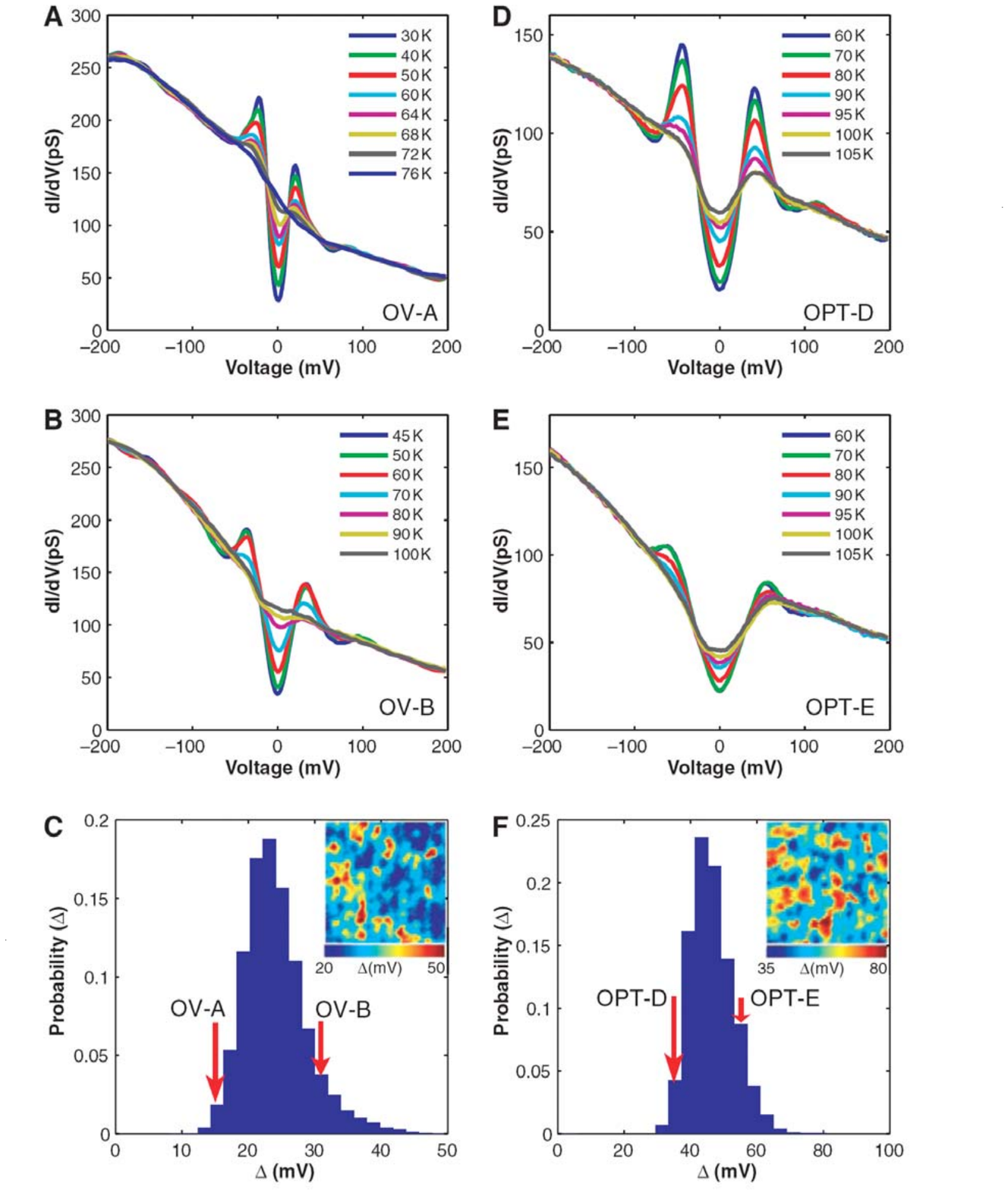}
         \vspace*{-1.1 cm}%
        \end{center}
\caption{(A and B) Spectra taken at two different atomic locations on an overdoped Bi$_2$Sr$_2$CaCu$_2$O$_{8+\delta}$ sample ($T_c = $ 68 K, OV68) at various temperatures from Pasupathy \etal (2008). The gaps in the spectra close at different temperatures leading to a temperature independent background conductance at high temperature. (C) Histogram of pairing gap values measured in the OV68 sample. (Inset) A typical pairing gap map (300 \AA) obtained at 30 K. (D and E), same as A and B but for an optimally doped sample ($T_c = $ 93 K, OPT). The background remains temperature dependent well above $T_c$. (F) Same as C but for the OPT sample.}
\label{fgTunnelling3}
\end{figure}

\begin{figure}
        \begin{center}
         \vspace*{-0.8 cm}%
                \leavevmode
                \includegraphics[origin=c, angle=0, width=15cm, clip]{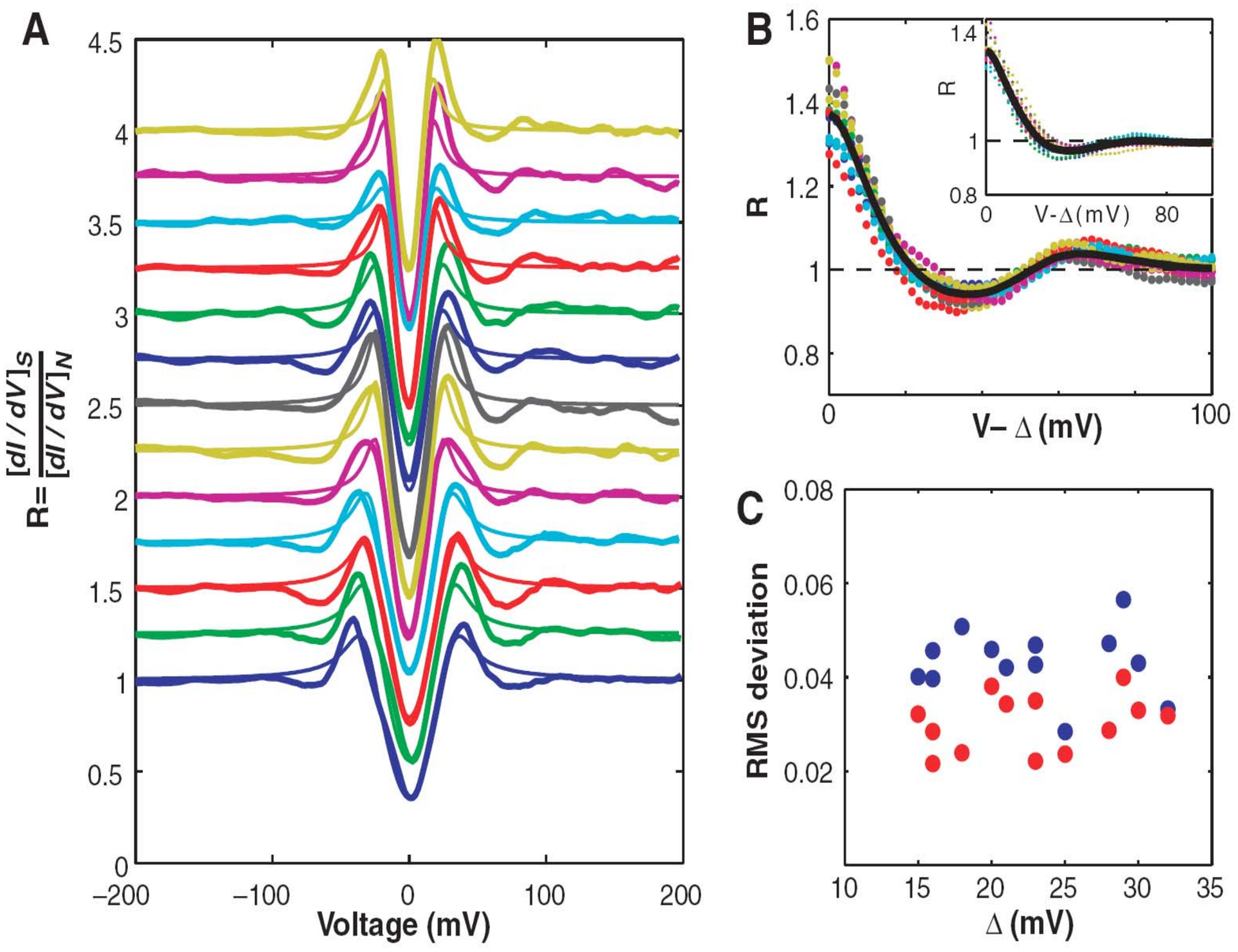}
         \vspace*{-1.5 cm}%
        \end{center}
\caption{(A) The low temperature ($T =$ 30 K) conductance ratio plotted for several different gaps  from Pasupathy \etal (2008). The conductance ratios deviate systematically from the d-wave model (thin lines) and go below unity over a range of voltages (50 - 80 mV) indicating strong coupling to bosonic modes. (B) The positive bias conductance ratios referenced to the local gap at different locations showing a common dip-hump feature. The heavy line is the average of all the locations. (Inset) Gap referenced conductance ratios for negative bias. (C) The RMS deviation of the conductance ratios from the d-wave model for positive (blue) and negative (red) bias over the energy range 20 - 120 mV, which shows a correlation with the size of the gap.}
\label{fgTunnelling4}
\end{figure}

\begin{figure}
        \begin{center}
         \vspace*{-0.5 cm}%
                \leavevmode
                \includegraphics[origin=c, angle=0, width=15cm, clip]{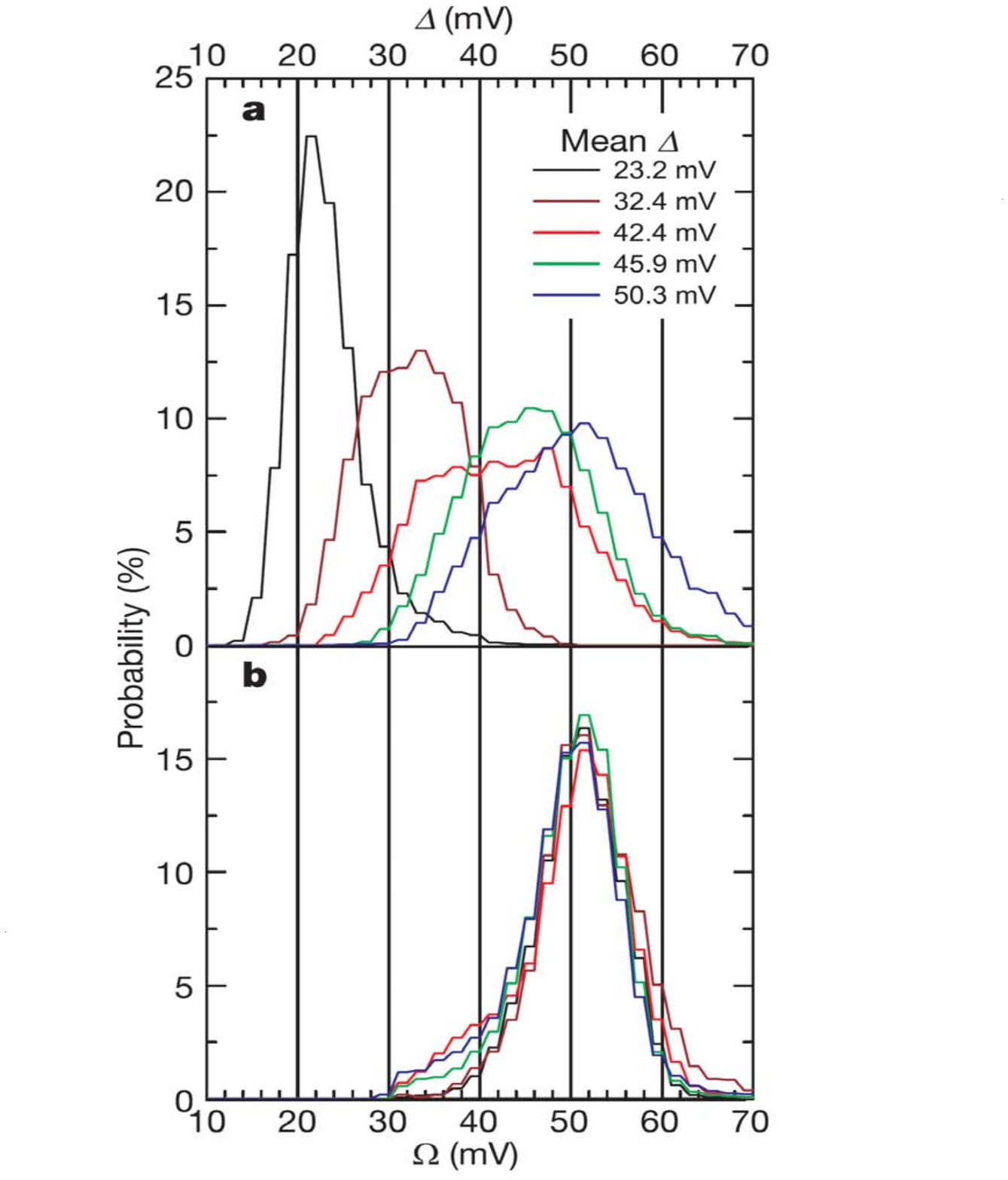}
         \vspace*{-1.1 cm}%
        \end{center}
\caption{Doping dependence of energy gap and boson energy histograms in \BISCCOa from Lee \etal (2006). {\bf a} Histograms of measured energy gaps $\Delta$ from a sequence of samples with different dopings, black being strongly overdoped and blue strongly underdoped. {\bf b} Histograms of measured boson energies $\Omega$, from $d^2I/dV^2$-imaging measurements performed simultaneously with {\bf a}. Within the uncertainty, neither the distribution nor the mean value of $\Omega = 52 \pm 1$ meV are influenced by doping.}
\label{fgTunnelling5}
\end{figure}

STM has revealed important gap inhomogeneities on the atomic scale. In figure  \ref{fgTunnelling3} we show results of Pasupathy \etal (2008) at two different atomic locations in an overdoped Bi-2212 sample with $T_c = 68$ K (A and B) and optimally doped with $T_c = 93$ K (D and E) at various temperatures. Note that the current voltage characteristics are asymmetric between positive and negative biases. To analyze the data in terms of  a boson structure they normalize their data to the normal state above $T_c$.  Frame (C and F) give respectively the gap distribution histogram found in each of the samples and the inset shows gap maps on a 300 \AA\   square  patch. The normalized conductances denoted $R=[dI/dV]_s/[dI/dV]_n$ are shown in figure  \ref{fgTunnelling4} frame A. All exhibit a peak-dip-hump structure. When the dip energy is referred to the gap, i.e. $V-\Delta$ is used on the horizontal axis rather than $V$, as shown in frame B all the dips align and fall around 35 meV. Also the dip size is uncorrelated with the size of the gap (see frame C) in contrast to what is expected of a conventional strong coupling superconductor where  the gap and the size of the bosonic dip would be correlated. Another closely related STM study is that of Lee \etal (2006).
These authors identify the boson energy involved from the second derivative $d^2I/dV^2$ imaging measurements and obtain a constant value of $\Omega = 52$ meV
independent of doping. This is shown in figure  \ref{fgTunnelling5} which is reproduced from their work. The top frame (a)
shows a histogram of the local gap variation for five samples characterized by the mean gap value for each sample as indicated in the figure. The lower frame (b) gives the histogram of boson energies obtained in each case. All are very similar and peak around 52 meV. This leads the authors to conclude that the boson involved is a phonon. This view is further reinforced in the work as the peak energy is found to scale with the inverse square root of the oxygen isotope mass on $^{16}$O $\rightarrow$ $^{18}$O substitution. A possible explanation is that the phonon is an oxygen vibration created during inelastic tunnelling of the charge carriers through the region between CuO$_2$ layers (Pilgrim \etal 2006, Hwang \etal 2007c and Scalapino \etal 2006).

\begin{figure}
        \begin{center}
         \vspace*{-0.5 cm}%
                \leavevmode
                \includegraphics[origin=c, angle=0, width=15cm, clip]{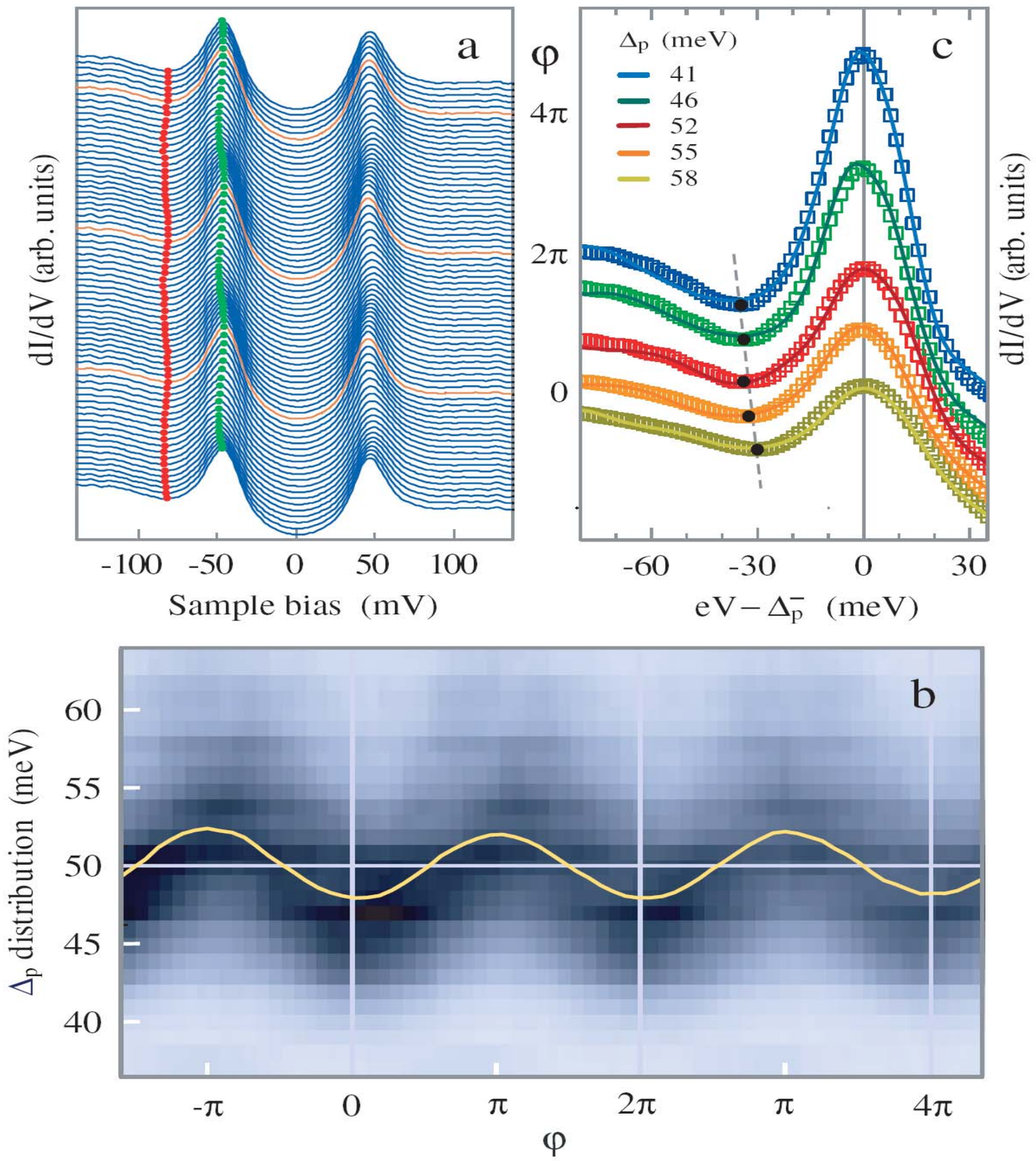}
         \vspace*{-1.1 cm}%
        \end{center}
\caption{Spatial modulation of the gap and dip energy in nearly optimally doped Bi-2223 from Jenkins \etal (2009). (a) Evolution of the $dI/dV$ spectra with the phase $\phi$. The green and red dots indicate the coherence peaks and the dip feature for negative bias, respectively. (b) Colour-scale representation of $\Delta_p$ distributions for spectra with the same $\phi$. The yellow curve depicts the average gap as a function of $\phi$. (c) Negative-bias part of the $\Delta_p$-averaged spectra, offset by $ \Delta_p^-$. Black dots indicate the energy location of the dip in the experimental spectra and the dotted grey line is a guide to the eye.}
\label{fgTunnelling6}
\end{figure}

\begin{figure}
        \begin{center}
         \vspace*{-0.5 cm}%
                \leavevmode
                \includegraphics[origin=c, angle=0, width=15cm, clip]{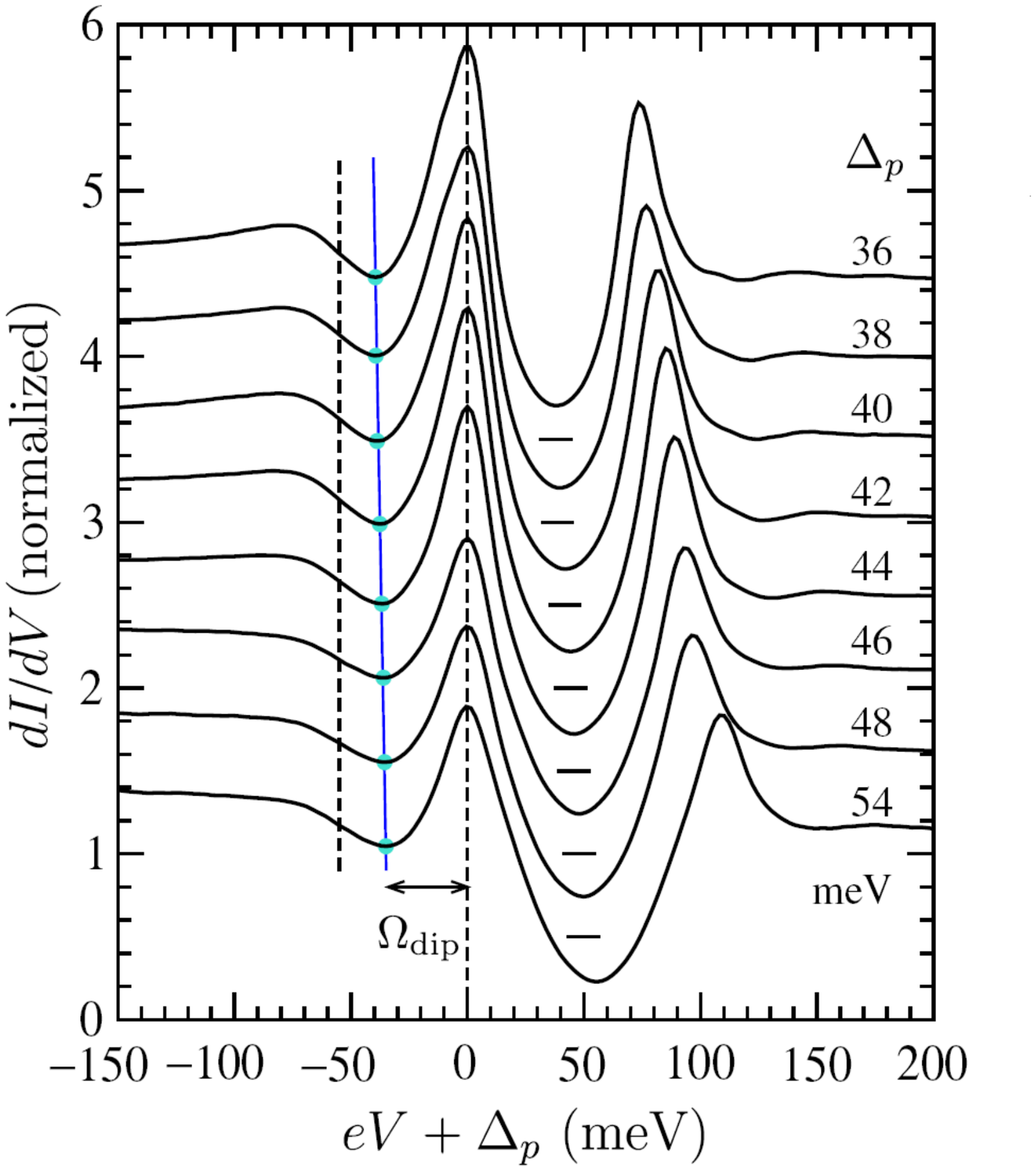}
         \vspace*{-1.1 cm}%
        \end{center}
\caption{STM conductance spectra of Bi-2223 ($T_c =$ 111 K) at $T =$ 2 K from de Castro \etal (2008). Each curves is an average of several spectra taken at different locations on the same sample, all having the indicated peak-to-peak $\Delta_p$. The energy $\Omega_{dip}$ is the energy difference between the dip minimum (dot) and the peak maximum at negative bias, relative to which voltages are measured.}
\label{fgTunnelling7}
\end{figure}

The work of Jenkins \etal (2009) provides additional insight into the effects  just described. These authors have made a detailed study of the effect of a boson mode of energy $\Omega_s$ peaked in the Brillouin zone around momentum $(\pi,\pi)$ on the average density of states for a d-wave superconductor with general tight binding band structure including the possibility that a van Hove singularity may fall near the Fermi energy.
They consider an optimally doped ($T_c = 109$ K) and an  overdoped ($T_c=108 $ K) of Bi-2223. Both show large boson structures and have
large values of the critical temperature. The cleaved surfaces have supermodulations of $\cong 5 a$ and the authors introduce an angle  $\phi$ to characterize this periodicity. In figure  \ref{fgTunnelling6} we reproduce the results for their optimally doped sample of Bi-2223. The conductance $dI/dV$ is shown in frame (a) for many values of the phase $\phi$. Frame (b) gives the distribution of gaps $\Delta_p$ as a function of  $\phi$ (the yellow curve is the average gap for a given $\phi$). The gaps display periodicity with $\phi$ as expected from the supermodulation on the cleaved surface of the sample. In frame (c) we show average spectra all having the same gap value $\Delta_p$ equal to the difference $(\Delta_p^+ - \Delta_p^-)/2$ where $\Delta_p^+$ and $\Delta_p^-$ are the energy of the positive and the negative bias coherence peaks. The solid curves through the data are fits obtained from calculations of the density of states including a boson at $\Omega_s$ and momentum peaking around $(\pi,\pi)$. It is found that the dip energy referred to $\Delta_p^-$ is within 1 meV of the input $\Omega_s$ and therefore the dip energies can be considered to be the signatures of the collective mode energies (CME) for these data. It is noted that the value of the gap and $\Omega_s$ are anticorrelated as emphasized by the dashed line with the average value of 33.5 meV (Jenkins \etal 2009) close to the value of 35 meV seen by Pasupathy \etal (2008). This is in contrast
the absence of doping dependence for optical phonons in Bi-2223 (Boris \etal 2002). Taken together these observations lead the authors to
favour coupling to spin fluctuations as the cause of the observed dip in the STM spectra. Additional results for a sample of Bi-2223 with $T_c = 111 $ K at $T = 2$ K by de Castro \etal (2008) are shown in figure  \ref{fgTunnelling7}. The one dashed line locates the coherence peak of $\Delta_p^-$, the solid line the dip position and the second dashed line the minimum in the $d^2I/dV^2$ spectrum as was used by Lee \etal (2006) to locate their phonon at 52 meV in Bi-2212. While the dip energy decreases from 39 to 35 meV corresponding in this case to a collective mode energy $\Omega_s$ decrease from 34 to 24 meV, the position of the dashed line is almost independent of the gap value and is at $\approx 57$ meV close to the value of 52 meV quoted above. Another recent study by Das \etal (2008) has further confirmed in hole doped NdBa$_2$Cu$_3$O$_{7-\delta}$ that the gap and the collective mode energies are anticorrelated favouring more a spin fluctuation rather than a phonon explanation.
\section{Raman spectroscopy}

There have been attempts to include inelastic scattering in the calculations of the Raman response of  d-wave superconductors. Jiang \etal (1996b) and Branch \etal (2000) based their work on a generalization of BCS theory of the nearly antiferromagnetic Fermi liquid model
(Branch \etal 1995) which included d-wave but no dynamics. Attempts to extract the inelastic contribution from Raman spectra have also
appeared, among them the work of Gallais \etal (2006) and Grilli \etal (2009). Muschler \etal (2010) write the Raman cross section for
polarization $\mu$ as
\begin{equation}
{\rm Im}\chi_{\mu}(\Omega)=\frac{\Omega \Gamma_{\mu}(\Omega)}{[\Omega(1+\lambda_{\mu}(\Omega)]^2+\Gamma_{\mu}^2(\Omega)}
\label{Ramanchi}
\end{equation}
and following the work of Sharapov and Carbotte (2005) show that the Raman scattering rate $\Gamma_{\mu}(\Omega)$ can be written in identical form to our equation (\ref{Shulga}) for the optical conductivity but with $\alpha^2F(\Omega)$ replaced by an appropriate electron-boson spectral density $I_{\mu}^2\chi(\Omega)$
which is dependent on the Raman polarization $\mu$. One can introduce an angular dependent version of this function as $I_{\mu}^2\chi(\Omega,\theta)$. Within a continuum model for the electronic structure the function entering the usual spectral
density would correspond to a straight angular average while for Raman there would be a further $\cos^2(2 \theta)$ weighting for B$_{1g}$ and $\sin^2(2 \theta)$ for B$_{2g}$ in the integral over angles. This provides new information on the spectral density. The weighting for B$_{1g}$ favours the antinodal direction while for B$_{2g}$ it favors the nodal direction.

\begin{figure}
        \begin{center}
         \vspace*{-0.7 cm}%
                \leavevmode
                \includegraphics[origin=c, angle=0, width=15cm, clip]{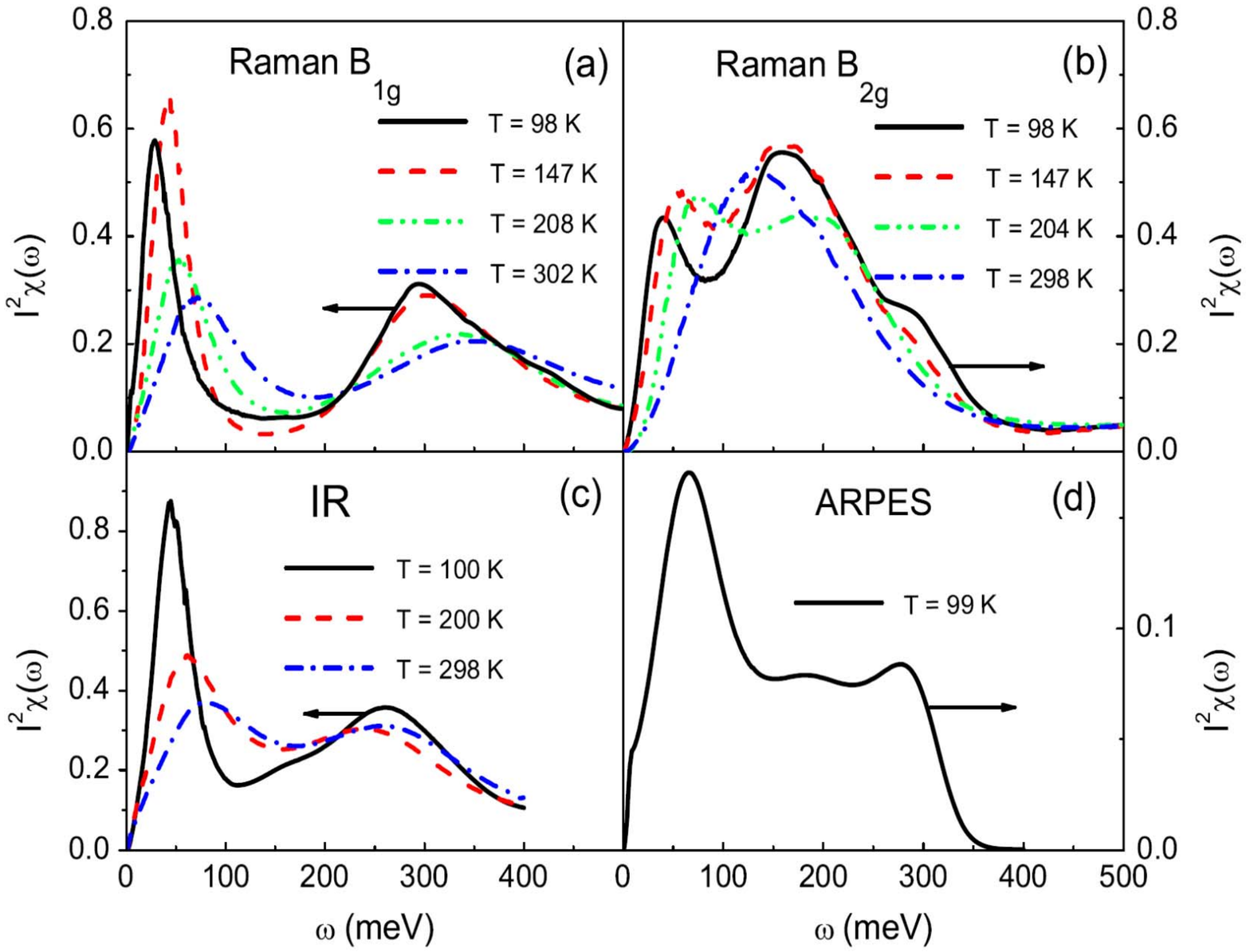}
         \vspace*{-1.3 cm}%
        \end{center}
\caption{The electron-boson spectral density (dimensionless) as a function of energy $\omega$ in meV from B$_{1g}$ (a) and B$_{2g}$ (b) Raman data from Muschler \etal (2010). We show in (c) results (Schachinger \etal 2006) obtained from optical data (Tu \etal 2002) and in (d) a result (Schachinger and Carbotte 2008) obtained from nodal direction ARPES (Zhang \etal 2008a).}
\label{fgRaman1}
\end{figure}

\begin{figure}
        \begin{center}
         \vspace*{-0.7 cm}%
                \leavevmode
                \includegraphics[origin=c, angle=0, width=15cm, clip]{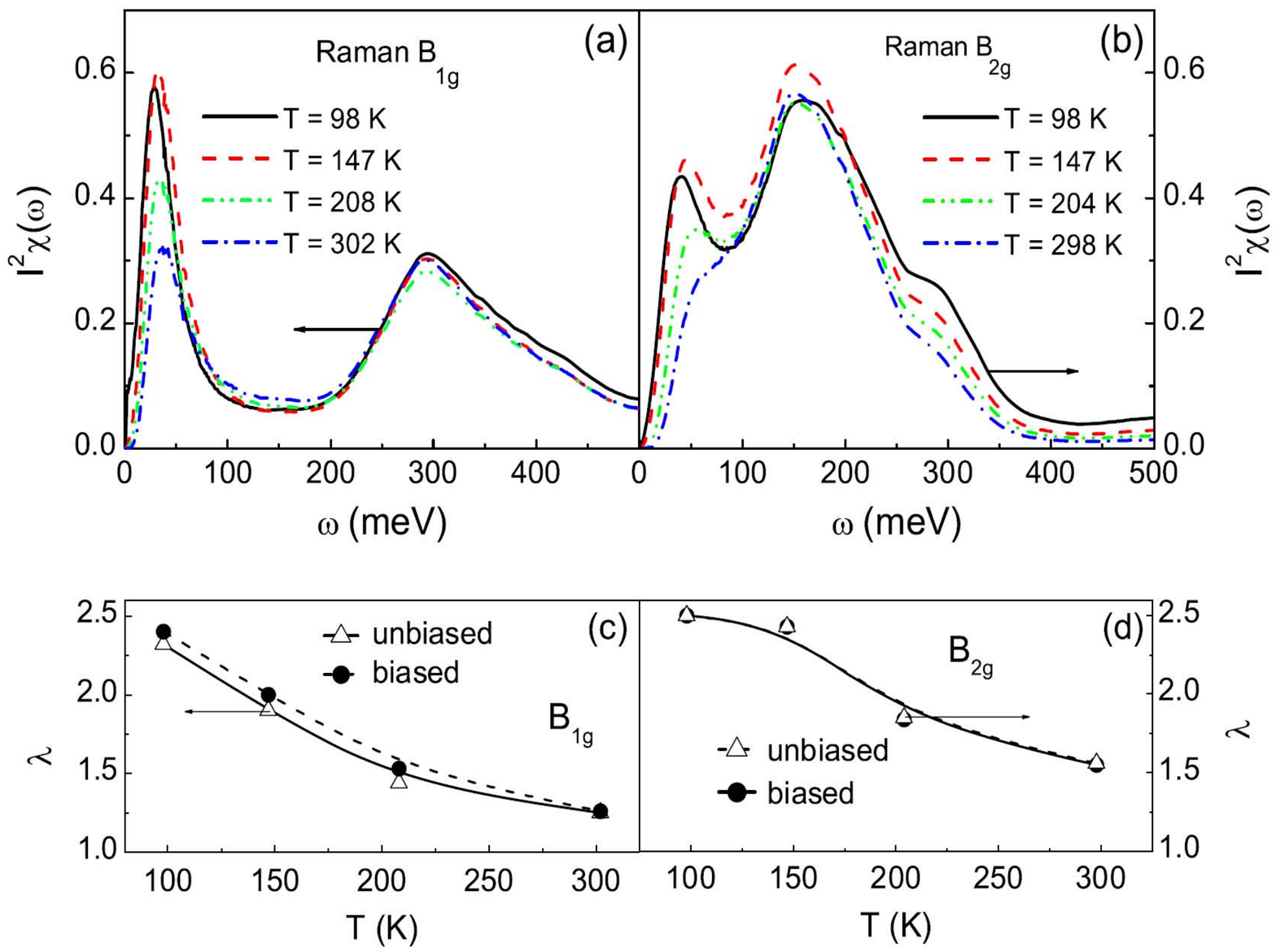}
         \vspace*{-1.2 cm}%
        \end{center}
\caption{(a) The electron-boson spectral density $I^2\chi(\omega)$ (dimensionless) as a function of energy $\omega$ in meV from B$_{1g}$ Raman scattering rates using a biased maximum entropy inversion as described in the text  from Muschler \etal (2010). (b) The same as (a)  but now for B$_{2g}$ symmetry. Frame (c) and (d) compare the temperature dependencies of the mass enhancement parameter $\lambda$ obtained from biased and unbiased inversion of the B$_{1g}$ and B$_{2g}$ Raman scattering rates, respectively.}
\label{fgRaman2}
\end{figure}

In figure  \ref{fgRaman1} we reproduce the results obtained by Muschler \etal (2010) for an optimally doped sample of Bi-2212 with $T_c =$ 94.5 K. The B$_{1g}$ (antinodal) spectrum has a large peak around 30 meV followed by a dip and then a second peak centred around 300 meV. As the temperature is increased there is a clear evolution in the spectra with the reduction in amplitude of the lower peak and its movement towards higher energy filling in the dip region. This distribution is very close to the optically derived spectrum found in Schachinger \etal (2006) for a comparable sample of Bi-2212 measured by Tu \etal (2002) with a peak around 40 meV and overall very similar shape (frame (c)). For the B$_{2g}$ (nodal direction) spectrum, the peak around 30 meV remains but is smaller in magnitude than for B$_{1g}$ 3 \% of the area under the curve as opposed to 23 \% in B$_{1g}$ but, as it has the same energy is probably of the same origin. Both could be associated with scattering about ($\pi$,$\pi$) which would favor antinodal as opposed to nodal direction scattering from
one point in the Fermi surface to another again on the Fermi surface. It is instructive to compare the B$_{2g}$ spectrum with that obtained from nodal direction ARPES described in the previous section (see panel (a) of figure  \ref{fgARPES14}) and reproduced in the lower right hand frame (d). This spectrum looks a lot like B$_{2g}$ but with a very important difference, it has a prominent peak at 65 meV, a much higher energy than found in the other case. It could be that this peak is related to scattering from a different boson, perhaps a phonon with scattering confined to the region of the nodal direction and so not sufficiently important in the Raman and optical cases because of averaging away from the nodal regions in momentum space. This possibility is supported by the recent work of Schachinger \etal (2009b) and further elaborated upon by Schachinger and Carbotte (2010). The recent isotope effect experiments of Iwasawa (2008) described in our previous section on ARPES can be understood on the assumption that only the sharp peak above the background in the $I^2\chi(\Omega)$ shown in the lower right hand frame of figure  \ref{fgRaman1} is shifted in substituting $^{16}$O by $^{18}$O and this involves only 10 \% of the area under the total spectral density, or a mass enhancement $\lambda$ of about 0.2.

In maximum entropy inversion one needs to specify an initial starting value for the spectral density. In the unbiased inversion mode this is taken as flat, while in the biased mode the next lowest temperature converged solution for $I^2\chi(\Omega)$ is used as the starting value for the next higher temperature run. Results for the biased mode are presented in the top two frames of figure  \ref{fgRaman2} and these are to be compared with the
top frames of figure  \ref{fgRaman1} obtained by unbiased inversion. We note that both methods give very similar results except for one important difference. The unbiased mode shows more temperature evolution than does the biased case. This does not translate into
important differences for the value of mass enhancement and its evolution with temperature. This is seen in the bottom frame of figure  \ref{fgRaman2} where we plot $\lambda$ in the two cases as a function of $T$. The differences do, however, have impact on the possible interpretation of the boson spectral density. For phonons one would expect little $T$ dependence of the peak frequencies and also little broadening, while for spin fluctuation the opposite holds. In that regard, the unbiased maximum entropy results differ from those obtained by least square fit of the formula for $I^2\chi(\Omega)$ as shown in figure \ref{fgOptics16}. This alternative method shows less temperature evolution than the unbiased maximum entropy method. The exact source of this difference is not fully understood. In maximum entropy noise always leads to some smearing of the resulting distributions. An alternate interpretation of boson structure seen in Raman has been presented very recently by Caprara \etal (2010). Their data is for the La$_{2-x}$Sr$_x$CuO$_4$ family and an aim of their work is to unravel the nature of the glue responsible for pairing. They do not provide inversions of their data. Instead they proceed with calculations of Raman response for B$_{1g}$ and B$_{2g}$ symmetry which they then compare with experiments. The calculations employ the equivalent of a Shulga \etal (1991) approximation adapted to the Raman cross section as opposed to optics. Separate spectral densities are introduced for
spin and charge fluctuations with different characteristic wave vectors $q_s \sim$ ($\pi$,$\pi$) for spins and
$g_c \sim$ ($\pm \pi/2$, 0), (0, $\pm \pi/2$) for charges with each having its own dynamics and collective modes (boson). While the spin
fluctuations correspond to the proximity of the antiferromagnetic state, the charge fluctuations correspond to the tendency towards stripes or checkerboard order. They find guidance from a selection rule in momentum space which emphasizes the spin in B$_{1g}$ and charge in B$_{2g}$ symmetry. At any doping they can separate each contribution and thus get additional information on the relative contribution of each process which goes beyond what has been provided in the main part of this review.

\section{Summary and Conclusion}

\begin{figure}
        \begin{center}
         \vspace*{-0.5 cm}%
                \leavevmode
                \includegraphics[origin=c, angle=0, width=12cm, clip]{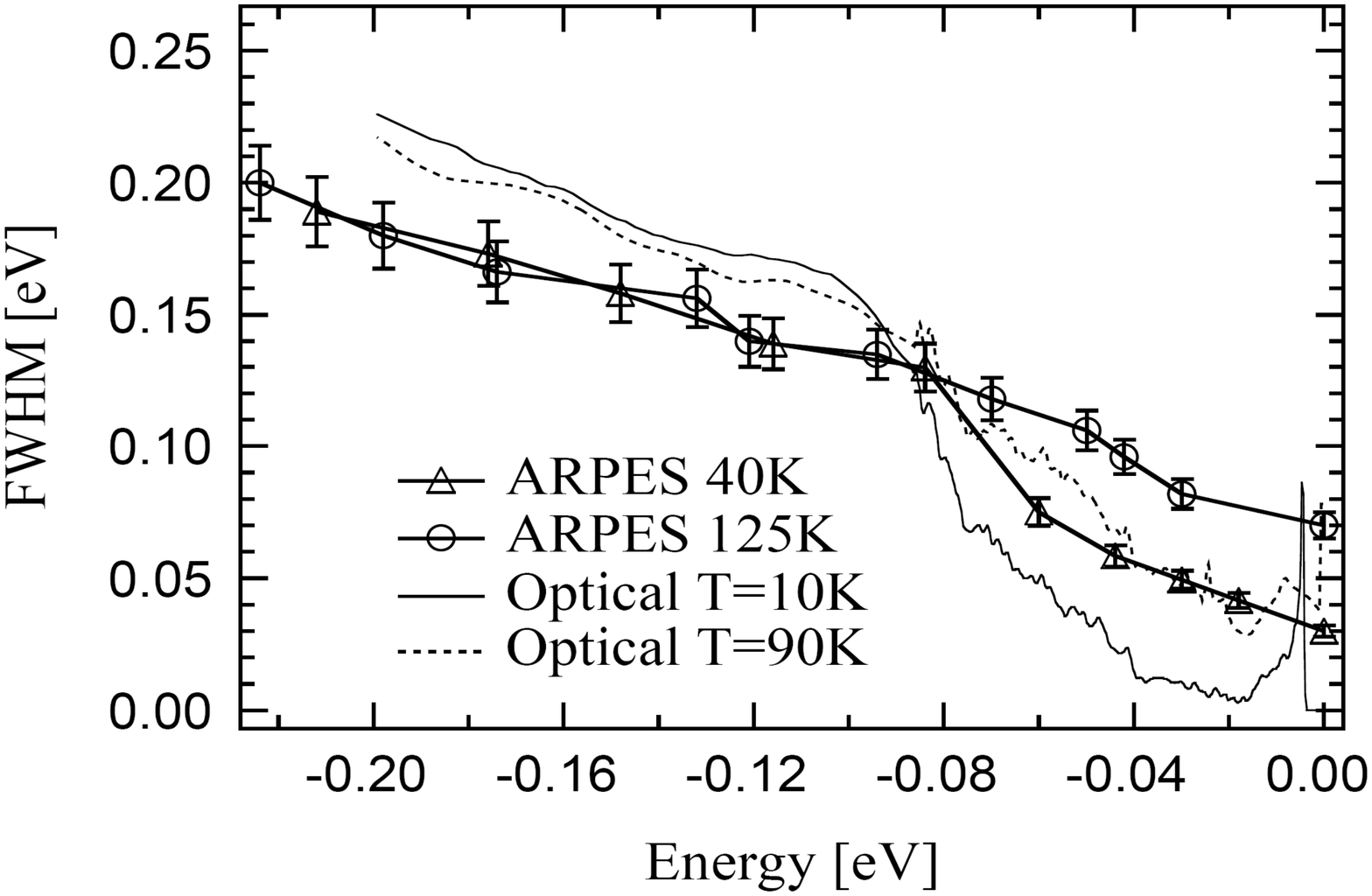}
         \vspace*{-1.0 cm}%
        \end{center}
\caption{Full width at half maximum of ARPES nodal spectral peaks versus the binding energy of the spectral peak (symbols), and the carrier scattering rate versus energy for Bi2212 ($T_c = 90$K) obtained from infrared reflectivity measurements (solid and dashed lines) [17]. The FWHM is defined by the horizontal arrow in the right panel of figure  1. After Kaminski \etal (2000)}
\label{fgConclusion1}
\end{figure}

The experiments presented in this review, primarily ARPES and optical conductivity, of the hole doped high temperature superconductors, lead us to the strong conclusion that the mobile charge carriers are scattered by a spectrum of bosonic excitations.  Although these two techniques are quite different they yield similar results. Confirming tunnelling and Raman experiments have also been published.  An early comparison between ARPES and optics was made by Kaminsky \etal (2000). As figure \ref{fgConclusion1} shows the quasiparticle lifetime, as measured by the ARPES MDC widths, tracks in both magnitude and structure the free carrier life time as determined from the optical conductivity in the same material (Bi-2212). This convergence of ARPES and optical self energies have been confirmed in many additional experiments as we have shown in this review. For example, figure \ref{fgOptics13} shows this for the real part of the quasiparticle self energy. But such comparisons are of limited use since, on theoretical grounds, one expects to see somewhat different results for the two techniques (Hwang et al. 2007a). First ARPES is momentum resolved whereas the optical self energy is a weighted average over the Fermi surface.  In both cases there are momentum dependent matrix elements that have to be taken into account.  More accurate experiments may identify these differences in the future.

Nevertheless, there is a remarkably uniform bosonic spectrum that emerges from the various experiments. At low temperature the spectrum has a peak whose frequency varies from 15 to 75 meV depending on the material and doping level. This frequency is generally proportional to the superconducting transition temperature $T_c$  as shown in figure \ref{fgOptics19}, but is totally absent in some materials with very low transition temperatures. The amplitude of the peak is strongly dependent on both temperature and doping.  It is strongest at low temperatures in the superconducting state but weakens at higher temperatures. In some cases, such as the optimally doped YBCO, it vanishes at $T_c$ but in other cases persists into the normal state such as underdoped YBCO. In addition to the peak there is a continuum in all the samples extending to 200 to 400 meV as shown in figure \ref{fgARPES14}.  The continuum seems to be fairly temperature independent but develops a gap at low temperature. The peak appears to be in the middle of this gap as shown in figure  \ref{fgOptics14} and \ref{fgOptics17}.

We next address the question of the origin of the bosonic spectrum.  The main candidates for this are spin fluctuations and phonons. While there seems to be a consensus among the ARPES groups that the spectrum is magnetic, arguments are made from time to time in favour of phonons.  The case for the spin fluctuation scenario is strong. First the bosonic function appears to change with temperature, the peak weakens as the temperature is increased and shifts in frequency to higher values.  As figure \ref{fgOptics19} shows the peak frequency is systematically related to the transition temperature of the superconductor.  Further, the spectrum of fluctuations extends to energies of the order of 300 meV. None of these properties are expected for a phonon.  Finally, where magnetic neutron spectra are available, the peak in the q-averaged susceptibility is close to the peak extracted from ARPES or optical conductivity spectra.

However, there remain compelling arguments for some role of phonons.  In particular, as recent high resolution ARPES data for nodal quasi particles show there is an oxygen isotope effect on the low frequency kink which is associated with the peak in the bosonic spectrum.  This effect could be understood in terms of a phonon contribution at the 10 percent level to the self energy of at least the nodal quasiparticles.

In summary, in our review of bosonic excitations in high temperature superconductors, we have identified these excitations as spin fluctuations. They are  responsible for the kinks in the ARPES dispersion curves and the spectral features in the optical conductivity.  Their contribution is strong enough to account for the superconducting transition temperature $T_c$.  But there remain several questions that need to be answered before we can claim to fully understand high temperature superconductivity.  What determines the nature of the magnetic fluctuation spectrum? For example, why is it so different in LSCO as compared to the three layer mercury material?  What is the role of the pseudogap?  Is it a coincidence that near the doping level where the pseudogap vanishes, $p \approx$ 0.19 -0.23, the peak in the magnetic spectrum is replaced by a uniform featureless background? We look to future experiments, particularly magnetic neutron scattering to address some of these issues.

\section{Acknowledgements}
We befitted from comments and discussions with many colleagues but in particular Dimitri Basov, Sergey Borisenko, Andrea Damascelli, Martin Greven, Dirk van der Marel and Ewald Schachinger. This work was supported in part by the Natural Science and Engineering Research Council and the Canadian Institute for Advanced Research. J.H. acknowledges financial support from the National Research Foundation of Korea (NRFK Grant No. 20100008552).

\eject

\end{document}